\documentclass[reprint,nofootinbib,superscriptaddress,amsmath,amssymb,aps,prd]{revtex4-1}
\usepackage{amsmath,mathrsfs,amsbsy,color,graphicx,bm,amsthm,amsfonts}
\usepackage{bbm}
\usepackage{times}
\usepackage{units}
\usepackage{dcolumn}
\usepackage{graphicx}
\usepackage{epsfig}
\usepackage{epstopdf}
\usepackage[colorlinks,linkcolor=red,anchorcolor=green,citecolor=blue,CJKbookmarks=True]{hyperref}
\DeclareMathSymbol{\shortminus}{\mathbin}{AMSa}{"39}
\usepackage{mathrsfs}
\usepackage{braket}
\usepackage{amssymb}
\usepackage{txfonts}
\usepackage{float}
\usepackage{enumitem}
\usepackage{multirow}
\usepackage{orcidlink}

\newcommand{\pp}{\partial}

\newcommand{\meq}[1]{(\ref{#1})}
\begin{document}

\title{
Quasinormal modes of scalar, electromagnetic, and gravitational perturbations in slowly rotating Kalb-Ramond black holes
}
\author{Weike Deng\orcidlink{0009-0002-5504-8151}}
\email[]{wkdeng@hnit.edu.cn} 
\affiliation{School of Science, Hunan Institute of Technology, Hengyang 421002, P. R. China}
\affiliation{Department of Physics, Key Laboratory of Low Dimensional Quantum Structures and Quantum Control of Ministry of Education, and Synergetic Innovation Center for Quantum Effects and Applications, Hunan Normal
University, Changsha, Hunan 410081, P. R. China}

\author{Wentao Liu\orcidlink{0009-0008-9257-8155}}
\email[]{wentaoliu@hunnu.edu.cn (Corresponding authors)} 
\affiliation{Lanzhou Center for Theoretical Physics, Key Laboratory of Theoretical Physics of Gansu Province, Key Laboratory of Quantum Theory and Applications of MoE, Gansu Provincial Research Center for Basic Disciplines of Quantum Physics, Lanzhou University, Lanzhou 730000, China}

\author{Kui Xiao\orcidlink{0000-0001-8429-8083}}
\email[]{xiaokui@hnit.edu.cn} 
\affiliation{School of Science, Hunan Institute of Technology, Hengyang 421002, P. R. China}

\author{Jiliang Jing\orcidlink{0000-0002-2803-7900}}
\email[]{jljing@hunnu.edu.cn (Corresponding authors)} 
\affiliation{Department of Physics, Key Laboratory of Low Dimensional Quantum Structures and Quantum Control of Ministry of Education, and Synergetic Innovation Center for Quantum Effects and Applications, Hunan Normal
University, Changsha, Hunan 410081, P. R. China}

\begin{abstract}

We investigate quasinormal modes (QNMs) of scalar, electromagnetic, and axial gravitational perturbations in slowly rotating Kalb–Ramond (KR) black holes, where an antisymmetric tensor field induces spontaneous Lorentz symmetry breaking. 
Working consistently to first order in the dimensionless spin parameter, we derive the corresponding master equations and compute the QNM spectrum using both the continued-fraction and matrix methods, finding excellent agreement. 
Lorentz violation modifies the oscillation and damping rates in a unified manner across all perturbative sectors: the real part of the QNM frequency increases monotonically with the Lorentz-violating parameter $\ell$, while the imaginary part becomes more negative. 
Axial gravitational modes exhibit the strongest response, revealing an intrinsic theoretical bound $\ell< 0.5$, beyond which the spectrum approaches an extremal behavior. 
Our results highlight the potential of gravitational-wave spectroscopy to probe Lorentz-violating signatures in KR gravity.

\end{abstract}

\maketitle

\section{Introduction}

The recent detection of the GW250114 event by the LVK Collaboration \cite{Akyuz:2025seg}, with a network matched-filter signal-to-noise ratio of nearly 80, represents the loudest binary black hole coalescence observed to date.
Such an exceptionally strong signal provides an unprecedented opportunity to probe gravitational theory in the highly dynamical regime and, in particular, to access the black hole quasinormal modes (QNMs) imprinted in the post merger ringdown phase \cite{LIGOScientific:2025rid}. 
The unusually high signal quality of GW250114 allows multiple QNMs to be extracted with statistical significance, demonstrating that QNM spectroscopy is not only theoretically essential but also observationally feasible with current gravitational-wave detectors \cite{Berti:2025hly,Cardoso:2025npr,DellaRocca:2025xwz,Berti2009,Konoplya:2011qq,Bian:2025ifp}.
Nowadays, QNM spectrum measurements play an increasingly central role in probing the nature of strong gravity, constraining deviations from general relativity, and exploring possible imprints of new fields or symmetry-breaking effects, such as those predicted in Lorentz-violating or modified gravity frameworks \cite{Allahyari:2025sbt,AouladLafkih:2025stw,Zhang:2024csc,Zhang:2025eqz,Zi:2025qos,Zi:2025lio,Deng:2025wzz,Jing:2025utt,Deng:2025hfn,Tan:2024qij,Tan:2024aym,Deng:2024ayh,Long:2024axi,Long:2023vph,Zhang:2024svj}.

Considering Lorentz-violating effects is important because breaking Lorentz symmetry provides a powerful probe of possible deviations from general relativity and offers insight into the fundamental nature of gravity \cite{Kostelecky1991,Casana2018,Ovgun2019,Gullu2020,Pan:2020zbl,Liu:2022xse,Maluf2021,Xu:2022frb,Ding2022,Poulis:2021nqh,Mai:2023ggs,Xu:2023xqh,Zhang:2023wwk,Lin:2023foj,Chen:2023cjd,Chen2020,Wang:2021gtd,Mai:2024lgk,Liang:2022gdk,Hosseinifar:2024wwe,Finke:2024ada,Liu:2024axg,Li:2025itp,Liu:2025oho,Gu:2025lyz,Lai:2025nyo,Li:2025tcd,Chen:2025ypx,Sekhmani:2025zen,Shi:2025tvu,AraujoFilho:2025zaj}.
By investigating the low-energy contributions of Lorentz symmetry breaking (LSB)  and analyzing the black hole QNM spectrum, especially the gravitational perturbation modes, we can explore the possible presence of LSB in spacetime and compare it with the predictions of general relativity.
A representative realization of Lorentz-symmetry breaking is provided by the Kalb–Ramond (KR) gravity theory, in which gravity is nonminimally coupled to an antisymmetric tensor field originating from the bosonic sector of string theory \cite{Kalb:1974yc,Kao:1996ea}.
In recent years, substantial effort has been devoted to uncovering exact black hole solutions in this framework \cite{Chakraborty:2014fva,Maluf:2021ywn,Duan:2023gng,Yang:2023wtu,Liu:2024oas,Liu:2025fxj,AraujoFilho:2025jcu}.
A major milestone was achieved by Yang \textit{et al.} \cite{Yang:2023wtu}, who first derived the correct Schwarzschild-like geometry supported by the KR field, and subsequently extended it to Schwarzschild–(A)dS configurations once the vacuum constraints were relaxed.
Following these advances, the physical and observational properties of KR black holes have been extensively investigated, spanning topics from stability and quasinormal modes to thermodynamics and astrophysical implications \cite{Guo:2023nkd,Filho:2023ycx,AraujoFilho:2024rcr,Jha:2024xtr,Junior:2024ety,Junior:2024vdk,Filho:2024kbq,Du:2024uhd,Filho:2024tgy}.

However, all of the aforementioned studies were carried out under the assumption of static and spherical symmetry.
Given that real astrophysical compact objects inevitably possess spin, it becomes essential to account for the influence of angular momentum on these results.
Recently, a slowly rotating KR black hole solution incorporating angular momentum was constructed \cite{Liu:2024lve}.
Yet, observational evidence from the LIGO-Virgo Collaboration shows that most binary black hole mergers detected in the \(O1\)-\(O2\) runs involve systems with low effective spins \cite{Deng:2024ayh}. 
A reanalysis of ten events by Roulet and Zaldarriaga \cite{Roulet:2018jbe}  further constrains typical spin magnitudes to \(a/M\!\lesssim\! 0.4\) for isotropic orientations and to \(a/M\!\sim\! 0.1\) for aligned configurations. 
The asymmetric GW190814 merger places an even stronger bound, limiting the primary black hole to \(a/M\!\lesssim\! 0.07\) \cite{LIGOScientific:2020zkf}.
These observational constraints align with perturbative results showing that the slow-rotation approximation remains accurate to within \(1\%\) in both the real and imaginary parts of Kerr QNMs for \(a/M\leq 0.3\) 
\cite{Pani2013IJMPA,Pani2011,Pani2012,Pani2013prd,Pani2012prl,Pani2013prl,Liu:2022dcn}. 
Collectively, these findings establish the slow-rotation regime as the physically relevant domain for modeling astrophysical black holes. 
Therefore, in the absence of an exact rotating KR black hole solution, exploring its rotational QNM properties within a controlled approximation provides a highly feasible and well-motivated approach.

In this manuscript, we aim to explore the impact of spontaneous Lorentz violation on the QNMs of  scalar, electromagnetic, and gravitational perturbations within this theoretical framework.
The rest of this paper is organized as follows.
In Sec.~\ref{Sec.2}, we present a concise overview of the spontaneously Lorentz-violating tensor field, the KR field, and outline the corresponding Einstein–Kalb–Ramond gravitational field equations.
In Sec. \ref{Sec.3}, using the slow-rotation approximation, we derive the master equations for scalar, electromagnetic, and axial gravitational perturbations.
In Sec. \ref{Sec.4}, we compute the eigenfrequencies of these perturbation equations using two independent numerical methods, the continued fraction method and the matrix method, and verify the convergence of our results.
Finally, Sec. \ref{Sec.5} summarizes our main conclusions and outlines possible directions for future research.

\section{Slowly rotating black holes in Kalb-Ramond gravity theory}\label{Sec.2}
Let us begin with an Einstein-Hilbert action in which gravity is nonminimally coupled to a self-interacting KR two-form field \cite{Altschul:2009ae}.
The action reads
\begin{equation}
\begin{aligned}\label{action}
\mathcal{S}=&\frac{1}{2\kappaup}\int d^4x\sqrt{-g}\bigg[R-2\Lambda-\frac{1}{6}H^{abc}H_{abc}-V(B^{ab}B_{ab})\\
&+\xi_2B^{ca}B^{b}{}_{a}R_{cb}+\xi_3B^{ab}B_{ab}R\bigg]+\int d^4x\sqrt{-g}\mathcal{L}_\text{M},
\end{aligned}
\end{equation}
where $\kappa=8\pi G_N$ denotes the gravitational coupling.
The constants $\xi_2$ and $\xi_3$ parameterize the nonminimal interaction between curvature and the KR field, and $\Lambda$ is the cosmological constant.
It should be emphasized that the present manuscript concentrates on the KR effects associated with the $\xi_2$ coupling; accordingly, we take $\xi_3=0$ in all subsequent computations.
Readers interested in the role of the $\xi_3$ term may consult Liu et al \cite{Liu:2025fxj}.
The antisymmetric field strength is defined as
\begin{align*}
H_{abc}\equiv\partial_aB_{bc}+\partial_b B_{ca}+\partial_c B_{ab}.
\end{align*}
The potential $V$ is chosen such that the KR tensor acquires a nontrivial VEV and vanishes at its minimum, thereby inducing spontaneous Lorentz symmetry breaking.

Variation of the action \eqref{action} with respect to $B_{ab}$ yields the KR field equation, leading to $ \Pi_{ab}=0 $, where
\begin{align}\label{EQPi}
\Pi_{ab} = \nabla^c H_{abc} + 3 \xi_2 R_{ca} B^c_{~b} - 3 \xi_2 R_{cb} B^c_{~a} - 6 V' B_{ab},
\end{align}
and varying with respect to the metric leads to the gravitational field equations,
\begin{align}\label{eqg}
R_{ab}-\frac{1}{2}g_{ab}R+\Lambda g_{ab}=T^\text{KR}_{ab}+T^\text{M}_{ab}.
\end{align}
Here, the KR contribution to the stress-energy tensor is
\begin{equation}
\begin{aligned}
\!\!\! T^\text{KR}_{ab}\!=&\frac{1}{2}H_{acd}H_{b}{}^{cd} \!-\! \frac{1}{\!12\!}g_{ab}H^{cde}H_{cde}\!+\! 2V'(\!X\!)B_{ca}B^{c}{}_{b} \!-\! g_{ab}\!V(\!X\!)\\ &+\xi_2\bigg[\frac{1}{2}g_{ab}B^{ce}B^{d}{}_{e}R_{cd} -B^{c}{}_{a}B^{d}{}_{b}R_{cd} -B^{cd}B_{bd}R_{ac}\\ &-B^{cd}B_{ad}R_{bc} +\frac{1}{2}\nabla_{c}\nabla_{a}(B^{cd}B_{bd}) +\frac{1}{2}\nabla_{c}\nabla_{b}(B^{cd}B_{ad})\\ &-\frac{1}{2}\nabla^c\nabla_c(B_{a}{}^{d}B_{bd}) -\frac{1}{2}g_{ab}\nabla_{c}\nabla_{d}(B^{ce}B^{d}{}_{e})\bigg],
\end{aligned}
\end{equation}
and $T^{\rm M}_{ab}$ denotes the stress tensor of the matter sector, and primes indicate differentiation with respect to $X$.

Motivated by the gravitational sector of the Standard-Model Extension, we introduce a self-interacting potential for the KR field such that it acquires a nonvanishing VEV,
\begin{equation}
\langle B_{ab}\rangle =\beta_{ab},
\end{equation}
as proposed in Ref. \cite{Altschul:2009ae}.
To implement this mechanism, we take the potential to depend on the invariant combination
\begin{equation}
V=V(B_{ab}B^{ab}\pm b^2),
\end{equation}
where the choice of sign ensures that $b^2$ is a positive constant.
At the minimum of the potential, the VEV satisfies the fixed-norm condition
\begin{equation}
\beta^{ab}\beta_{ab}=\mp b^2,
\end{equation}
so that the KR field develops a constant background configuration and spontaneously breaks Lorentz symmetry through its self-interaction \cite{Lessa:2019bgi}.

In the vacuum, i.e., for $B_{ab}B^{ab} = \beta_{ab}\beta^{ab}$, we adopt the following 2-form ansatz for the background field \cite{Lessa:2019bgi}:
\begin{align}
\beta_2=-\tilde{E}(r)dt \wedge dr.
\end{align}
This implies that the only nonvanishing components of $\beta_{ab}$ in the VEV are
\begin{equation}
\beta_{rt}=-\beta_{tr}=\tilde{E}(r).
\end{equation}
Equivalently, in matrix form one has
\begin{equation}
\beta_{ab}=\left(
\begin{array}{cccc}
0 & -\tilde{E}(r) & 0 & 0 \\
\tilde{E}(r) & 0 & 0 & 0 \\
0 & 0 & 0 & 0 \\
0 & 0 & 0 & 0
\end{array}\right),
\end{equation}
which corresponds to a pseudoelectric-type vacuum configuration.
For this choice, the associated KR field strength vanishes identically, $ H_{abc}=0 $ or $ H_3=d\beta_2=0 $, as discussed in Ref. \cite{Yang:2023wtu}.
Then, we can define the efficient gravitational field equation, satisfying $\mathcal{G}_{ab}=0$, as follows:
\begin{equation}\label{EQG}
\mathcal{G}_{ab}=R_{ab}-\Lambda g_{ab}+Vg_{ab}-\xi_2\mathcal{B}_{ab}-(2\beta_{ac} \beta_{b}{}^{c}+b^2g_{ab})V',
\end{equation}
where $\mathcal{B}_{ab}$ denotes the derivative coupling between the curvature and the antisymmetric KR field.
The explicit tensorial form of $\mathcal{B}_{ab}$ and the associated covariant derivative structures are lengthy, and for completeness we present them in Appendix \ref{AppA}.

By solving Eqs. \eqref{EQPi} and \eqref{EQG} simultaneously and expanding both the metric and the field equations to first order in the dimensionless spin parameter $\tilde{a}=a/M$, we obtain a black hole solution describing a slowly rotating KR spacetime, hereafter referred to as the slowly rotating KR black hole \cite{Liu:2024lve}.
The corresponding line element reads
\begin{equation}\label{caseB}
\begin{aligned}
ds^2=&-f(r)dt^2+\frac{1}{f(r)}dr^2+r^2d\theta^2+r^2\sin^2\theta d\varphi^2\\
&-2\tilde{a}M\left(\frac{2M}{r}+\frac{\Lambda_e}{3}r^2\right)\sin^2\theta dtd\varphi+\mathcal{O}(\tilde{a}^2).
\end{aligned}
\end{equation}
Here, the metric function takes the form
\begin{align}
f(r)=\frac{1}{1-\ell}-\frac{2M}{r}-\frac{\Lambda_e}{3}r^2,
\end{align}
where $\ell$ denotes the Lorentz-violation parameter and $\Lambda_e$ represents the effective cosmological constant.
Their explicit definitions are
\begin{align}
\ell\equiv \xi_2 b^2/2,\quad\quad   \Lambda_e\equiv\lambda/\xi_2.
\end{align}
The $\lambda$ is interpreted as a Lagrange multiplier field \cite{Bluhm:2007bd}, originating from the potential of the KR field \cite{Liu:2024lve}.
Although $\lambda$ acts as a cosmological constant like contribution \cite{Liu:2024lve}, our analysis focuses on the $\Lambda = 0$ sector, which in turn requires the Lagrange-multiplier field to vanish, which in turn implies $\Lambda_e=\lambda=\Lambda=0$.
This corresponds to the asymptotically flat, slowly rotating vacuum case, in which the spacetime admits a single event horizon located at $ r_h=2(1-\ell)M $.
When the Lorentz violation parameter $ \ell $ vanishes, the solutions reduce to the slowly rotating Kerr black hole cases \cite{Pani2012}; when $\tilde{a}=0$, they reduce to the static neutral KR black hole cases \cite{Liu:2024oas,Yang:2023wtu}.

\section{PERTURBATIONS OF SLOWLY ROTATION KALB-RAMOND BLACK HOLEs}\label{Sec.3}

To explore the dynamical response and stability of the slowly rotating KR black hole, we consider small perturbations of both test fields and the spacetime geometry itself.
Scalar and electromagnetic perturbations propagate as test fields on the fixed background, whereas gravitational perturbations are treated as linear fluctuations of the metric.
The evolution of these perturbations encodes how the spacetime reacts to external disturbances and determines the characteristic oscillation spectrum, namely the QNMs.
These modes play a crucial role in assessing the stability of the black hole and provide potential observational signatures that may distinguish KR induced Lorentz-violating effects from general relativity in the ringdown regime.
In the following, we derive the master equations governing scalar, electromagnetic, and gravitational perturbations in the slowly rotating KR spacetime.

\subsection{Scalar perturbation}
Considering that the scalar field coupling to the KR field is neglected, then the massive Klein-Gordon equation reads
\begin{equation}
\nabla^a\nabla_a\phi=0,
\end{equation}
we express the field as a sum of spherical harmonics:
\begin{equation}
\phi=\sum_{lm}\frac{1}{r}\psi_s(r)e^{-i\omega t}Y^{lm}(\theta,\varphi),
\end{equation}
and perform a first-order expansion of $ \tilde{a} $.
Subsequently, by utilizing the eigenvalue equation for the angular part:
\begin{equation}\label{eq177}
\left[\frac{1}{\sin\theta}\frac{\pp}{\pp\theta}\left(\sin\theta \frac{\pp}{\pp \theta} \right)
+\frac{1}{\sin^2\theta}\frac{\pp^2}{\pp \varphi^2}\right]Y^{lm}=l(l+1)Y^{lm},
\end{equation}
the scalar equation is fully decoupled, and it can be cast in the form:
\begin{equation}\label{meq1}
\left[\frac{d^2}{dr^2_*}+(1-\ell)^2\omega^2-V_s\right]\psi_s(r)=0,
\end{equation}
where the effective potential is given by
\begin{equation}
\!\!V_s\!=\! F(r)\bigg[\frac{2(1\!-\! \ell)M}{r^3}\!+\!\frac{l(l+1)(1\!-\!\ell)}{r^2} \bigg]\!+\!\frac{4\tilde{a}\omega m M^2(1\!-\!\ell)^2}{r^3},
\end{equation}
with $F(r)\!\equiv\! 1-2(1\!-\!\ell)M/r$ and $r_*$ is the tortoise coordinate, defined through
\begin{equation}
r_*=\int \frac{1}{F(r)}dr= r + 2M(1 - \ell) \ln \bigg[ \frac{r}{2M(1 - \ell)} - 1 \bigg].
\end{equation}
Here, the couplings between modes with angular indices $ l\pm1 $ vanish for a simple reason.
In general, a polar Klein-Gordon field would couple, at first order in rotation, to axial perturbations with $ l\pm1 $, following a selection rule analogous to the Laporte rule \cite{Pani2013IJMPA}.
In the scalar case, however, no axial sector exists, and therefore the corresponding $l\pm1$ couplings identically disappear.

\subsection{Electromagnetic perturbation}

Electromagnetic perturbations on the slowly rotating KR background satisfy the Maxwell equations on a fixed curved spacetime.
Since the background KR configuration does not carry an electromagnetic field, the Maxwell sector remains decoupled from the scalar and gravitational perturbations, and therefore evolves solely according to the geometry of the slowly rotating KR black hole.
For Maxwell equations
\begin{equation}
\nabla_a\mathcal{F}^{ab}=0,
\end{equation}
where $\mathcal{F}_{ab}=2\nabla_{[a}A_{b]}$ and $A^a$ is the electromagnetic four-potential, we linearize the vector potential by writing
\begin{equation}
A_a=A^{(0)}_a+\delta A^{(0)}_a.
\end{equation}
Here, $A^{(0)}_a$ denotes the background electromagnetic four-potential, and $\delta A_a$ represents the electromagnetic perturbation, which can be expanded in terms of vector spherical harmonics as follows:
\begin{align}\label{deltavector}
\delta A^{(0)}_{a}(t,r,\theta ,\varphi)=
\left[
\begin{array}{c}
0 \\ 0  \\ u^{lm}_{(4)} \mathbf{S}^{lm}_b/\lambda
\end{array}
\right]
+\left[
\begin{array}{c}
u^{lm}_{(1)}Y^{lm} /r\\u^{lm}_{(2)}Y^{lm}/\left(rf\right)  \\ u^{lm}_{(3)} \mathbf{Y}^{lm}_b/\lambda
\end{array}
\right],
\end{align}
where $\lambda=l(l+1)$, and the index $b$ refers to the angular components $b=(\theta, \varphi)$.
The functions
\begin{equation}
\mathbf{S}^{lm}_{\theta}\equiv-\csc\theta \frac{\pp}{\pp\varphi }Y^{lm},\quad\quad
\mathbf{S}^{lm}_{\varphi}\equiv\sin\theta \frac{\pp}{\pp \theta}Y^{lm},
\end{equation}
are the axial  vector spherical harmonics, which acquire a factor $(-1)^{l+1}$ under the parity transformation $(\theta,\varphi)\rightarrow(\pi-\theta,\pi+\varphi)$.
In contrast, $\mathbf{Y}^{lm}_{b}\equiv\frac{\partial}{\partial b}Y^{lm}$ denotes the polar vector spherical harmonics, transforming as $(-1)^{l}$ under parity.
Since electromagnetic fields have no mass, the axial and polar sectors remain isospectral.
For the axial sector, after eliminating the angular couplings by means of equation \eqref{eq177}, the perturbation equation can be simplified to
\begin{equation}
\left[\frac{d^2}{dr^2_*}+(1-\ell)^2\omega^2-V_e\right] \psi_e(r)=0,
\end{equation}
where the effective potential is given by
\begin{equation}\label{meq2}
\!V_e= F(r)\bigg[\frac{l(l+1)(1\!-\!\ell)}{r^2} \bigg]+\frac{4\tilde{a}\omega m M^2(1\!-\!\ell)^2}{r^3}.
\end{equation}
Note that the variable is defined as $u_{(4)}(t,r)=e^{-i\omega t}\psi_e(r)$.

\subsection{Gravitational perturbation}

Before turning to the metric perturbations, let us comment on the role of fluctuations of the KR field.
In constructing the background solution, we have explicitly checked that the nontrivial components of the KR field equation $\Pi_{ab}=0$ are not independent from the modified Einstein equations: once the gravitational field equations are satisfied, the KR equations are automatically fulfilled.
In other words, at the background level the KR sector does not introduce additional independent constraints beyond those already encoded in the effective gravitational equations.
Motivated by this redundancy and by our main goal of characterizing the gravitational QNM spectrum of the slowly rotating KR black hole, we adopt a simplified perturbative scheme in which the KR 2-form is kept fixed at its vacuum configuration $\beta_{ab}$, and only the metric is perturbed. 
This approximation should be viewed as a first step towards understanding the gravitational response of the slowly rotating KR black hole; a more complete analysis including coupled KR metric perturbations is left for future work.

In the framework of linearized gravity, the spacetime metric $g_{ab}$ is decomposed into a background part $g^{(0)}_{ab}$ and a small perturbation $h_{ab}$,
\begin{align}\label{eq27}
g_{\mu\nu} = g^{(0)}_{ab} + h_{ab}, \quad\quad h_{ab}=\delta g_{ab}^{(0)}.
\end{align}
In the slowly rotating KR background, following the approach of Pani et al. \cite{Pani2011,Pani2012,Pani2013prd,Pani2012prl,Pani2013prl}, we assume that the rotation is sufficiently small so that the underlying spherical symmetry is only mildly broken.
Under this approximation, the metric perturbations can still be expanded in tensor spherical harmonics, as in the spherically symmetric case.
Adopting the Regge-Wheeler notation and imposing the Regge-Wheeler gauge to eliminate four nonphysical degrees of freedom \cite{ReggeWheeler1957,Zerilli1970PRD,Liu:2022csl}, the perturbation can be written as
\begin{equation}
h_{ab} = \left(\begin{array}{cccc}
H_0Y^{lm} & H_1 Y^{lm} & h_0 S_\theta^{lm} & h_0 S_{\varphi}^{lm} \\
H_1 Y^{lm} & H_2 Y^{lm} & h_1 S_\theta^{lm} & h_1 S_{\varphi}^{lm} \\
h_0 S_\theta^{lm} & h_1 S_{\theta}^{lm} & r^2 K Y^{lm} & 0 \\
h_0 S_\varphi^{lm} & h_1 S_{\varphi}^{lm} & 0 & r^2 \sin^2\theta KY^{lm}
\end{array}\right)e^{-i\omega t},
\end{equation}
where all perturbation functions $h_{0,1}(r), H_{0,1,2}(r)$ and $K(r)$ depend solely on the radial coordinate.
The angular structure is contained in the spherical harmonics $Y^{lm}(\theta,\varphi)$ together with 
\begin{align}
S_{\theta}^{lm} = -\frac{1}{\sin \theta} \frac{\partial}{\partial \theta} Y^{lm}, \quad
S_{\varphi}^{lm} = \sin \theta \frac{\partial}{\partial \varphi} Y^{lm}.
\end{align}
Substituting Eq. \eqref{eq27} into the gravitational field equation \eqref{EQG}, we obtain the ten independent components of the tensor $\mathcal{G}_{ab}$, in which the radial and angular dependencies remain coupled.

To obtain a purely radial set of perturbation equations, the angular dependence must be separated out.
This can be achieved by following the method developed by Kojima \cite{Kojima}.
The structure of the linearized gravitational field equations allows the ten independent components of $\delta\mathcal{G}_{ab}(h)\!=\!0$ to be organized into three groups.
The first group consists of the $(t,t),(t,r),(r,r) $, and $ (\theta,\theta)+(\varphi,\varphi)/\sin^2\theta $ components, and these can be written in the generic form
\begin{equation}
\begin{aligned}\label{FCZ1}
\delta \mathcal{E}_{(I)}\equiv&\left(A^{(I)}_{l}+\tilde{A}^{(I)}_{l}\cos\theta\right)Y^{lm}\\
&+B^{(I)}_{l}\sin\theta\frac{\pp}{\pp \theta}Y^{lm}+C^{(I)}_l\frac{\pp }{\pp \varphi}Y^{lm}=0,
\end{aligned}
\end{equation}
where $I\!=\!0,1,2,3$ corresponds respectively to $\delta\mathcal{G}_{tt}\!=\!0$, $\delta\mathcal{G}_{tr}\!=\!0$, $\delta\mathcal{G}_{rr}\!=\!0$ and $\delta\mathcal{G}_{\theta\theta}+\delta\mathcal{G}_{\varphi\varphi}/\sin^2\theta\!=\!0$.
The second group consists of the $(t\theta)$, $(r\theta)$, $(t\varphi)$ and $(r\varphi)$ components, and these equations take the form
\begin{align}\label{FCZ21}
\delta \mathcal{E}_{(L\theta)}\! \equiv
&\left(\alpha^{(L)}_l\!+\!\tilde{\alpha}^{(L)}_l\cos\theta\right)\frac{\pp}{\pp \theta}Y^{lm}
\!-\!\left(\beta^{(L)}_l\!+\! \tilde{\beta}^{(L)}_l\cos\theta\right)\frac{\pp }{\sin\theta \pp \varphi}\!Y^{lm} \nonumber \\
&+\eta^{(L)}_l\sin\theta Y^{lm}+\xi^{(L)}_lX^{lm}+\chi^{(L)}_l\sin\theta W^{lm}=0, 
\end{align}
\begin{align}\label{FCZ22}
\delta \mathcal{E}_{(L\varphi)}\!\equiv
&\left(\beta^{(L)}_l\!+\!\tilde{\beta}^{(L)}_l\cos\theta\right)\frac{\pp}{\pp \theta}Y^{lm}
\!+\!\left(\alpha^{(L)}_l\!+\!\tilde{\alpha}^{(L)}_l\cos\theta\right)\frac{\pp }{\sin\theta \pp \varphi}\!Y^{lm} \nonumber \\
&+\zeta^{(L)}_l\sin\theta Y^{lm}+\chi^{(L)}_lX^{lm}-\xi^{(L)}_l\sin\theta W^{lm}=0,
\end{align}
where $L\!=\! \{ t,r \}$.
The first of these corresponds to $\delta\mathcal{G}_{t\theta}\!=\!0$ and $\delta\mathcal{G}_{r\theta}\!=\!0$, while the second originates from $\delta\mathcal{G}_{t\varphi}\!=\!0$ and $\delta\mathcal{G}_{r\varphi}\!=\!0$.
Finally, the $ (\theta,\varphi) $ and $ (\theta,\theta)-(\varphi,\varphi)/\sin^2\theta $ components form the third group, and these can be written as
\begin{equation}\label{FCZ31}
\!\! \delta \mathcal{E}_{(\theta\varphi)}\equiv f_l\sin\theta \frac{\pp }{\pp \theta}Y^{lm}+g_l\frac{\pp }{\pp \varphi}Y^{lm}+s_l\frac{X^{lm}}{\sin\theta}+t_lW^{lm}=0,
\end{equation}
\begin{equation}\label{FCZ32}
\delta \mathcal{E}_{(-)}\equiv g_l\sin\theta \frac{\pp }{\pp \theta}Y^{lm}-f_l\frac{\pp }{\pp \varphi}Y^{lm}-t_l\frac{X^{lm}}{\sin\theta}+s_lW^{lm}=0.
\end{equation}
The functions $ X^{lm} $ and $ W^{lm} $ are the standard tensor spherical harmonics:
\begin{equation}
\begin{aligned}
&X^{lm}=2\left( \pp_\theta \pp_\varphi Y^{lm}-\cot\theta \pp_\varphi Y^{lm} \right),\\
&W^{lm}=\pp_\theta \pp_\theta Y^{lm}-\cot\theta\pp_\theta Y^{lm}-\csc^2\theta \pp_\varphi \pp_\varphi Y^{lm} . 
\end{aligned}
\end{equation}  
All coefficients appearing in Eqs. \eqref{FCZ1}-\eqref{FCZ32} are purely radial functions, each linear in the perturbations.
These coefficients naturally split into axial and polar sectors:
\begin{align*}
&\text{axial coefficients}: \tilde{A}^{(I)}_l, B^{(I)}_l, \beta^{(L)}_l, \tilde{a}^{(L)}_l, \eta^{(L)}_l, \chi^{(L)}_l, t_l, g_l,\\
&\text{polar coefficients}: A^{(I)}_l, C^{I}_l, \alpha^{(L)}_l, \tilde{\beta}^{(L)}_l,\zeta^{(L)}_l,\xi^{(L)}_l, \zeta_l^{(L)}, s_l,f_l.
\end{align*}
The essential point is that each of these coefficients is purely radial.
Once this is established, all angular dependence in the linearized efficient gravitational equations becomes completely separable, allowing the perturbation equations to be reduced to a system of radial ordinary differential equations.

For slowly rotating stars, Kojima achieved the separation of angular dependence in the Einstein equations \cite{Kojima} and derived a set of important identities, now known as the Kojima identities, which are given by
\begin{equation}\label{Kojimaid}
\begin{aligned}
\cos\theta Y^{l}=&\mathcal{Q}_{l+1}Y^{l+1}+\mathcal{Q}_{l}Y^{l-1}, \\
\sin\theta \pp_\theta Y^{l}=&\mathcal{Q}_{l+1}lY^{l+1}-\mathcal{Q}_{l}(l+1)Y^{l-1}, \\
\cos^2\theta Y^{l}=&\left(\mathcal{Q}^2_{l+1}+\mathcal{Q}^2_{l}\right)Y^{l} \\
& +\mathcal{Q}_{l+1}\mathcal{Q}_{l+2}Y^{l+2} +\mathcal{Q}_{l}\mathcal{Q}_{l-1}Y^{l-2},\\
\cos\theta \sin\theta \pp_\theta Y^{l}=&\left[l\mathcal{Q}^2_{l+1}-(l+1)\mathcal{Q}^2_{l}\right]Y^{l}\\
&+\mathcal{Q}_{l+1}\mathcal{Q}_{l+2}lY^{l+2}\\
&-\mathcal{Q}_{l}\mathcal{Q}_{l-1}(l+1)Y^{l-2}.
\end{aligned}
\end{equation}
Here the azimuthal index $m$ has been omitted, and the coefficient $\mathcal{Q}_{l}$ is defined as
\begin{align}
\mathcal{Q}_{l} = \sqrt{\frac{l^2-m^2}{4l^2-1}},
\end{align}
These mathematical identities provide the essential toolkit for separating the angular dependence and form the basis of the subsequent perturbation analysis.
The method has later been extended to slowly rotating black holes, as discussed in Ref. \cite{Pani2011}.
In this work, we adopt the same technique.

By multiplying equation \eqref{FCZ1} by $Y^{lm}_*$ and integrating over the two-sphere, and by making use of the Kojima identities together with the orthogonality relations of the spherical harmonics, one obtains
\begin{equation}
0=\int d\Omega Y_{*}^{l'm'}\delta\mathcal{E}_{(I)}=A^{(I)}_l+im \mathcal{C}^{(I)}_l+\mathcal{L}^{\pm1}_0 \tilde{A}^{(I)}_l
+\mathcal{L}^{\pm1}_1 B^{(I)}_l.
\end{equation}
Similarly, the system of equations \eqref{FCZ21}-\eqref{FCZ22} can be decoupled by proceeding in an analogous manner:

\begin{equation}
\begin{aligned}
0=&\int d\Omega \bigg[\pp_\theta Y_*^{l'm'}\delta\mathcal{E}_{(L\theta)}+\csc\theta\pp_\varphi Y_*^{l'm'}\delta\mathcal{E}_{(L\varphi)} \bigg]\\
=&l(l+1)\alpha^{(L)}_l+im\left[ (l-1)(l+2)\xi^{(L)}_l-\tilde{\beta}^{(L)}_l-\zeta^{(L)}_l \right]\\
&+\mathcal{L}^{\pm1}_2\eta^{(L)}_l+\mathcal{L}^{\pm1}_3\tilde{\alpha}^{(L)}_l+\mathcal{L}^{\pm1}_4\chi^{(L)}_l,
\end{aligned}
\end{equation}
\begin{equation}
\begin{aligned}
0=&\int d\Omega \bigg[\pp_\theta Y_*^{l'm'}\delta\mathcal{E}_{(L\varphi)}+\csc\theta\pp_\varphi Y_*^{l'm'}\delta\mathcal{E}_{(L\theta)} \bigg]\\
=&l(l+1)\beta^{(L)}_l+im\left[ (l-1)(l+2)\chi^{(L)}_l-\tilde{\alpha}^{(L)}_l-\eta^{(L)}_l \right]\\
&+\mathcal{L}^{\pm1}_2\zeta^{(L)}_l+\mathcal{L}^{\pm1}_3\tilde{\beta}^{(L)}_l+\mathcal{L}^{\pm1}_4\xi^{(L)}_l.
\end{aligned}
\end{equation}
Finally, for the system of equations \eqref{FCZ31}-\eqref{FCZ32}, the angular dependence can be eliminated by constructing the following combinations:
\begin{align}
\begin{aligned}
0=&\int  \frac{d\Omega}{l(l+1)-2}\left[ W_*^{l'm'}\delta\mathcal{E}_{(-)}+\csc\theta X_*^{l'm'}\delta \mathcal{E}_{(\theta\varphi)} \right]\\
=&l(l+1)s_l-imf_l+\mathcal{L}^{\pm1}_2 g_l,
\end{aligned}
\\
\begin{aligned}
0&=\int  \frac{d\Omega}{l(l+1)-2}\left[ W_*^{l'm'}\delta\mathcal{E}_{(\theta\varphi)}+\csc\theta X_*^{l'm'}\delta \mathcal{E}_{(-)} \right]\\
&=l(l+1)t_l-img_l+\mathcal{L}^{\pm1}_2 f_l.
\end{aligned}
\end{align}
It is worth noting that the operators $\mathcal{L}^{\pm1}_{i}$ appearing in the above equations act on a generic function $A_{lm}$ according to the following definitions:
\begin{equation*}
\begin{aligned}
\mathcal{L}^{\pm1}_0A_{lm}\equiv& A_{l'm'}\int d\Omega Y_*^{lm}\cos\theta Y^{l'm'}
=Q_{lm}A_{l-1m}+Q_{l+1m}A_{l+1m},\\
\mathcal{L}^{\pm1}_1A_{lm}\equiv& A_{l'm'}\int d\Omega Y_*^{lm}\sin\theta \pp_\theta Y^{l'm'}\\
=&(l-1)Q_{lm}A_{l-1m}-(l+2)Q_{l+1m}A_{l+1m},\\
\mathcal{L}^{\pm1}_2A_{lm}\equiv& \left[\!-2\mathcal{L}^{\pm1}_0\!-\!\mathcal{L}^{\pm1}_1\!  \right]A_{lm}
\!=\!-(l+1)Q_{lm}A_{l-1m}\!+\! lQ_{l+1m}A_{l+1m},\\
\mathcal{L}^{\pm1}_3 A_{lm}\equiv& \left[l(l+1)\mathcal{L}^{\pm1}_0-\mathcal{L}^{\pm 1}_1  \right]A_{lm}\\
=&(l-1)(l+1)Q_{lm}A_{l-1m}+l(l+2)Q_{l+1m}A_{l+1m},\\
\mathcal{L}^{\pm1}_4A_{lm}\equiv&\left[ -2(l(l+1)-2)\mathcal{L}^{\pm1}_0+(l(l+1)+2)\mathcal{L}^{\pm1}_1 \right]A_{lm}\\
=&(l^2-1)(l-2)Q_{lm}A_{l-1m}-l(l+2)(l+3)Q_{l+1m}A_{l+1m}.
\end{aligned}
\end{equation*}
At first order in the slow-rotation parameter, the mixing of different angular functions in Eqs. \eqref{FCZ1}-\eqref{FCZ32} causes the radial perturbation equations to couple functions with multipoles $l$ and $l\pm1$.
As shown by Pani et al. \cite{Pani2012}, the terms involving the $l\pm1$ couplings do not affect the QNM spectrum at linear order in $\tilde{a}$.
For this reason, we shall neglect these terms in the following.

These perturbation equations naturally split into two decoupled sectors: the axial and polar modes.
In the axial sector, we obtain the following pair of coupled differential equations:
\begin{align}\label{BB1}
&\!\!\! 2i\tilde{a}mM^2\!\left[ i\omega(\lambda\!-\!2)(1\!-\!\ell)^3\!/\!F\!(r)h_0(r)\!+\!(\lambda\ell\!+\!2\ell\!-\!2)F(r)h_1'(r) \right]\nonumber \\
&\!\!\!-\!2i\tilde{a}mM^2\!/\!r^{2}\!\left[ r(1\!-\!2\ell)(6\!+\!\lambda)\!+\!2M(1\!-\!\ell)(14\ell\!+\!\lambda\ell\!-\!8) \right]\!h_1\!(r)\nonumber \\
&\!\!\!\lambda\left[ (\ell^2-2\ell+1)r\lambda-2r\ell^2-4M(1-\ell)(1-2\ell) \right]h_0(r)\\
&\!\!\!-r^2\lambda(1-2\ell)F(r)\left[ 2i\omega h_1(r)+r(i\omega h_1'(r)+h_0''(r)) \right]=0, \nonumber
\end{align}
and
\begin{equation}\label{BB2}
\begin{aligned}
&\!\!\! (\lambda\!-\!2)r^2F\!(r)h_1\!(r)\!-\!i\omega r^3(1\!-\!2\ell)[2h_0(r)\!-\!rh_0'(r)\!-\!i\omega r h_1\!(r)]\\
&\!\!\!+\!2i\tilde{a}mM^2(1\!-\!2\ell)[ (2\!-\!6/\lambda)h_0(r)\!-\!2i\omega rh_1\!(r)\!-\!rh_0'(r)  ]\!=\!0.
\end{aligned}
\end{equation}
To decouple the above equations, using equation \meq{BB1}, we first solve for $h_{0}(r)$ and then differentiate the result with respect to the radial coordinate. 
Substituting $h_{0}(r)$, $h_{0}’(r)$,  and $h_0''(r) $ into equation \meq{BB2} and expand to the first-order approximation of rotation, we obtain a second-order differential equation for the single perturbation function $h_{1}(r)$.
However, this equation is not yet in a Schr\"odinger-like form, as it still contains first-derivative terms of the perturbation function. 
To remove these terms, we perform the following field redefinition:
\begin{align}
h_{1}(r)=\frac{r}{F(r)}\left(1-\frac{2m\tilde{a}M^2}{r^3\omega}\right)Z^{(-)}(r).
\end{align}
After this transformation, Eq. (A2) simplifies to
\begin{align}
\left[F^2\frac{d^2}{dr^2}+F'F\frac{d}{dr}+(1-\ell)^2\omega^2-\mathcal{V}^{(-)}\right]Z^{(-)}(r)=0.
\end{align}
By further introducing the tortoise coordinate, via $ r\rightarrow r_*$,
the radial equation can be cast into the standard Schr\"odinger-type wave equation
\begin{equation}\label{meq3}
\left[\frac{d^2}{dr^2_*}+(1-\ell)^2\omega^2-V^{(-)}\right]Z^{(-)}(r)=0,
\end{equation}
where the effective potential $V^{(-)}$ takes the form
\begin{equation}
\begin{aligned}
\!\! V^{(-)}=&\frac{4\tilde{a}\omega m M^2(1\!-\! \ell)^2}{r^3}\!+\! F(r)\frac{24\tilde{a}mM^2\left[3r\!-\!7M(1\!-\! \ell)\right]}{l(l\!+\! 1)r^6\omega}\\
&+F(r)\bigg[\frac{l(l+1)(1\!-\!\ell)^2-2\ell^2}{r^2(1\!-\! 2\ell)}-\frac{2M(1\!-\!\ell)}{r^3} \bigg].
\end{aligned}
\end{equation}
For the polar sector, the structure of the perturbation equations is considerably more intricate.
The seven polar equations contain four perturbation functions, and in the static limit $\tilde{a}=0$ the system can be reduced to a single master equation, as shown in Ref. \cite{Guo:2023nkd}.
In the slow-rotation approximation, and following the standard method developed by Pani for Kerr black holes \cite{Pani2013IJMPA}, we find that it is not possible to construct an appropriate linear combination of the polar variables that removes the spin dependent couplings and yields a Schr\"odinger-type master equation.
In other words, the usual decoupling strategy that succeeds in the GR spacetime fails in the KR background, and the polar sector remains irreducibly coupled at the radial level.
Given this obstruction, we leave the decoupling of the polar perturbations as an open problem for future investigation.
Since the axial perturbations do admit a Regge–Wheeler–type master equation suitable for QNM analysis, in the present work we therefore restrict our attention to the axial sector.

\section{The Eigenvalue Problem For Quasinormal Modes}\label{Sec.4}
\subsection{Boundary conditions}

In this study, we investigate how the Lorentz-violating parameter $\ell$ influences the QNMs of of scalar, electromagnetic, and gravitational perturbations in slowly rotating spacetimes.  
The perturbation equations \eqref{meq1}, \eqref{meq2}, and \eqref{meq3} can be recast into a unified eigenvalue problem of the form
\begin{equation}
\mathcal{D}_s\,\psi_s = -(1-\ell)^2\omega^{2}\psi_s ,
\end{equation}
where the differential operator is defined as $ \mathcal{D}_s \equiv \pp^2_{r_*} - V_s $, and ${s=0,1,2}$ correspond respectively to scalar, electromagnetic, and gravitational perturbations.
Near the horizon $r_h$ and at spatial infinity, the master equation admits the asymptotic expansions
\begin{align}
&\pp^2_{r_*}\psi_s=-\left[  (1-\ell)\omega-\tfrac{m\tilde{a}}{4M(1-\ell)^2} \right]^2\psi_s,\\
&\pp^2_{r_*}\psi_s=- (1-\ell)^2 \omega^2  \psi_s,
\end{align}
which, together with the appropriate boundary conditions at the horizon and at infinity, define an eigenvalue problem for the complex frequency spectrum.
Thus, for our asymptotically flat spacetime, the QNM frequencies, denoted by $\omega = \omega_R - i \omega_I$, are determined by imposing the boundary conditions
\begin{equation}
\psi_s\sim
\begin{cases}
e^{-i(1-\ell)[\omega-m\Omega_H]r_*} &\text{for}~~r_*\rightarrow~-\infty ,\\
e^{i(1-\ell)\omega r_*} &\text{for}~~r_*\rightarrow~+\infty,
\end{cases}
\end{equation}
which require the perturbation to be purely ingoing at the event horizon and purely outgoing at spatial infinity.
Here, the horizon angular frequency is given by
\begin{equation}
\Omega_H \equiv -\left.\frac{g_{t\varphi}}{g_{\varphi\varphi}}\right|_{r\to r_h}
= \frac{\tilde{a}}{4M(1-\ell)^3},
\end{equation}
evaluated to first order in the dimensionless spin parameter $\tilde a$ within the slow-rotation approximation.
To implement the physical boundary conditions, we introduce an ansatz for the radial function $\psi_s$ that correctly captures the asymptotic behavior at the event horizon and at spatial infinity.
A suitable choice is
\begin{equation}\label{psia2}
\!\!\psi_{s} = e^{i (1 - \ell) \omega r} \left( r - r_h \right)^{-i \frac{ r_h^2 \omega (1 - \ell) - \tilde{a} M m}{r_h}} r^{i r_h \omega (1 - \ell) + i \frac{ r_h^2 \omega (1 - \ell) - \tilde{a} M m}{r_h}}R_{s},
\end{equation}
which ensures that the solution behaves as an ingoing wave at the horizon and as an outgoing wave at infinity.
In the following subsections, we compute the QNMs using the continued fraction method.
For verification, we also employ the matrix method (MM) \cite{Lin:2016sch,Lin:2017oag,Lin:2019mmf,Lin:2022ynv,Shen:2022xdp,Lei2021,
Liu:2024oeq}, whose detailed implementation is presented in Appendix \ref{AppB}.

\subsection{Continued fraction method}
Following Leaver’s pioneering work \cite{Leaver:1985ax}, the continued fraction method (CFM) has become one of the most accurate techniques for computing QNMs.
In this approach, the eigenfunction is expanded as a power series whose coefficients satisfy a finite-term recurrence relation.
By inserting the ansatz \eqref{psia2} into the master perturbation equation, the radial solution $R_s(r)$ can be expressed as a Frobenius-type series around the event horizon:
\begin{align}
R_s=\sum_{n=0}^{\infty}d_n\left(1-\frac{r_h}{r} \right)^n.
\end{align}
A substitution of the series expansion into the master perturbation equation yields a three-term recurrence relation for scalar and electromagnetic perturbations, whereas in the gravitational sector a six-term recurrence relation is obtained.
These recurrence relations fully determine the QNM spectrum in each perturbative sector.
For scalar and electromagnetic perturbations, we have
\begin{equation}\label{56}
d_{n+1}\alpha_n+d_{n}\beta_n+d_{n+1}\gamma_n=0,
\end{equation}
The recurrence coeﬃcients $\alpha_n$, $\beta_n$, and $\gamma_n$ are simple functions consisting of $n$ and other diﬀerential equation parameters, the explicit forms are as follows
\begin{align}
&\begin{aligned}
\!\alpha_n=2i\omega n r_h^3(1-\ell)-n^2r_h^2-2i\tilde{a}mMmr_h,
\end{aligned}
\\
&\begin{aligned}
\!\beta_n=&2mMr_h\tilde{a}\left[ i+2in+3r_h(1\!-\!\ell)\omega  \right]\!+\!r_h^2(1\!-\!s\!+\!2n^2\!+\!2n)\\
&+(1\!-\!\ell)r_h^2\left[ (l^2\!+\!l)\!-\!4i(1\!+\!2n)r_h\omega-8r_h^2(1\!-\!\ell)\omega^2 \right],
\end{aligned}\\
&\begin{aligned}
\!\gamma_n=&r_h^2 s\!-\!r_h^2\left[ 1\!-\!2i\omega r_h (1\!-\!\ell) \right]^2\!-\!nr_h^2\left[2\!+\!n\!-\!4i\omega r_h(1\!-\!\ell) \right]\\
&-i2r_h\tilde{a}mM\left[ 2i\omega r_h (\ell-1)+1+n \right],
\end{aligned}
\end{align}
where $s=0,1$ correspond to the scalar and electromagnetic perturbations, respectively.
For gravitational perturbation, they satisfy the six-term recurrence relation
\begin{equation}\label{6oder}
d_{n+1}\alpha_n+d_n\beta_n+d_{n-1}\gamma_n
+d_{n-2}\sigma_n+d_{n-3}\tau_n+d_{n-4}\delta_n=0 ,
\end{equation}
where 
\begin{align}
&\begin{aligned}
\!\!\alpha_n=&\lambda n r_h^3(2\ell-1)\omega\left[ n-2i\omega r_h(1-\ell)  \right]\\
&- 2i\omega \tilde{a}m Mr_h^2(1-2\ell) \lambda n,
\end{aligned}\\
&\begin{aligned}
\!\!\beta_n\!=
&4i\omega^2r_h^4\lambda (1\!+\!2n)(3\ell\!-\!2\ell^2\!-\!1)\!-\!8\omega^3r_h^5\lambda(1\!-\!\ell)^2(1\!-\!2\ell)\\
&+\lambda\omega r_h^3\big[ (2\!-\!4\ell)(n^2\!+\!n)\!+\!(1\!-\!\ell)^2\lambda\!+\!6\ell\!-\!2\ell^2\!\!-\!\!3  \big]\\
&+2mM\tilde{a}(1\!-\!2\ell)\big[ i\omega r_h^2\lambda(1\!+\!2n\!-\!3ir_h(1\!-\!\ell)\omega)\!-\!6M \big],
\end{aligned}\\
&\begin{aligned}
\!\!\!\!\gamma_n\!=&2mM\tilde{a}(2\ell\!-\!1)\big[ \lambda \omega r_h^2(i\!+\!in\!+\!2\omega r_h (1\!-\!\ell))\!-\!60\!M\! \big]\!+\!\lambda \omega r_h^3\\
&\!\!\times\!(1\!-\!2\ell)\big[ 3\!-\!n^2\!-\!2n\!+\!4i\omega r_h(1\!-\!\ell)(1\!+\!n\!-\!i\omega r_h(1\!-\!\ell))  \big],
\end{aligned}
\end{align}
and
\begin{align}
&\begin{aligned}
\sigma_n=&-288mM^2\tilde{a}(1-2\ell), 
\end{aligned}\\
&\begin{aligned}
 \tau_n=&264mM^2\tilde{a}(1-2\ell),
\end{aligned}\\
&\begin{aligned}
 \delta_n=&-86mM^2\tilde{a}(1-2\ell).
\end{aligned}
\end{align}
Using Gaussian elimination \cite{Leaver:1985ax,Leaver:1990zz,Percival:2020skc,Guo:2022rms}, the six-term recurrence relation \eqref{6oder} can be systematically reduced to a standard three-term recurrence relation, in analogy with equation \eqref{56}.
With these recurrence relations at hand, the QNM frequencies can be computed using the standard continued-fraction convergence procedure.

\subsection{Numerical results}
In this section, we display the QNMs results obtained from our numerical analysis, without loss of generality, we set $ M=1 $.
First, we need to verify the convergence of the QNMs computed with the CFM. 
As the truncation order $n$ in the recurrence relation increases, the relative difference between two successive frequencies,
\begin{equation}
\delta_n 
= 100\,\frac{\bigl||\omega_n|-|\omega_{n-1}|\bigr|}{|\omega_{n-1}|},
\end{equation}
is expected to decrease monotonically and approach zero.
Here we take the fundamental gravitational mode with $l\!=\!m\!=\!2$, under the representative choice $\tilde a\!=\!\ell\!=\!0.2$, as a benchmark case.
The choice $l\!=\!m\!=\!2$ is motivated by the fact that it dominates the gravitational wave signal and is the most relevant mode for QNM phenomenology.

\begin{figure}[h]
\centering
\includegraphics[width=1\linewidth]{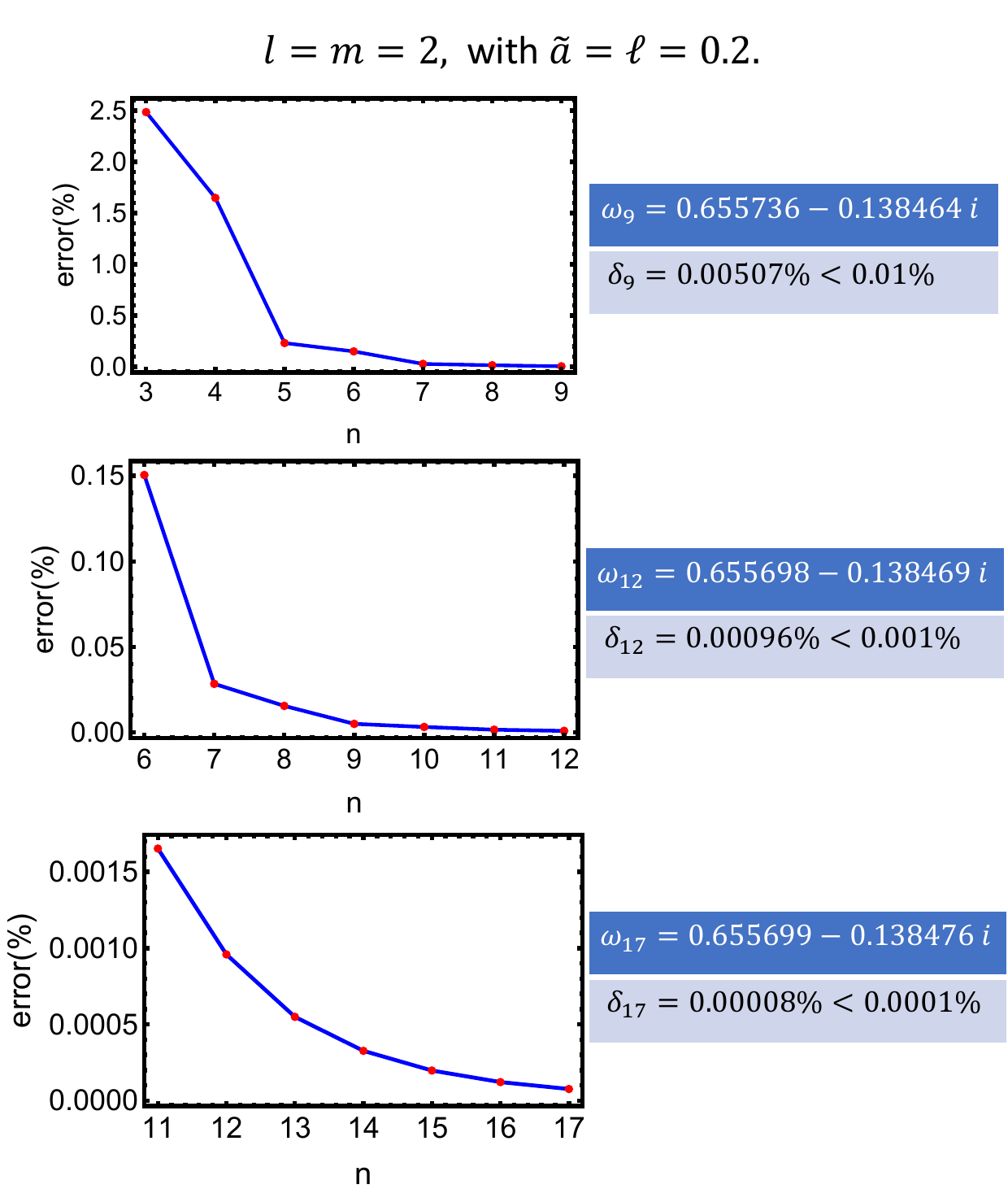}
\caption{Convergence of the CFM for the $l\!=\!m\!=\!2$ axial gravitational mode at $\tilde{a}\!=\!\ell\!=\!0.2$.
The relative error $\delta_n$ decreases monotonically with the recurrence order $n$.
}
\label{fig1}
\end{figure}
In Fig. \ref{fig1}, we present the convergence test for the CFM computed fundamental axial gravitational mode with $l\!=\!m\!=\!2$.
By truncating the continued fraction at orders corresponding to relative errors of $10^{-2}\%$, $10^{-3}\%$, and $10^{-4}\%$, we obtain stable results at$ n=9$, $n=12$, and $n=17$, respectively.
The rapid decrease of the relative difference $\delta_n$ between successive truncation orders demonstrates that the continued fraction method converges efficiently for the slowly rotating KR spacetime in the parameter regime considered.
By comparing these results, we conclude that the choice $n=12$ yields sufficiently accurate QNM frequencies, and we adopt this truncation throughout the remainder of our computations.

Next, we present the QNM spectra of scalar, electromagnetic, and gravitational perturbations under the combined influence of the dimensionless spin parameter $\tilde{a}$ and the Lorentz violating parameter $\ell$.
\begin{figure}[h]
\centering
\includegraphics[width=0.45\linewidth]{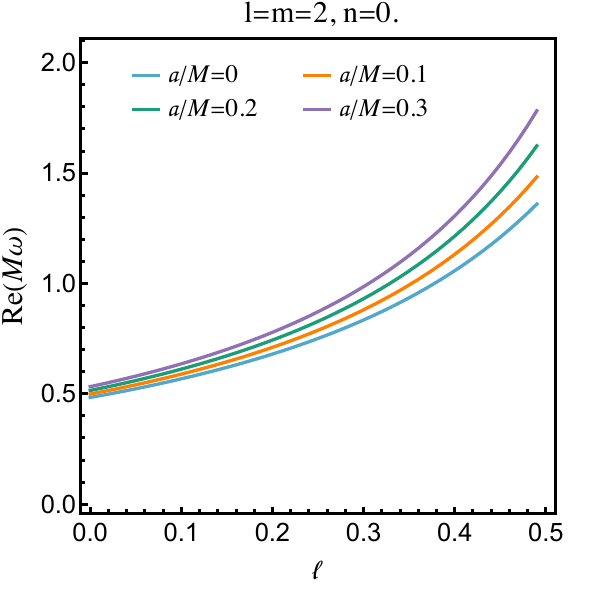}
\includegraphics[width=0.48\linewidth]{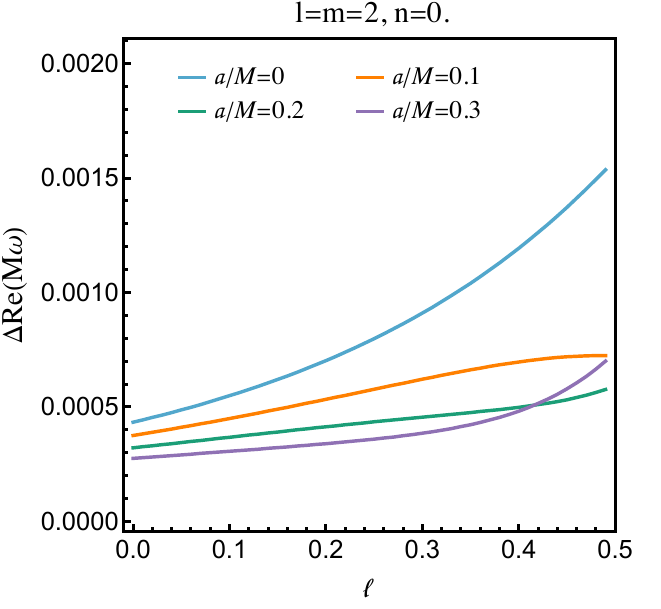}
\includegraphics[width=0.46\linewidth]{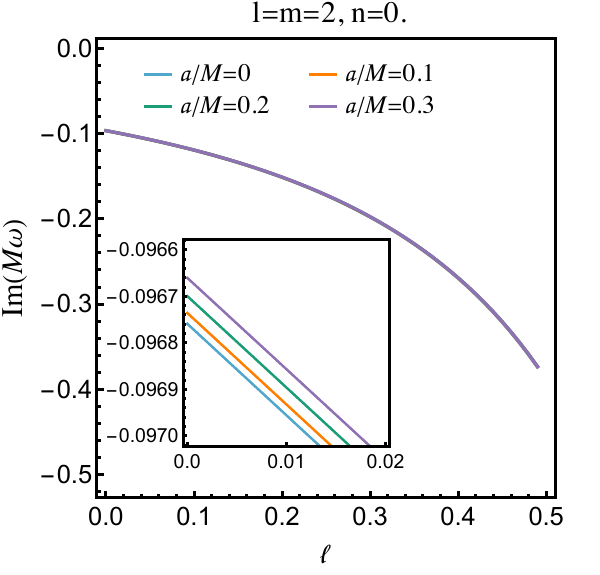}
\includegraphics[width=0.49\linewidth]{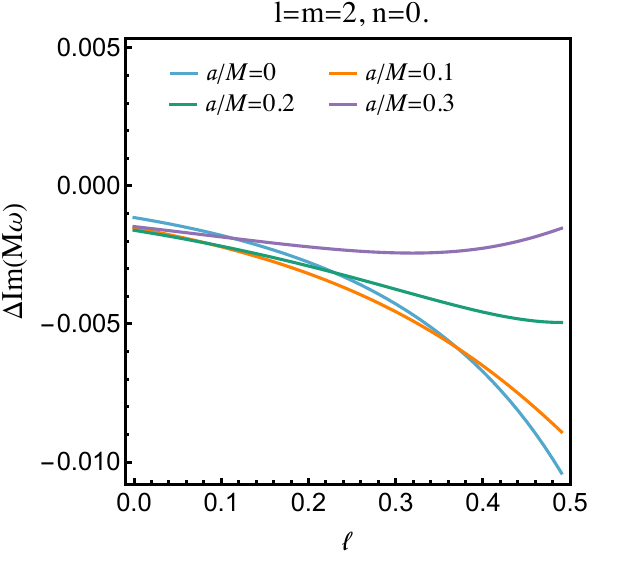}
\caption{
The scalar frequencies in the $l=2, n=0$ mode are shown, where the upper and lower panels on the left correspond to the real and imaginary parts of the results, while the right panel presents the discrepancy between the matrix method and the CFM.
}
\label{fig2}
\end{figure}
In Figs. \ref{fig2}, we present the scalar QNM frequencies for the $l=2$, $n=0$ fundamental mode. 
The left panels show the real and imaginary parts of $M\omega$ as functions of the Lorentz-violating parameter $\ell$ for several values of the dimensionless spin parameter $\tilde{a}$. 
As the Lorentz-violating parameter $\ell$ increases, the real part of the QNM frequencies grows, while the imaginary part becomes more negative.  
This implies that Lorentz violation enhances the oscillation frequency and accelerates the damping of the perturbations.  
Moreover, Lorentz violation amplifies the influence of the spin parameter $\tilde{a}$ on the real part of the frequency, whereas its effect on the imaginary part remains extremely weak, consistent with expectations from the first-order slow-rotation approximation.
The right panels display the corresponding deviations, $\Delta\mathrm{Re}(M\omega)$ and $\Delta\mathrm{Im}(M\omega)$, defined as the differences between the results obtained from the MM with $	N=15$ and the CFM with $n=12$. 
These comparisons indicate that the two numerical schemes agree to high precision across the full parameter range considered.
Here, the percentage error of the real part, for example, is defined as \cite{Liu:2023uft}
\begin{equation}
\Delta\text{Re}(\omega)=100\times\frac{\text{Re}\left(\omega_\text{CFM}\right)-\text{Re}\left(\omega_\text{MM}\right)}
{\text{Re}\left(\omega_\text{CFM}\right)}.
\end{equation}

\begin{figure}[h]
\centering
\includegraphics[width=0.45\linewidth]{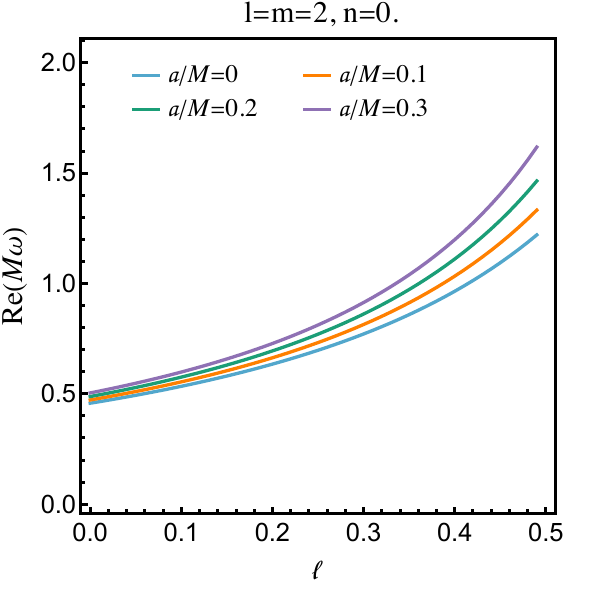}
\includegraphics[width=0.48\linewidth]{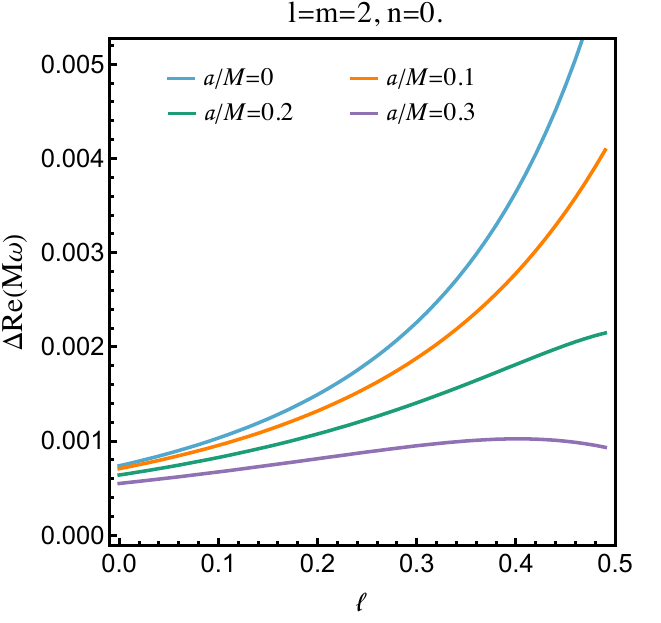}
\includegraphics[width=0.46\linewidth]{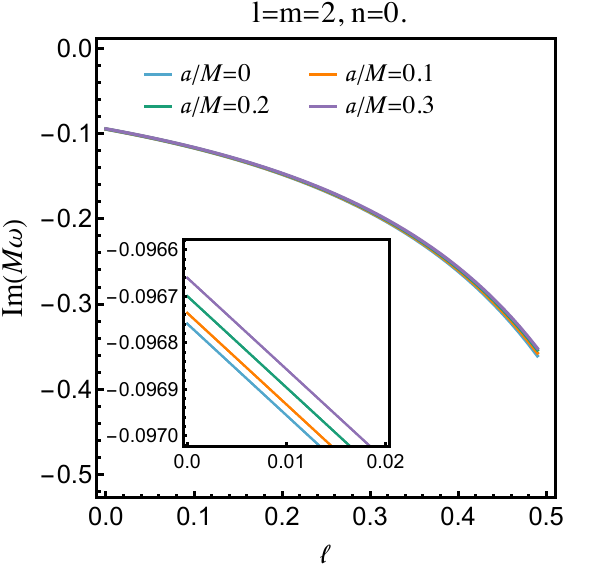}
\includegraphics[width=0.49\linewidth]{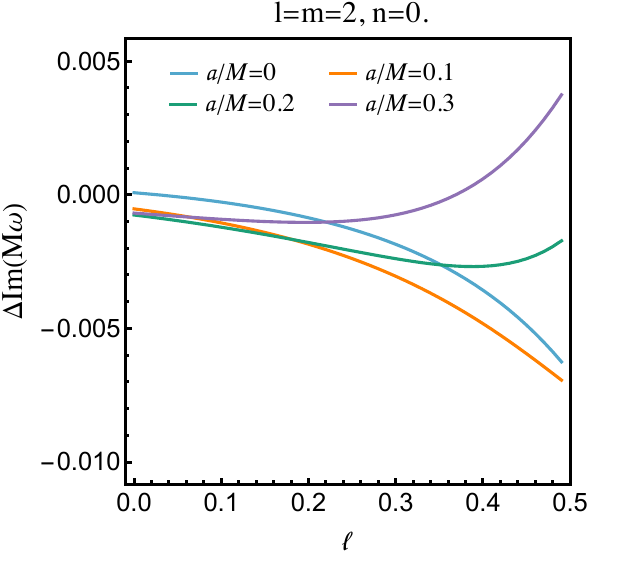}
\caption{The axial electromagnetic frequencies in the $l=2, n=0$ mode are shown, where the upper and lower panels on the left correspond to the real and imaginary parts of the results, while the right panel presents the discrepancy between the matrix method and the CFM.
}
\label{fig3}
\end{figure}

In Figs. \ref{fig3}, we present the axial electromagnetic QNM frequencies for the $l=2$, $n=0$ fundamental mode.
Overall, the behavior closely parallels that of the scalar perturbations.
Both the real and imaginary parts of$ M\omega$ display the same qualitative dependence on the Lorentz-violating parameter $\ell $and the spin parameter $\tilde{a}$ as in the scalar case.
The only notable difference is that the electromagnetic QNM frequencies are systematically smaller in magnitude, effectively resembling a uniformly rescaled version of the scalar spectrum (in a purely qualitative sense).
\begin{figure}[h]
\centering
\includegraphics[width=0.43\linewidth]{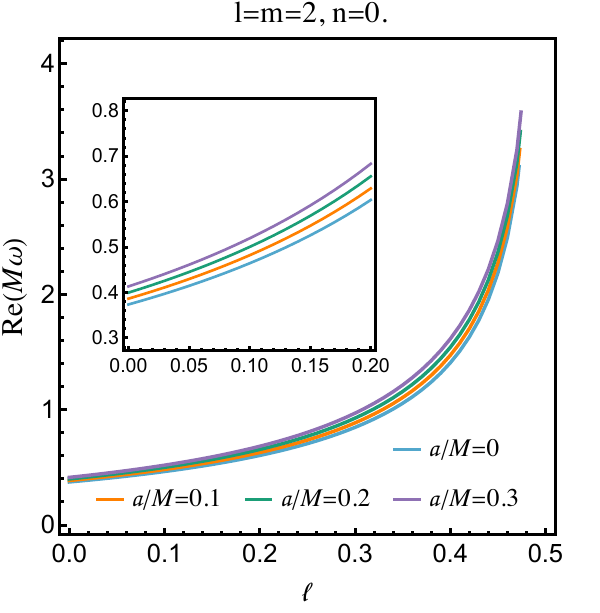}
\includegraphics[width=0.49\linewidth]{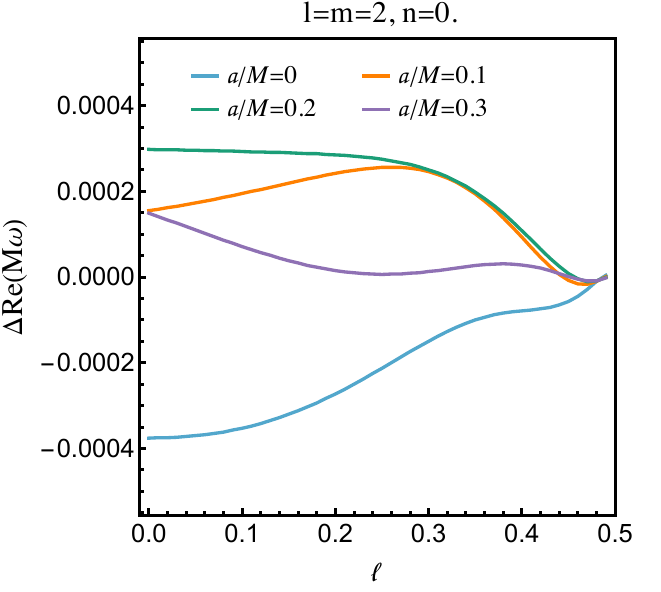}
\includegraphics[width=0.47\linewidth]{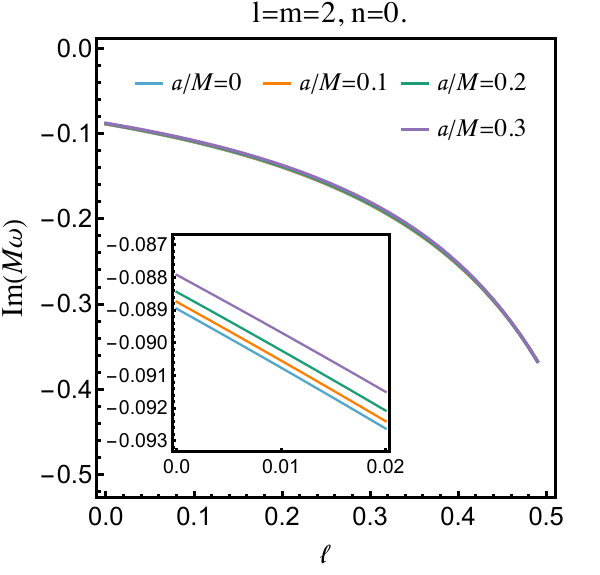}
\includegraphics[width=0.48\linewidth]{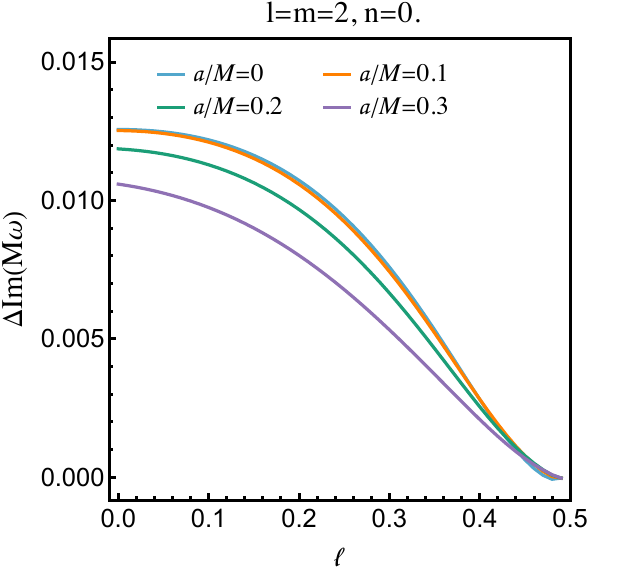}
\caption{The axial gravitational frequencies in the $l=2, n=0$ mode are shown, where the upper and lower panels on the left correspond to the real and imaginary parts of the results, while the right panel presents the discrepancy between the matrix method and the CFM.
}
\label{fig4}
\end{figure}

In Figs. \ref{fig4}, we present the axial gravitational QNM frequencies for the fundamental mode, which constitute the most significant results of this work.
Compared with the scalar and electromagnetic sectors, the Lorentz-violating parameter $\ell$ produces a markedly stronger impact on both the real and imaginary parts of the frequency.
As $\ell$ approaches its upper bound $\ell\to 0.5$, the oscillation frequency grows rapidly while the damping rate also increases substantially, indicating an extreme behavior of the mode in this limit.
This trend confirms that physically viable configurations require $\ell$ to lie within the interval $[0,0.5)$; notably, the first-order rotation parameter $\tilde{a}$ does not modify this bound.
The right panels show the discrepancies between the CFM and MM results.
A behavior consistent with the frequency itself is observed: as $\ell\to 0.5$, the percentage differences of both the real and imaginary parts decrease rapidly toward zero.
This convergence reflects the enhanced numerical stability of the equations in the large-$\ell$ regime and indicates that the two methods remain in excellent agreement throughout the entire parameter space.
Finally, in Appendix \ref{AppC} we provide a collection of representative QNM frequency data for reference. 
These tables include, but are not limited to, the modes discussed in the main text. 
Specifically, we list the scalar modes with $l=m=1,2$ in Tables \ref{tab1} and \ref{tab2},  the electromagnetic modes with $l=m=1,2$ in Tables \ref{tab3} and \ref{tab4},  and the axial gravitational modes with $l=m=2,3$ in Tables \ref{tab5} and \ref{tab6}.

\section{CONCLUSIONS AND OUTLOOKS}\label{Sec.5}

The KR gravity model provides a compelling framework in which an antisymmetric tensor field couples non-minimally to the metric, leading to spontaneous Lorentz symmetry breaking.
In this work, we have derived the master perturbation equations for scalar, electromagnetic, and axial gravitational fields in the slowly rotating KR black hole spacetime, retaining all Lorentz-violating effects consistently to first order in the dimensionless spin parameter.
To compute the associated QNMs, we have employed two complementary numerical techniques, the continued fraction method and the matrix method, and demonstrated excellent agreement between them.
The convergence analysis further confirms the reliability and robustness of the CFM across different truncation orders, ensuring the accuracy of our results.
Our computations reveal several noteworthy physical features.
For all three types of perturbations, scalar, electromagnetic, and axial gravitational, the Lorentz-violating parameter $\ell$ produces a consistent qualitative trend: the real part of the QNM frequency increases, while the imaginary part becomes more negative, indicating higher oscillation frequency and stronger damping as Lorentz symmetry is increasingly broken.
The effect of rotation is also uniform across the three sectors, enhancing the real part of the frequency and causing only a very mild increase in the imaginary part, as expected in the first-order slow-rotation approximation.
In the gravitational sector, however, the response to Lorentz violation is much more pronounced, enabling us to identify a theoretical upper bound on the Lorentz-violating parameter, $\ell<0.5$, beyond which the axial gravitational QNMs approach an extremal behavior.

Beyond the QNM analysis presented here, several promising directions remain open for future exploration within the framework of KR gravity.
First, exploring independent constraints on the Lorentz-violating parameter from other observational channels, such as black hole shadows and horizon-scale imaging \cite{Liu:2025wwq,Liu:2024soc,Zhang:2024jrw,Zhao:2025ouq,Wang:2025btn,Liu:2025lwj}, would provide an important point of comparison.
In addition, KR gravity may exhibit nontrivial thermodynamic topological properties \cite{Wei:2022dzw,Wei:2021vdx,Wu:2023fcw,Wu:2023meo,Wu:2024rmv,Wu:2024asq,Zhu:2024zcl,Liu:2025iyl} as well as rich quantum information features of the spacetime \cite{Fuentes-Schuller:2004iaz,Liu:2024wpa,Liu:2025bpp,Wu:2022xwy,Wu:2023spa,Wu:2023sye,Wu:2023sye,Wu:2025euf,Li:2025jlu,Liu:2025hcx,Liu:2025bzv,Tang:2025eew,Tang:2025mtc}, suggesting that these aspects could provide further insight into the structure and physical implications of the theory.
Finally, extending the present analysis to higher-order rotation or fully dynamical perturbations, including potential couplings between metric and KR-field fluctuations, remains an important challenge for achieving a more complete understanding of Lorentz-violating black hole physics.

\acknowledgments

This work was supported by the National Natural Science Foundation of China Grant Nos. 12035005, and the Natural Science Foundation of Hunan Province Grant Nos. 2022JJ30220, the Fundamental Research Funds for the Central Universities (Grant No. lzujbky-2025-jdzx07), the Natural Science Foundation of Gansu Province (No. 22JR5RA389, No.25JRRA799), and the `111 Center' under Grant No. B20063.

\appendix
\begin{widetext}
\section{Explicit expansions of the tensor $\mathcal{B}_{ab}$}\label{AppA}
In this appendix, we provide the explicit tensorial form of the Kalb-Ramond contribution $\mathcal{B}_{ab}$ to the effective gravitational field equation (\ref{EQG}), which is defined as
\begin{equation}\label{EQB}
\begin{aligned}
\mathcal{B}_{ab} = g_{ab}\beta^{ce}\beta^{d}{}_{e}R_{cd} -\beta^{c}{}_{a}\beta^{d}{}_{b}R_{cd} -\beta^{cd}\beta_{ad}R_{bc} -\beta^{cd}\beta_{bd}R_{ac}
+\frac{1}{2}\nabla_c\nabla_{a}(\beta^{cd}\beta_{bd}) +\frac{1}{2}\nabla_c\nabla_{b}(\beta^{cd}\beta_{ad}) -\frac{1}{2}\nabla^c\nabla_c(\beta_a{}^{d}\beta_{bd}),
\end{aligned}
\end{equation}
where the covariant derivative can be expanded using the Christoffel symbols as
\begin{equation}
\begin{aligned}
\nabla_c\nabla_{a}\beta^{cd}\beta_{bd}=&\Gamma^{c}{}_{ad} \Gamma^{e}{}_{ce} \beta_{bf} \beta^{df} -  \Gamma^{c}{}_{ad} \Gamma^{d}{}_{ce} \beta_{bf} \beta^{ef} + \Gamma^{c}{}_{ad} \Gamma^{d}{}_{be} \beta_{cf} \beta^{ef}  -  \Gamma^{c}{}_{ab} \Gamma^{d}{}_{de} \beta_{cf} \beta^{ef} + \Gamma^{c}{}_{cd} \beta^{de} \partial_{a}\beta_{be}\\
& -  \Gamma^{c}{}_{bd} \beta^{de} \partial_{a}\beta_{ce} + \Gamma^{c}{}_{cd} \beta_{be} \partial_{a}\beta^{de} -  \Gamma^{c}{}_{bd} \beta_{ce} \partial_{a}\beta^{de} 
-  \beta_{ce} \beta^{cd} \partial_{d}\Gamma^{e}{}_{ab} -  \partial_{a}\beta^{cd} \partial_{d}\beta_{bc} -  \partial_{a}\beta_{bc} \partial_{d}\beta^{cd} \\
& -  \beta^{cd} \partial_{d}\partial_{a}\beta_{bc} -  \beta_{bc} \partial_{d}\partial_{a}\beta^{cd} -  \beta_{bc} \beta^{cd} \partial_{e}\Gamma^{e}{}_{ad} 
+ \Gamma^{c}{}_{ab} \beta^{de} \partial_{e}\beta_{cd} + \Gamma^{c}{}_{ab} \beta_{cd} \partial_{e}\beta^{de}.
\end{aligned}
\end{equation}
Similarly,
\begin{equation}
\begin{aligned}
\nabla^c\nabla_c\beta_a^{~d}\beta_{bd}=&\Gamma^{c}{}_{be} \Gamma^{ed}{}_{d} \beta_{a}{}^{f} \beta_{cf} + \Gamma^{c}{}_{ae} \Gamma^{ed}{}_{d} \beta_{bf} \beta_{c}{}^{f}
 + \Gamma^{c}{}_{b}{}^{e} \Gamma^{d}{}_{ce} \beta_{a}{}^{f} \beta_{df} + \Gamma^{c}{}_{a}{}^{e} \Gamma^{d}{}_{be} \beta_{c}{}^{f} \beta_{df} + \Gamma^{c}{}_{ae} \Gamma^{d}{}_{b}{}^{e} \beta_{c}{}^{f} \beta_{df} \\
&+ \Gamma^{c}{}_{a}{}^{e} \Gamma^{d}{}_{ce} \beta_{bf} \beta_{d}{}^{f} -  \Gamma^{ce}{}_{e} \beta_{bd} \partial_{c}\beta_{a}{}^{d} -  \Gamma^{ce}{}_{e} \beta_{a}{}^{d} \partial_{c}\beta_{bd} 
 + \beta_{bc} \beta^{c}{}_{e} \partial^{d}\Gamma^{e}{}_{ad} + \beta_{a}{}^{c} \beta_{ce} \partial^{d}\Gamma^{e}{}_{bd} -  \Gamma^{c}{}_{b}{}^{e} \beta_{cd} \partial_{e}\beta_{a}{}^{d} \\
&-  \Gamma^{c}{}_{a}{}^{e} \beta_{c}{}^{d} \partial_{e}\beta_{bd} -  \Gamma^{c}{}_{b}{}^{e} \beta_{a}{}^{d} \partial_{e}\beta_{cd} -  \Gamma^{c}{}_{a}{}^{e} \beta_{bd} \partial_{e}\beta_{c}{}^{d}
 + \partial_{e}\beta_{bc} \partial^{e}\beta_{a}{}^{c} -  \Gamma^{c}{}_{be} \beta_{cd} \partial^{e}\beta_{a}{}^{d} + \partial_{e}\beta_{a}{}^{c} \partial^{e}\beta_{bc} \\
&-  \Gamma^{c}{}_{ae} \beta_{c}{}^{d} \partial^{e}\beta_{bd} -  \Gamma^{c}{}_{be} \beta_{a}{}^{d} \partial^{e}\beta_{cd} -  \Gamma^{c}{}_{ae} \beta_{bd} \partial^{e}\beta_{c}{}^{d}
 + \beta_{bc} \partial^{e}\partial_{e}\beta_{a}{}^{c} + \beta_{a}{}^{c} \partial^{e}\partial_{e}\beta_{bc}.
\end{aligned}
\end{equation}
The Christoffel symbols used above are
\begin{equation}
\begin{aligned}
\!\!\!
\Gamma^{a}{}_{bc}\!=\!\frac{1}{2}g^{ad}\left(\pp_b g_{cd}+\pp_c g_{bd}-\pp_d g_{bc}\right),\quad
\Gamma^{ab}{}_{c}\!=\!\frac{1}{2}g^{ad}g^{be}\left(\pp_cg_{ed}+\pp_eg_{cd}-\pp_dg_{ec}\right),\quad
\Gamma^{a}{}_{b}{}^{c}\!=\!\frac{1}{2}g^{ad}g^{ce}\left(\pp_bg_{ed}+\pp_eg_{bd}-\pp_d g_{be} \right),
\end{aligned}
\end{equation}
complete the explicit construction of the tensor $\mathcal{B}_{ab}$ entering the effective field equation (\ref{EQG}).



\section{Matrix method}\label{AppB}

In this appendix, we present the numerical techniques employed to compute the QNMs of the slowly rotating KR black hole.

To verify the accuracy of the continued fraction method in computing the QNMs of scalar, electromagnetic, and gravitational
perturbations in the Lorentz-violating black hole spacetime, we further employ the matrix method (MM) for comparison.
This numerical scheme was developed by Lin and collaborators \cite{Lin:2016sch,Lin:2017oag,Lin:2019mmf,Lin:2022ynv,Shen:2022xdp,Lei2021,
Liu:2024oeq}, and is known for requiring only the correct boundary behavior of the solution while imposing no additional constraints on the trial function.
Following the boundary conditions discussed in the previous subsection, we introduce the compactified coordinate
\begin{align}\label{transyr}
y=\frac{r-r_h}{r},
\end{align}
which maps the QNM domain to the finite interval $y\in[0,1]$.
To implement the boundary conditions at the horizon and at spatial infinity, we define the rescaled wave function
\begin{align}\label{transchipsi}
\chi(y)=y(1-y)\,\psi_l(y),
\end{align}
which satisfies the homogeneous conditions
\begin{align}\label{chi01}
\chi(0)=\chi(1)=0.
\end{align}
With this transformation, all perturbation equations can be written in
the generic second–order form
\begin{align}\label{qicieq}
\tilde{\mathcal{C}}_2(y,\omega)\,\chi''(y)
+\tilde{\mathcal{C}}_1(y,\omega)\,\chi'(y)
+\tilde{\mathcal{C}}_0(y,\omega)\,\chi(y)=0,
\end{align}
where the coefficient functions $\tilde{\mathcal{C}}_j$ depend
linearly on the frequency,
\begin{align}
\tilde{\mathcal{C}}_j(y,\omega)
=\tilde{\mathcal{C}}_{j,0}(y)+\omega\,\tilde{\mathcal{C}}_{j,1}(y),
\qquad j=0,1,2.
\end{align}
These coefficients follow directly by inserting the ansatz \eqref{psia2} into the master perturbation equations \eqref{meq1}, \meq{meq2} and \eqref{meq3} and rewriting them using the coordinate transformation \eqref{transyr} together with \eqref{transchipsi}.

To discretize Eq. \eqref{qicieq}, we choose $N$ uniformly spaced grid points in the interval $[0,1]$ and perform a Taylor expansion of $\chi(y)$ around each point.  
This procedure yields the differential matrices associated with $\chi$, $\chi'$, and $\chi''$, allowing Eq. \eqref{qicieq} to be recast into the algebraic matrix equation
\begin{align}
\left(\mathcal{M}_0+\omega\,\mathcal{M}_1\right)\chi = 0,
\end{align}
where $\mathcal{M}_0$ and $\mathcal{M}_1$ are $N\times N$ matrices constructed from $\tilde{\mathcal{C}}_{j,0}$ and $\tilde{\mathcal{C}}_{j,1}$ and their derivatives.
Once the matrices are assembled following the prescriptions in Refs. \cite{Lin:2016sch,Lin:2017oag,Lin:2019mmf}, the QNMs frequencies are obtained by solving the resulting generalized eigenvalue problem.

\section{QNM FREQUENCY TABLES}\label{AppC}

In this appendix, we present the QNM frequency tables for the most representative modes considered in this work.  
Specifically, we list the fundamental ($n=0$) scalar modes with $l=m=1,2$, the fundamental electromagnetic modes with $l=m=1,2$, and the fundamental axial gravitational mode with $l=m=2$.  
These tables provide a useful reference for comparing the behavior of different perturbation sectors under variations of the Lorentz-violating parameter $\ell$ and the dimensionless spin parameter $\tilde{a}$.
\begin{table}[H]
\renewcommand{\arraystretch}{1.2}
\centering
\caption{Comparison of the $n=0$, $l=m=1$ mode massless scalar QNM frequencies calculated by the matrix method and the continued fraction method under the first order slow rotation approximation.}
\label{tab1}
\setlength\tabcolsep{5mm}{
\begin{tabular}{cccccccc}
\hline\hline
\multirow{2}{*}{~\bf }	&
\multirow{3}{*}{~\bf $ a $~}  &
\multicolumn{2}{c}{Matrix method} & \multicolumn{2}{c}{Continued fraction method} & \multicolumn{2}{c}{\% error } \\
\cline{3-4}  \cline{5-6} \cline{7-8}
&~&{Re$(M\omega)$ }&{-Im$(M\omega)$}&{Re$(M\omega)$}&{-Im$(M\omega)$}&{Re$(M\omega)$}&{-Im$(M\omega)$}\\
\hline
\multirow{5}{*}{$ \ell=0 $}
&0      &0.292934 &0.097649 &0.292927 &0.097677 &0.002425 &-0.028136 \\
&0.05   &0.296833 &0.097631 &0.296827 &0.097658 &0.001990 &-0.027609 \\
&0.10   &0.300813 &0.097609 &0.300808 &0.097636 &0.001624 &-0.026888 \\
&0.15   &0.304876 &0.097586 &0.304872 &0.097611 &0.001322 &-0.025996 \\
&0.20   &0.309023 &0.097560 &0.309020 &0.097585 &0.001078 &-0.024957 \\
\\
\multirow{5}{*}{$ \ell=0.1 $}
&0      &0.345701 &0.120714 &0.345692 &0.120757 &0.002539 &-0.035106 \\
&0.05   &0.351077 &0.120685 &0.351070 &0.120726 &0.001942 &-0.034150 \\
&0.10   &0.356583 &0.120651 &0.356578 &0.120691 &0.001459 &-0.032918 \\
&0.15   &0.362221 &0.120614 &0.362217 &0.120652 &0.001081 &-0.031449 \\
&0.20   &0.367993 &0.120574 &0.367990 &0.120610 &0.000796 &-0.029784 \\
\\
\multirow{5}{*}{$ \ell=0.2 $}
&0      &0.416372 &0.153022 &0.416361 &0.153090 &0.002534 &-0.044504 \\
&0.05   &0.424076 &0.152973 &0.424069 &0.153038 &0.001698 &-0.042799 \\
&0.10   &0.431998 &0.152916 &0.431993 &0.152979 &0.001062 &-0.040689 \\
&0.15   &0.440143 &0.152855 &0.440140 &0.152913 &0.000602 &-0.038249 \\
&0.20   &0.448514 &0.152790 &0.448513 &0.152844 &0.000297 &-0.035555 \\
\hline\hline
\end{tabular}}
\end{table}

\begin{table}[H]
\renewcommand{\arraystretch}{1.2}
\centering
\caption{Comparison of the $n=0$, $l=m=2$ mode massless scalar QNM frequencies calculated by the matrix method and the continued fraction method under the first order slow rotation approximation.}
\label{tab2}
\setlength\tabcolsep{5mm}{
\begin{tabular}{cccccccc}
\hline\hline
\multirow{2}{*}{~\bf }	&
\multirow{3}{*}{~\bf $ a $~}  &
\multicolumn{2}{c}{Matrix method} & \multicolumn{2}{c}{Continued fraction method} & \multicolumn{2}{c}{\% error } \\
\cline{3-4}  \cline{5-6} \cline{7-8}
&~&{Re$(M\omega)$ }&{-Im$(M\omega)$}&{Re$(M\omega)$}&{-Im$(M\omega)$}&{Re$(M\omega)$}&{-Im$(M\omega)$}\\
\hline
\multirow{5}{*}{$ \ell=0 $}
&0      &0.483644 &0.096758 &0.483642 &0.096759 &0.000433 &-0.001145 \\
&0.05   &0.491268 &0.096748 &0.491266 &0.096749 &0.000404 &-0.001371 \\
&0.10   &0.499094 &0.096734 &0.499092 &0.096736 &0.000375 &-0.001519 \\
&0.15   &0.507125 &0.096717 &0.507123 &0.096718 &0.000347 &-0.001595 \\
&0.20   &0.515368 &0.096698 &0.515366 &0.096699 &0.000321 &-0.001607 \\
\\
\multirow{5}{*}{$ \ell=0.1 $}
&0      &0.568025 &0.119522 &0.568022 &0.119524 &0.000549 &-0.001789 \\
&0.05   &0.578523 &0.119507 &0.578520 &0.119509 &0.000497 &-0.002059 \\
&0.10   &0.589344 &0.119484 &0.589341 &0.119487 &0.000449 &-0.002205 \\
&0.15   &0.600498 &0.119457 &0.600495 &0.119460 &0.000406 &-0.002241 \\
&0.20   &0.611991 &0.119427 &0.611989 &0.119430 &0.000367 &-0.002182 \\
\\
\multirow{5}{*}{$ \ell=0.2 $}
&0      &0.680134 &0.151377 &0.680129 &0.151381 &0.000702 &-0.002762 \\
&0.05   &0.695153 &0.151350 &0.695148 &0.151355 &0.000612 &-0.003070 \\
&0.10   &0.710721 &0.151312 &0.710718 &0.151317 &0.000533 &-0.003175 \\
&0.15   &0.726857 &0.151267 &0.726853 &0.151272 &0.000467 &-0.003108 \\
&0.20   &0.743574 &0.151220 &0.743571 &0.151225 &0.000413 &-0.002901 \\
\hline\hline
\end{tabular}}
\end{table}

\begin{table}[H]
\renewcommand{\arraystretch}{1.2}
\centering
\caption{Comparison of the $n=0$, $l=m=1$ mode massless electromagnetic QNM frequencies calculated by the matrix method and the continued fraction method under the first order slow rotation approximation.}
\label{tab3}
\setlength\tabcolsep{5mm}{
\begin{tabular}{cccccccc}
\hline\hline
\multirow{2}{*}{~\bf }	&
\multirow{3}{*}{~\bf $ a $~}  &
\multicolumn{2}{c}{Matrix method} & \multicolumn{2}{c}{Continued fraction method} & \multicolumn{2}{c}{\% error } \\
\cline{3-4}  \cline{5-6} \cline{7-8}
&~&{Re$(M\omega)$ }&{-Im$(M\omega)$}&{Re$(M\omega)$}&{-Im$(M\omega)$}&{Re$(M\omega)$}&{-Im$(M\omega)$}\\
\hline
\multirow{5}{*}{$ \ell=0 $}
&0      &0.248279 &0.092473 &0.248221 &0.092490 &0.023255 &-0.018457 \\
&0.05   &0.251614 &0.092286 &0.251558 &0.092304 &0.022133 &-0.019607 \\
&0.10   &0.255040 &0.092088 &0.254986 &0.092107 &0.020957 &-0.020206 \\
&0.15   &0.258560 &0.091881 &0.258509 &0.091899 &0.019735 &-0.020302 \\
&0.20   &0.262177 &0.091665 &0.262128 &0.091683 &0.018475 &-0.019945 \\
\\
\multirow{5}{*}{$ \ell=0.1 $}
&0      &0.287689 &0.113625 &0.287595 &0.113650 &0.032673 &-0.022198 \\
&0.05   &0.292216 &0.113321 &0.292126 &0.113348 &0.030820 &-0.023622 \\
&0.10   &0.296890 &0.112998 &0.296805 &0.113025 &0.028863 &-0.024177 \\
&0.15   &0.301719 &0.112657 &0.301638 &0.112684 &0.026819 &-0.023959 \\
&0.20   &0.306708 &0.112302 &0.306632 &0.112328 &0.024712 &-0.023070 \\
\\
\multirow{5}{*}{$ \ell=0.2 $}
&0      &0.338667 &0.142943 &0.338505 &0.142980 &0.047888 &-0.025750 \\
&0.05   &0.345031 &0.142420 &0.344877 &0.142459 &0.044640 &-0.027538 \\
&0.10   &0.351651 &0.141858 &0.351506 &0.141897 &0.041172 &-0.027909 \\
&0.15   &0.358540 &0.141264 &0.358406 &0.141302 &0.037532 &-0.027072 \\
&0.20   &0.365711 &0.140646 &0.365587 &0.140681 &0.033781 &-0.025246 \\
\hline\hline
\end{tabular}}
\end{table}
~\\
\begin{table}[H]
\renewcommand{\arraystretch}{1.2}
\centering
\caption{Comparison of the $n=0$, $l=m=2$ mode massless electromagnetic QNM frequencies calculated by the matrix method and the continued fraction method under the first order slow rotation approximation.}
\label{tab4}
\setlength\tabcolsep{5mm}{
\begin{tabular}{cccccccc}
\hline\hline
\multirow{2}{*}{~\bf }	&
\multirow{3}{*}{~\bf $ a $~}  &
\multicolumn{2}{c}{Matrix method} & \multicolumn{2}{c}{Continued fraction method} & \multicolumn{2}{c}{\% error } \\
\cline{3-4}  \cline{5-6} \cline{7-8}
&~&{Re$(M\omega)$ }&{-Im$(M\omega)$}&{Re$(M\omega)$}&{-Im$(M\omega)$}&{Re$(M\omega)$}&{-Im$(M\omega)$}\\
\hline
\multirow{5}{*}{$ \ell=0 $}
&0      &0.457596 &0.095003 &0.457593 &0.095003 &0.000737 &0.000080 \\
&0.05   &0.464822 &0.094933 &0.464819 &0.094934 &0.000727 &-0.000273 \\
&0.10   &0.472256 &0.094856 &0.472253 &0.094856 &0.000706 &-0.000525 \\
&0.15   &0.479905 &0.094772 &0.479902 &0.094772 &0.000676 &-0.000683 \\
&0.20   &0.487774 &0.094683 &0.487771 &0.094684 &0.000639 &-0.000755 \\
\\
\multirow{5}{*}{$ \ell=0.1 $}
&0      &0.534148 &0.117119 &0.534142 &0.117119 &0.001032 &-0.000272 \\
&0.05   &0.544041 &0.117005 &0.544036 &0.117006 &0.001001 &-0.000747 \\
&0.10   &0.554271 &0.116879 &0.554266 &0.116880 &0.000954 &-0.001050 \\
&0.15   &0.564849 &0.116742 &0.564844 &0.116744 &0.000894 &-0.001199 \\
&0.20   &0.575786 &0.116599 &0.575781 &0.116600 &0.000825 &-0.001212 \\
\\
\multirow{5}{*}{$ \ell=0.2 $}
&0      &0.634690 &0.147960 &0.634680 &0.147961 &0.001493 &-0.000860 \\
&0.05   &0.648746 &0.147766 &0.648737 &0.147768 &0.001418 &-0.001503 \\
&0.10   &0.663378 &0.147547 &0.663369 &0.147549 &0.001319 &-0.001846 \\
&0.15   &0.678609 &0.147311 &0.678601 &0.147313 &0.001203 &-0.001925 \\
&0.20   &0.694463 &0.147066 &0.694455 &0.147068 &0.001075 &-0.001784 \\
\hline\hline
\end{tabular}}
\end{table}

\begin{table}[h]
\renewcommand{\arraystretch}{1.2}
\centering
\caption{Comparison of the $n=0$, $l=m=2$ mode gravitational QNM frequencies calculated by the matrix method and the continued fraction method under the first order slow rotation approximation.}
\label{tab5}
\setlength\tabcolsep{5mm}{
\begin{tabular}{cccccccc}
\hline\hline
\multirow{2}{*}{~\bf }	&
\multirow{3}{*}{~\bf $ a $~}  &
\multicolumn{2}{c}{Matrix method} & \multicolumn{2}{c}{Continued fraction method} & \multicolumn{2}{c}{\% error } \\
\cline{3-4}  \cline{5-6} \cline{7-8}
&~&{Re$(M\omega)$ }&{-Im$(M\omega)$}&{Re$(M\omega)$}&{-Im$(M\omega)$}&{Re$(M\omega)$}&{-Im$(M\omega)$}\\
\hline
\multirow{5}{*}{$ \ell=0 $}
&0      &0.373675 &0.088965 &0.373677 &0.088953 &-0.000376 &0.012565 \\
&0.05   &0.380024 &0.088864 &0.380024 &0.088853 &-0.000056 &0.012628 \\
&0.10   &0.386490 &0.088753 &0.386489 &0.088742 &0.000155  &0.012527 \\
&0.15   &0.393067 &0.088620 &0.393066 &0.088609 &0.000269  &0.012271 \\
&0.20   &0.399749 &0.088451 &0.399748 &0.088440 &0.000298  &0.011864 \\
\\
\multirow{5}{*}{$ \ell=0.1 $}
&0      &0.464301 &0.109907 &0.464303 &0.109893 &-0.000353 &0.012190 \\
&0.05   &0.473039 &0.109771 &0.473039 &0.109758 &-0.000016 &0.012244 \\
&0.10   &0.481956 &0.109619 &0.481955 &0.109605 &0.000195  &0.012106 \\
&0.15   &0.491043 &0.109429 &0.491041 &0.109416 &0.000292  &0.011785 \\
&0.20   &0.500289 &0.109177 &0.500288 &0.109165 &0.000293  &0.011288 \\
\\
\multirow{5}{*}{$ \ell=0.2 $}
&0      &0.603688 &0.139492 &0.603690 &0.139477 &-0.000273 &0.010725 \\
&0.05   &0.616261 &0.139315 &0.616261 &0.139300 &0.000053  &0.010755 \\
&0.10   &0.629130 &0.139111 &0.629128 &0.139096 &0.000242  &0.010582 \\
&0.15   &0.642281 &0.138848 &0.642279 &0.138833 &0.000313  &0.010219 \\
&0.20   &0.655700 &0.138482 &0.655698 &0.138469 &0.000285  &0.009673 \\
\hline\hline
\end{tabular}}
\end{table}

\begin{table}[h]
\renewcommand{\arraystretch}{1.2}
\centering
\caption{Comparison of the $n=0$, $l=m=3$ mode gravitational QNM frequencies calculated by the matrix method and the continued fraction method under the first order slow rotation approximation.}
\label{tab6}
\setlength\tabcolsep{5mm}{
\begin{tabular}{cccccccc}
\hline\hline
\multirow{2}{*}{~\bf }	&
\multirow{3}{*}{~\bf $ a $~}  &
\multicolumn{2}{c}{Matrix method} & \multicolumn{2}{c}{Continued fraction method} & \multicolumn{2}{c}{\% error } \\
\cline{3-4}  \cline{5-6} \cline{7-8}
&~&{Re$(M\omega)$ }&{-Im$(M\omega)$}&{Re$(M\omega)$}&{-Im$(M\omega)$}&{Re$(M\omega)$}&{-Im$(M\omega)$}\\
\hline
\multirow{5}{*}{$ \ell=0 $}
&0      &0.599444 &0.092703 &0.599444 &0.092702 &-0.000067 &0.001134 \\
&0.05   &0.609701 &0.092599 &0.609701 &0.092598 &-0.000027 &0.001216 \\
&0.10   &0.620272 &0.092483 &0.620272 &0.092482 &0.000001  &0.001260 \\
&0.15   &0.631166 &0.092353 &0.631166 &0.092351 &0.000018  &0.001269 \\
&0.20   &0.642392 &0.092208 &0.642392 &0.092207 &0.000026  &0.001247 \\
\\
\multirow{5}{*}{$ \ell=0.1 $}
&0      &0.744782 &0.114496 &0.744783 &0.114494 &-0.000067 &0.001093 \\
&0.05   &0.758893 &0.114355 &0.758893 &0.114354 &-0.000024 &0.001183 \\
&0.10   &0.773481 &0.114195 &0.773481 &0.114194 &0.000004  &0.001226 \\
&0.15   &0.788562 &0.114015 &0.788562 &0.114014 &0.000020  &0.001228 \\
&0.20   &0.804151 &0.113813 &0.804151 &0.113811 &0.000024  &0.001194 \\
\\
\multirow{5}{*}{$ \ell=0.2 $}
&0      &0.968128 &0.145154 &0.968128 &0.145152 &-0.000064 &0.000936 \\
&0.05   &0.988353 &0.144968 &0.988353 &0.144967 &-0.000023 &0.001035 \\
&0.10   &1.009333 &0.144754 &1.009333 &0.144753 &0.000003  &0.001080 \\
&0.15   &1.031094 &0.144510 &1.031094 &0.144509 &0.000015  &0.001078 \\
&0.20   &1.053662 &0.144234 &1.053662 &0.144232 &0.000017  &0.001038 \\
\hline\hline
\end{tabular}}
\end{table}

\end{widetext}


\begin{thebibliography}{127}%
\makeatletter
\providecommand \@ifxundefined [1]{%
 \@ifx{#1\undefined}
}%
\providecommand \@ifnum [1]{%
 \ifnum #1\expandafter \@firstoftwo
 \else \expandafter \@secondoftwo
 \fi
}%
\providecommand \@ifx [1]{%
 \ifx #1\expandafter \@firstoftwo
 \else \expandafter \@secondoftwo
 \fi
}%
\providecommand \natexlab [1]{#1}%
\providecommand \enquote  [1]{``#1''}%
\providecommand \bibnamefont  [1]{#1}%
\providecommand \bibfnamefont [1]{#1}%
\providecommand \citenamefont [1]{#1}%
\providecommand \href@noop [0]{\@secondoftwo}%
\providecommand \href [0]{\begingroup \@sanitize@url \@href}%
\providecommand \@href[1]{\@@startlink{#1}\@@href}%
\providecommand \@@href[1]{\endgroup#1\@@endlink}%
\providecommand \@sanitize@url [0]{\catcode `\\12\catcode `\$12\catcode
  `\&12\catcode `\#12\catcode `\^12\catcode `\_12\catcode `\%12\relax}%
\providecommand \@@startlink[1]{}%
\providecommand \@@endlink[0]{}%
\providecommand \url  [0]{\begingroup\@sanitize@url \@url }%
\providecommand \@url [1]{\endgroup\@href {#1}{\urlprefix }}%
\providecommand \urlprefix  [0]{URL }%
\providecommand \Eprint [0]{\href }%
\providecommand \doibase [0]{https://doi.org/}%
\providecommand \selectlanguage [0]{\@gobble}%
\providecommand \bibinfo  [0]{\@secondoftwo}%
\providecommand \bibfield  [0]{\@secondoftwo}%
\providecommand \translation [1]{[#1]}%
\providecommand \BibitemOpen [0]{}%
\providecommand \bibitemStop [0]{}%
\providecommand \bibitemNoStop [0]{.\EOS\space}%
\providecommand \EOS [0]{\spacefactor3000\relax}%
\providecommand \BibitemShut  [1]{\csname bibitem#1\endcsname}%
\let\auto@bib@innerbib\@empty
\bibitem [{\citenamefont {Aky{\"u}z}\ \emph {et~al.}(2025)\citenamefont
  {Aky{\"u}z}, \citenamefont {Correia}, \citenamefont {Garofalo}, \citenamefont
  {Kacanja}, \citenamefont {Roy}, \citenamefont {Soni}, \citenamefont {Tan},
  \citenamefont {Y}, \citenamefont {Nitz},\ and\ \citenamefont
  {Capano}}]{Akyuz:2025seg}%
  \BibitemOpen
  \bibfield  {author} {\bibinfo {author} {\bibfnamefont {A.}~\bibnamefont
  {Aky{\"u}z}}, \bibinfo {author} {\bibfnamefont {A.}~\bibnamefont {Correia}},
  \bibinfo {author} {\bibfnamefont {J.}~\bibnamefont {Garofalo}}, \bibinfo
  {author} {\bibfnamefont {K.}~\bibnamefont {Kacanja}}, \bibinfo {author}
  {\bibfnamefont {L.}~\bibnamefont {Roy}}, \bibinfo {author} {\bibfnamefont
  {K.}~\bibnamefont {Soni}}, \bibinfo {author} {\bibfnamefont {H.}~\bibnamefont
  {Tan}}, \bibinfo {author} {\bibfnamefont {V.~J.}\ \bibnamefont {Y}}, \bibinfo
  {author} {\bibfnamefont {A.~H.}\ \bibnamefont {Nitz}},\ and\ \bibinfo
  {author} {\bibfnamefont {C.~D.}\ \bibnamefont {Capano}},\ }\bibfield  {title}
  {\bibinfo {title} {{Potential science with GW250114 -- the loudest binary
  black hole merger detected to date}},\ }\href@noop {} {\  (\bibinfo {year}
  {2025})},\ \Eprint {https://arxiv.org/abs/2507.08789} {arXiv:2507.08789
  [gr-qc]} \BibitemShut {NoStop}%
\bibitem [{\citenamefont {Abac}\ \emph {et~al.}(2025)\citenamefont {Abac} \emph
  {et~al.}}]{LIGOScientific:2025rid}%
  \BibitemOpen
  \bibfield  {author} {\bibinfo {author} {\bibfnamefont {A.~G.}\ \bibnamefont
  {Abac}} \emph {et~al.} (\bibinfo {collaboration} {LIGO Scientific, Virgo,
  KAGRA}),\ }\bibfield  {title} {\bibinfo {title} {{GW250114: Testing
  Hawking{\textquoteright}s Area Law and the Kerr Nature of Black Holes}},\
  }\href {https://doi.org/10.1103/kw5g-d732} {\bibfield  {journal} {\bibinfo
  {journal} {Phys. Rev. Lett.}\ }\textbf {\bibinfo {volume} {135}},\ \bibinfo
  {pages} {111403} (\bibinfo {year} {2025})},\ \Eprint
  {https://arxiv.org/abs/2509.08054} {arXiv:2509.08054 [gr-qc]} \BibitemShut
  {NoStop}%
\bibitem [{\citenamefont {Berti}\ \emph {et~al.}(2025)\citenamefont {Berti}
  \emph {et~al.}}]{Berti:2025hly}%
  \BibitemOpen
  \bibfield  {author} {\bibinfo {author} {\bibfnamefont {E.}~\bibnamefont
  {Berti}} \emph {et~al.},\ }\bibfield  {title} {\bibinfo {title} {{Black hole
  spectroscopy: from theory to experiment}},\ }\href@noop {} {\  (\bibinfo
  {year} {2025})},\ \Eprint {https://arxiv.org/abs/2505.23895}
  {arXiv:2505.23895 [gr-qc]} \BibitemShut {NoStop}%
\bibitem [{\citenamefont {Cardoso}\ \emph {et~al.}(2025)\citenamefont
  {Cardoso}, \citenamefont {Biswas},\ and\ \citenamefont
  {Sarkar}}]{Cardoso:2025npr}%
  \BibitemOpen
  \bibfield  {author} {\bibinfo {author} {\bibfnamefont {V.}~\bibnamefont
  {Cardoso}}, \bibinfo {author} {\bibfnamefont {S.}~\bibnamefont {Biswas}},\
  and\ \bibinfo {author} {\bibfnamefont {S.}~\bibnamefont {Sarkar}},\
  }\bibfield  {title} {\bibinfo {title} {{The Physics of Black Holes and Their
  Environments: Consequences for Gravitational Wave Science}}\ }(\bibinfo
  {year} {2025})\ \Eprint {https://arxiv.org/abs/2511.14841} {arXiv:2511.14841
  [gr-qc]} \BibitemShut {NoStop}%
\bibitem [{\citenamefont {Della~Rocca}\ \emph {et~al.}(2025)\citenamefont
  {Della~Rocca}, \citenamefont {Spieksma}, \citenamefont {Duque}, \citenamefont
  {Gualtieri},\ and\ \citenamefont {Cardoso}}]{DellaRocca:2025xwz}%
  \BibitemOpen
  \bibfield  {author} {\bibinfo {author} {\bibfnamefont {M.}~\bibnamefont
  {Della~Rocca}}, \bibinfo {author} {\bibfnamefont {T.~F.~M.}\ \bibnamefont
  {Spieksma}}, \bibinfo {author} {\bibfnamefont {F.}~\bibnamefont {Duque}},
  \bibinfo {author} {\bibfnamefont {L.}~\bibnamefont {Gualtieri}},\ and\
  \bibinfo {author} {\bibfnamefont {V.}~\bibnamefont {Cardoso}},\ }\bibfield
  {title} {\bibinfo {title} {{Gravitational Atom Spectroscopy}},\ }\href@noop
  {} {\  (\bibinfo {year} {2025})},\ \Eprint {https://arxiv.org/abs/2511.13848}
  {arXiv:2511.13848 [gr-qc]} \BibitemShut {NoStop}%
\bibitem [{\citenamefont {Berti}\ \emph {et~al.}(2009)\citenamefont {Berti},
  \citenamefont {Cardoso},\ and\ \citenamefont {Starinets}}]{Berti2009}%
  \BibitemOpen
  \bibfield  {author} {\bibinfo {author} {\bibfnamefont {E.}~\bibnamefont
  {Berti}}, \bibinfo {author} {\bibfnamefont {V.}~\bibnamefont {Cardoso}},\
  and\ \bibinfo {author} {\bibfnamefont {A.~O.}\ \bibnamefont {Starinets}},\
  }\bibfield  {title} {\bibinfo {title} {{Quasinormal modes of black holes and
  black branes}},\ }\href {https://doi.org/10.1088/0264-9381/26/16/163001}
  {\bibfield  {journal} {\bibinfo  {journal} {Class. Quant. Grav.}\ }\textbf
  {\bibinfo {volume} {26}},\ \bibinfo {pages} {163001} (\bibinfo {year}
  {2009})},\ \Eprint {https://arxiv.org/abs/0905.2975} {arXiv:0905.2975
  [gr-qc]} \BibitemShut {NoStop}%
\bibitem [{\citenamefont {Konoplya}\ and\ \citenamefont
  {Zhidenko}(2011)}]{Konoplya:2011qq}%
  \BibitemOpen
  \bibfield  {author} {\bibinfo {author} {\bibfnamefont {R.~A.}\ \bibnamefont
  {Konoplya}}\ and\ \bibinfo {author} {\bibfnamefont {A.}~\bibnamefont
  {Zhidenko}},\ }\bibfield  {title} {\bibinfo {title} {{Quasinormal modes of
  black holes: From astrophysics to string theory}},\ }\href
  {https://doi.org/10.1103/RevModPhys.83.793} {\bibfield  {journal} {\bibinfo
  {journal} {Rev. Mod. Phys.}\ }\textbf {\bibinfo {volume} {83}},\ \bibinfo
  {pages} {793} (\bibinfo {year} {2011})},\ \Eprint
  {https://arxiv.org/abs/1102.4014} {arXiv:1102.4014 [gr-qc]} \BibitemShut
  {NoStop}%
\bibitem [{\citenamefont {Bian}\ \emph {et~al.}(2026)\citenamefont {Bian} \emph
  {et~al.}}]{Bian:2025ifp}%
  \BibitemOpen
  \bibfield  {author} {\bibinfo {author} {\bibfnamefont {L.}~\bibnamefont
  {Bian}} \emph {et~al.},\ }\bibfield  {title} {\bibinfo {title}
  {{Gravitational wave cosmology}},\ }\href
  {https://doi.org/10.1007/s11433-025-2740-8} {\bibfield  {journal} {\bibinfo
  {journal} {Sci. China Phys. Mech. Astron.}\ }\textbf {\bibinfo {volume}
  {69}},\ \bibinfo {pages} {210401} (\bibinfo {year} {2026})},\ \Eprint
  {https://arxiv.org/abs/2505.19747} {arXiv:2505.19747 [gr-qc]} \BibitemShut
  {NoStop}%
\bibitem [{\citenamefont {Allahyari}\ \emph {et~al.}(2026)\citenamefont
  {Allahyari}, \citenamefont {Davari},\ and\ \citenamefont
  {Mota}}]{Allahyari:2025sbt}%
  \BibitemOpen
  \bibfield  {author} {\bibinfo {author} {\bibfnamefont {A.}~\bibnamefont
  {Allahyari}}, \bibinfo {author} {\bibfnamefont {M.}~\bibnamefont {Davari}},\
  and\ \bibinfo {author} {\bibfnamefont {D.~F.}\ \bibnamefont {Mota}},\
  }\bibfield  {title} {\bibinfo {title} {{Lorentz violation with gravitational
  waves: Constraints from NANOGrav and IPTA data}},\ }\href
  {https://doi.org/10.1016/j.jheap.2025.100448} {\bibfield  {journal} {\bibinfo
   {journal} {JHEAp}\ }\textbf {\bibinfo {volume} {49}},\ \bibinfo {pages}
  {100448} (\bibinfo {year} {2026})},\ \Eprint
  {https://arxiv.org/abs/2505.22736} {arXiv:2505.22736 [astro-ph.CO]}
  \BibitemShut {NoStop}%
\bibitem [{\citenamefont {Aoulad~Lafkih}\ \emph {et~al.}(2025)\citenamefont
  {Aoulad~Lafkih}, \citenamefont {Angonin}, \citenamefont {Le~Poncin-Lafitte},\
  and\ \citenamefont {Nilsson}}]{AouladLafkih:2025stw}%
  \BibitemOpen
  \bibfield  {author} {\bibinfo {author} {\bibfnamefont {S.}~\bibnamefont
  {Aoulad~Lafkih}}, \bibinfo {author} {\bibfnamefont {M.-C.}\ \bibnamefont
  {Angonin}}, \bibinfo {author} {\bibfnamefont {C.}~\bibnamefont
  {Le~Poncin-Lafitte}},\ and\ \bibinfo {author} {\bibfnamefont {N.~A.}\
  \bibnamefont {Nilsson}},\ }\bibfield  {title} {\bibinfo {title}
  {{Gravitational-wave generation in the presence of Lorentz invariance
  violation}},\ }\href@noop {} {\  (\bibinfo {year} {2025})},\ \Eprint
  {https://arxiv.org/abs/2506.08859} {arXiv:2506.08859 [gr-qc]} \BibitemShut
  {NoStop}%
\bibitem [{\citenamefont {Zhang}\ \emph
  {et~al.}(2025{\natexlab{a}})\citenamefont {Zhang}, \citenamefont {Fu},\ and\
  \citenamefont {Gong}}]{Zhang:2024csc}%
  \BibitemOpen
  \bibfield  {author} {\bibinfo {author} {\bibfnamefont {C.}~\bibnamefont
  {Zhang}}, \bibinfo {author} {\bibfnamefont {G.}~\bibnamefont {Fu}},\ and\
  \bibinfo {author} {\bibfnamefont {Y.}~\bibnamefont {Gong}},\ }\bibfield
  {title} {\bibinfo {title} {{The constraint on modified black holes with
  extreme mass ratio inspirals}},\ }\href
  {https://doi.org/10.1140/epjc/s10052-025-14100-5} {\bibfield  {journal}
  {\bibinfo  {journal} {Eur. Phys. J. C}\ }\textbf {\bibinfo {volume} {85}},\
  \bibinfo {pages} {385} (\bibinfo {year} {2025}{\natexlab{a}})},\ \Eprint
  {https://arxiv.org/abs/2408.15064} {arXiv:2408.15064 [gr-qc]} \BibitemShut
  {NoStop}%
\bibitem [{\citenamefont {Zhang}\ \emph
  {et~al.}(2025{\natexlab{b}})\citenamefont {Zhang}, \citenamefont {Cai},
  \citenamefont {Fu}, \citenamefont {Gong}, \citenamefont {Lu},\ and\
  \citenamefont {Zhou}}]{Zhang:2025eqz}%
  \BibitemOpen
  \bibfield  {author} {\bibinfo {author} {\bibfnamefont {C.}~\bibnamefont
  {Zhang}}, \bibinfo {author} {\bibfnamefont {R.-g.}\ \bibnamefont {Cai}},
  \bibinfo {author} {\bibfnamefont {G.}~\bibnamefont {Fu}}, \bibinfo {author}
  {\bibfnamefont {Y.}~\bibnamefont {Gong}}, \bibinfo {author} {\bibfnamefont
  {X.}~\bibnamefont {Lu}},\ and\ \bibinfo {author} {\bibfnamefont
  {W.}~\bibnamefont {Zhou}},\ }\bibfield  {title} {\bibinfo {title} {{Generic
  effective source for gravitational self-force calculations in Schwarzschild
  spacetime}},\ }\href@noop {} {\  (\bibinfo {year} {2025}{\natexlab{b}})},\
  \Eprint {https://arxiv.org/abs/2505.19732} {arXiv:2505.19732 [gr-qc]}
  \BibitemShut {NoStop}%
\bibitem [{\citenamefont {Zi}\ and\ \citenamefont {Shu}(2025)}]{Zi:2025qos}%
  \BibitemOpen
  \bibfield  {author} {\bibinfo {author} {\bibfnamefont {T.}~\bibnamefont
  {Zi}}\ and\ \bibinfo {author} {\bibfnamefont {F.-W.}\ \bibnamefont {Shu}},\
  }\bibfield  {title} {\bibinfo {title} {{Eccentric extreme-mass-ratio
  inspirals: a new window into ultra-light vector fields}},\ }\href
  {https://doi.org/10.1140/epjc/s10052-025-14990-5} {\bibfield  {journal}
  {\bibinfo  {journal} {Eur. Phys. J. C}\ }\textbf {\bibinfo {volume} {85}},\
  \bibinfo {pages} {1251} (\bibinfo {year} {2025})},\ \Eprint
  {https://arxiv.org/abs/2510.22275} {arXiv:2510.22275 [gr-qc]} \BibitemShut
  {NoStop}%
\bibitem [{\citenamefont {Zi}\ and\ \citenamefont {Kumar}(2025)}]{Zi:2025lio}%
  \BibitemOpen
  \bibfield  {author} {\bibinfo {author} {\bibfnamefont {T.}~\bibnamefont
  {Zi}}\ and\ \bibinfo {author} {\bibfnamefont {S.}~\bibnamefont {Kumar}},\
  }\bibfield  {title} {\bibinfo {title} {{Probing scalar field with generic
  extreme mass-ratio inspirals around Kerr black holes}},\ }\href@noop {} {\
  (\bibinfo {year} {2025})},\ \Eprint {https://arxiv.org/abs/2508.00516}
  {arXiv:2508.00516 [gr-qc]} \BibitemShut {NoStop}%
\bibitem [{\citenamefont {Deng}\ \emph
  {et~al.}(2025{\natexlab{a}})\citenamefont {Deng}, \citenamefont {Long},
  \citenamefont {Tan},\ and\ \citenamefont {Jing}}]{Deng:2025wzz}%
  \BibitemOpen
  \bibfield  {author} {\bibinfo {author} {\bibfnamefont {W.}~\bibnamefont
  {Deng}}, \bibinfo {author} {\bibfnamefont {S.}~\bibnamefont {Long}}, \bibinfo
  {author} {\bibfnamefont {Q.}~\bibnamefont {Tan}},\ and\ \bibinfo {author}
  {\bibfnamefont {J.}~\bibnamefont {Jing}},\ }\bibfield  {title} {\bibinfo
  {title} {{Gravitational waveforms from periodic orbits around a charged black
  hole with scalar hair}},\ }\href@noop {} {\  (\bibinfo {year}
  {2025}{\natexlab{a}})},\ \Eprint {https://arxiv.org/abs/2510.24468}
  {arXiv:2510.24468 [gr-qc]} \BibitemShut {NoStop}%
\bibitem [{\citenamefont {Jing}\ \emph {et~al.}(2025)\citenamefont {Jing},
  \citenamefont {Long}, \citenamefont {Deng},\ and\ \citenamefont
  {Wang}}]{Jing:2025utt}%
  \BibitemOpen
  \bibfield  {author} {\bibinfo {author} {\bibfnamefont {J.}~\bibnamefont
  {Jing}}, \bibinfo {author} {\bibfnamefont {S.}~\bibnamefont {Long}}, \bibinfo
  {author} {\bibfnamefont {W.}~\bibnamefont {Deng}},\ and\ \bibinfo {author}
  {\bibfnamefont {J.}~\bibnamefont {Wang}},\ }\bibfield  {title} {\bibinfo
  {title} {{Effective one-body theory of spinless binary evolution dynamics}},\
  }\href {https://doi.org/10.1007/s11433-025-2758-5} {\bibfield  {journal}
  {\bibinfo  {journal} {Sci. China Phys. Mech. Astron.}\ }\textbf {\bibinfo
  {volume} {68}},\ \bibinfo {pages} {120411} (\bibinfo {year} {2025})},\
  \Eprint {https://arxiv.org/abs/2509.06448} {arXiv:2509.06448 [gr-qc]}
  \BibitemShut {NoStop}%
\bibitem [{\citenamefont {Deng}\ \emph
  {et~al.}(2025{\natexlab{b}})\citenamefont {Deng}, \citenamefont {Long},
  \citenamefont {Tan}, \citenamefont {Chen},\ and\ \citenamefont
  {Jing}}]{Deng:2025hfn}%
  \BibitemOpen
  \bibfield  {author} {\bibinfo {author} {\bibfnamefont {W.}~\bibnamefont
  {Deng}}, \bibinfo {author} {\bibfnamefont {S.}~\bibnamefont {Long}}, \bibinfo
  {author} {\bibfnamefont {Q.}~\bibnamefont {Tan}}, \bibinfo {author}
  {\bibfnamefont {Z.-C.}\ \bibnamefont {Chen}},\ and\ \bibinfo {author}
  {\bibfnamefont {J.}~\bibnamefont {Jing}},\ }\bibfield  {title} {\bibinfo
  {title} {{Scalar-gravitational quasinormal modes and echoes in a five
  dimensional thick brane}},\ }\href@noop {} {\  (\bibinfo {year}
  {2025}{\natexlab{b}})},\ \Eprint {https://arxiv.org/abs/2508.20937}
  {arXiv:2508.20937 [gr-qc]} \BibitemShut {NoStop}%
\bibitem [{\citenamefont {Tan}\ \emph {et~al.}(2025{\natexlab{a}})\citenamefont
  {Tan}, \citenamefont {Long}, \citenamefont {Deng},\ and\ \citenamefont
  {Jing}}]{Tan:2024qij}%
  \BibitemOpen
  \bibfield  {author} {\bibinfo {author} {\bibfnamefont {Q.}~\bibnamefont
  {Tan}}, \bibinfo {author} {\bibfnamefont {S.}~\bibnamefont {Long}}, \bibinfo
  {author} {\bibfnamefont {W.}~\bibnamefont {Deng}},\ and\ \bibinfo {author}
  {\bibfnamefont {J.}~\bibnamefont {Jing}},\ }\bibfield  {title} {\bibinfo
  {title} {{Quasinormal modes and echoes of a double braneworld}},\ }\href
  {https://doi.org/10.1007/JHEP02(2025)055} {\bibfield  {journal} {\bibinfo
  {journal} {JHEP}\ }\textbf {\bibinfo {volume} {02}},\ \bibinfo {pages}
  {055}},\ \Eprint {https://arxiv.org/abs/2410.06945} {arXiv:2410.06945
  [gr-qc]} \BibitemShut {NoStop}%
\bibitem [{\citenamefont {Tan}\ \emph {et~al.}(2025{\natexlab{b}})\citenamefont
  {Tan}, \citenamefont {Long}, \citenamefont {Deng},\ and\ \citenamefont
  {Jing}}]{Tan:2024aym}%
  \BibitemOpen
  \bibfield  {author} {\bibinfo {author} {\bibfnamefont {Q.}~\bibnamefont
  {Tan}}, \bibinfo {author} {\bibfnamefont {S.}~\bibnamefont {Long}}, \bibinfo
  {author} {\bibfnamefont {W.}~\bibnamefont {Deng}},\ and\ \bibinfo {author}
  {\bibfnamefont {J.}~\bibnamefont {Jing}},\ }\bibfield  {title} {\bibinfo
  {title} {{Graviscalar quasinormal modes and asymptotic tails of a thick
  brane}},\ }\href {https://doi.org/10.1016/j.physletb.2025.139667} {\bibfield
  {journal} {\bibinfo  {journal} {Phys. Lett. B}\ }\textbf {\bibinfo {volume}
  {868}},\ \bibinfo {pages} {139667} (\bibinfo {year} {2025}{\natexlab{b}})},\
  \Eprint {https://arxiv.org/abs/2409.06947} {arXiv:2409.06947 [gr-qc]}
  \BibitemShut {NoStop}%
\bibitem [{\citenamefont {Deng}\ \emph {et~al.}(2024)\citenamefont {Deng},
  \citenamefont {Long},\ and\ \citenamefont {Jing}}]{Deng:2024ayh}%
  \BibitemOpen
  \bibfield  {author} {\bibinfo {author} {\bibfnamefont {W.}~\bibnamefont
  {Deng}}, \bibinfo {author} {\bibfnamefont {S.}~\bibnamefont {Long}},\ and\
  \bibinfo {author} {\bibfnamefont {J.}~\bibnamefont {Jing}},\ }\bibfield
  {title} {\bibinfo {title} {{Energy flux and waveform of gravitational wave
  generated by coalescing slow-spinning binary system in effective one-body
  theory}},\ }\href@noop {} {\  (\bibinfo {year} {2024})},\ \Eprint
  {https://arxiv.org/abs/2405.16423} {arXiv:2405.16423 [gr-qc]} \BibitemShut
  {NoStop}%
\bibitem [{\citenamefont {Long}\ \emph {et~al.}(2024)\citenamefont {Long},
  \citenamefont {Deng},\ and\ \citenamefont {Jing}}]{Long:2024axi}%
  \BibitemOpen
  \bibfield  {author} {\bibinfo {author} {\bibfnamefont {S.}~\bibnamefont
  {Long}}, \bibinfo {author} {\bibfnamefont {W.}~\bibnamefont {Deng}},\ and\
  \bibinfo {author} {\bibfnamefont {J.}~\bibnamefont {Jing}},\ }\bibfield
  {title} {\bibinfo {title} {{Energy flux and waveforms by coalescing spinless
  binary system in effective one-body theory}},\ }\href
  {https://doi.org/10.1007/s11433-023-2354-1} {\bibfield  {journal} {\bibinfo
  {journal} {Sci. China Phys. Mech. Astron.}\ }\textbf {\bibinfo {volume}
  {67}},\ \bibinfo {pages} {260412} (\bibinfo {year} {2024})},\ \Eprint
  {https://arxiv.org/abs/2403.09211} {arXiv:2403.09211 [gr-qc]} \BibitemShut
  {NoStop}%
\bibitem [{\citenamefont {Long}\ \emph {et~al.}(2023)\citenamefont {Long},
  \citenamefont {Zou},\ and\ \citenamefont {Jing}}]{Long:2023vph}%
  \BibitemOpen
  \bibfield  {author} {\bibinfo {author} {\bibfnamefont {S.}~\bibnamefont
  {Long}}, \bibinfo {author} {\bibfnamefont {Y.}~\bibnamefont {Zou}},\ and\
  \bibinfo {author} {\bibfnamefont {J.}~\bibnamefont {Jing}},\ }\bibfield
  {title} {\bibinfo {title} {{Reconstruction of gravitational waveforms of
  coalescing spinless binaries in EOB theory based on PM approximation}},\
  }\href {https://doi.org/10.1088/1361-6382/acfdee} {\bibfield  {journal}
  {\bibinfo  {journal} {Class. Quant. Grav.}\ }\textbf {\bibinfo {volume}
  {40}},\ \bibinfo {pages} {225006} (\bibinfo {year} {2023})}\BibitemShut
  {NoStop}%
\bibitem [{\citenamefont {Zhang}\ and\ \citenamefont
  {Wang}(2024)}]{Zhang:2024svj}%
  \BibitemOpen
  \bibfield  {author} {\bibinfo {author} {\bibfnamefont {C.}~\bibnamefont
  {Zhang}}\ and\ \bibinfo {author} {\bibfnamefont {A.}~\bibnamefont {Wang}},\
  }\bibfield  {title} {\bibinfo {title} {{Quasi-normal modes of loop quantum
  black holes formed from gravitational collapse}},\ }\href
  {https://doi.org/10.1088/1475-7516/2024/10/070} {\bibfield  {journal}
  {\bibinfo  {journal} {JCAP}\ }\textbf {\bibinfo {volume} {10}},\ \bibinfo
  {pages} {070}},\ \Eprint {https://arxiv.org/abs/2407.19654} {arXiv:2407.19654
  [gr-qc]} \BibitemShut {NoStop}%
\bibitem [{\citenamefont {Kostelecky}\ and\ \citenamefont
  {Potting}(1991)}]{Kostelecky1991}%
  \BibitemOpen
  \bibfield  {author} {\bibinfo {author} {\bibfnamefont {V.~A.}\ \bibnamefont
  {Kostelecky}}\ and\ \bibinfo {author} {\bibfnamefont {R.}~\bibnamefont
  {Potting}},\ }\bibfield  {title} {\bibinfo {title} {{CPT and strings}},\
  }\href {https://doi.org/10.1016/0550-3213(91)90071-5} {\bibfield  {journal}
  {\bibinfo  {journal} {Nucl. Phys. B}\ }\textbf {\bibinfo {volume} {359}},\
  \bibinfo {pages} {545} (\bibinfo {year} {1991})}\BibitemShut {NoStop}%
\bibitem [{\citenamefont {Casana}\ \emph {et~al.}(2018)\citenamefont {Casana},
  \citenamefont {Cavalcante}, \citenamefont {Poulis},\ and\ \citenamefont
  {Santos}}]{Casana2018}%
  \BibitemOpen
  \bibfield  {author} {\bibinfo {author} {\bibfnamefont {R.}~\bibnamefont
  {Casana}}, \bibinfo {author} {\bibfnamefont {A.}~\bibnamefont {Cavalcante}},
  \bibinfo {author} {\bibfnamefont {F.~P.}\ \bibnamefont {Poulis}},\ and\
  \bibinfo {author} {\bibfnamefont {E.~B.}\ \bibnamefont {Santos}},\ }\bibfield
   {title} {\bibinfo {title} {{Exact Schwarzschild-like solution in a bumblebee
  gravity model}},\ }\href {https://doi.org/10.1103/PhysRevD.97.104001}
  {\bibfield  {journal} {\bibinfo  {journal} {Phys. Rev. D}\ }\textbf {\bibinfo
  {volume} {97}},\ \bibinfo {pages} {104001} (\bibinfo {year} {2018})},\
  \Eprint {https://arxiv.org/abs/1711.02273} {arXiv:1711.02273 [gr-qc]}
  \BibitemShut {NoStop}%
\bibitem [{\citenamefont {\"Ovg\"un}\ \emph {et~al.}(2019)\citenamefont
  {\"Ovg\"un}, \citenamefont {Jusufi},\ and\ \citenamefont
  {Sakall\i{}}}]{Ovgun2019}%
  \BibitemOpen
  \bibfield  {author} {\bibinfo {author} {\bibfnamefont {A.}~\bibnamefont
  {\"Ovg\"un}}, \bibinfo {author} {\bibfnamefont {K.}~\bibnamefont {Jusufi}},\
  and\ \bibinfo {author} {\bibfnamefont {I.}~\bibnamefont {Sakall\i{}}},\
  }\bibfield  {title} {\bibinfo {title} {{Exact traversable wormhole solution
  in bumblebee gravity}},\ }\href {https://doi.org/10.1103/PhysRevD.99.024042}
  {\bibfield  {journal} {\bibinfo  {journal} {Phys. Rev. D}\ }\textbf {\bibinfo
  {volume} {99}},\ \bibinfo {pages} {024042} (\bibinfo {year} {2019})},\
  \Eprint {https://arxiv.org/abs/1804.09911} {arXiv:1804.09911 [gr-qc]}
  \BibitemShut {NoStop}%
\bibitem [{\citenamefont {G\"ull\"u}\ and\ \citenamefont
  {\"Ovg\"un}(2022)}]{Gullu2020}%
  \BibitemOpen
  \bibfield  {author} {\bibinfo {author} {\bibfnamefont {I.}~\bibnamefont
  {G\"ull\"u}}\ and\ \bibinfo {author} {\bibfnamefont {A.}~\bibnamefont
  {\"Ovg\"un}},\ }\bibfield  {title} {\bibinfo {title} {{Schwarzschild-like
  black hole with a topological defect in bumblebee gravity}},\ }\href
  {https://doi.org/10.1016/j.aop.2021.168721} {\bibfield  {journal} {\bibinfo
  {journal} {Annals Phys.}\ }\textbf {\bibinfo {volume} {436}},\ \bibinfo
  {pages} {168721} (\bibinfo {year} {2022})},\ \Eprint
  {https://arxiv.org/abs/2012.02611} {arXiv:2012.02611 [gr-qc]} \BibitemShut
  {NoStop}%
\bibitem [{\citenamefont {Pan}\ \emph {et~al.}(2020)\citenamefont {Pan},
  \citenamefont {Qi}, \citenamefont {Cao}, \citenamefont {Liu}, \citenamefont
  {Liu}, \citenamefont {Geng}, \citenamefont {Lian},\ and\ \citenamefont
  {Zhu}}]{Pan:2020zbl}%
  \BibitemOpen
  \bibfield  {author} {\bibinfo {author} {\bibfnamefont {Y.}~\bibnamefont
  {Pan}}, \bibinfo {author} {\bibfnamefont {J.}~\bibnamefont {Qi}}, \bibinfo
  {author} {\bibfnamefont {S.}~\bibnamefont {Cao}}, \bibinfo {author}
  {\bibfnamefont {T.}~\bibnamefont {Liu}}, \bibinfo {author} {\bibfnamefont
  {Y.}~\bibnamefont {Liu}}, \bibinfo {author} {\bibfnamefont {S.}~\bibnamefont
  {Geng}}, \bibinfo {author} {\bibfnamefont {Y.}~\bibnamefont {Lian}},\ and\
  \bibinfo {author} {\bibfnamefont {Z.-H.}\ \bibnamefont {Zhu}},\ }\bibfield
  {title} {\bibinfo {title} {{Model-independent constraints on Lorentz
  invariance violation: implication from updated Gamma-ray burst
  observations}},\ }\href {https://doi.org/10.3847/1538-4357/ab6ef5} {\bibfield
   {journal} {\bibinfo  {journal} {Astrophys. J.}\ }\textbf {\bibinfo {volume}
  {890}},\ \bibinfo {pages} {169} (\bibinfo {year} {2020})},\ \Eprint
  {https://arxiv.org/abs/2001.08451} {arXiv:2001.08451 [astro-ph.CO]}
  \BibitemShut {NoStop}%
\bibitem [{\citenamefont {Liu}\ \emph {et~al.}(2022)\citenamefont {Liu},
  \citenamefont {Zhang},\ and\ \citenamefont {Meng}}]{Liu:2022xse}%
  \BibitemOpen
  \bibfield  {author} {\bibinfo {author} {\bibfnamefont {Z.-K.}\ \bibnamefont
  {Liu}}, \bibinfo {author} {\bibfnamefont {B.-B.}\ \bibnamefont {Zhang}},\
  and\ \bibinfo {author} {\bibfnamefont {Y.-Z.}\ \bibnamefont {Meng}},\
  }\bibfield  {title} {\bibinfo {title} {{Spectral Lag Transition of 32 Fermi
  Gamma-Ray Bursts and Their Application on Constraining Lorentz Invariance
  Violation}},\ }\href {https://doi.org/10.3847/1538-4357/ac81b9} {\bibfield
  {journal} {\bibinfo  {journal} {Astrophys. J.}\ }\textbf {\bibinfo {volume}
  {935}},\ \bibinfo {pages} {79} (\bibinfo {year} {2022})},\ \Eprint
  {https://arxiv.org/abs/2202.09999} {arXiv:2202.09999 [astro-ph.HE]}
  \BibitemShut {NoStop}%
\bibitem [{\citenamefont {Maluf}\ and\ \citenamefont
  {Neves}(2021)}]{Maluf2021}%
  \BibitemOpen
  \bibfield  {author} {\bibinfo {author} {\bibfnamefont {R.~V.}\ \bibnamefont
  {Maluf}}\ and\ \bibinfo {author} {\bibfnamefont {J.~C.~S.}\ \bibnamefont
  {Neves}},\ }\bibfield  {title} {\bibinfo {title} {{Black holes with a
  cosmological constant in bumblebee gravity}},\ }\href
  {https://doi.org/10.1103/PhysRevD.103.044002} {\bibfield  {journal} {\bibinfo
   {journal} {Phys. Rev. D}\ }\textbf {\bibinfo {volume} {103}},\ \bibinfo
  {pages} {044002} (\bibinfo {year} {2021})},\ \Eprint
  {https://arxiv.org/abs/2011.12841} {arXiv:2011.12841 [gr-qc]} \BibitemShut
  {NoStop}%
\bibitem [{\citenamefont {Xu}\ \emph {et~al.}(2023{\natexlab{a}})\citenamefont
  {Xu}, \citenamefont {Liang},\ and\ \citenamefont {Shao}}]{Xu:2022frb}%
  \BibitemOpen
  \bibfield  {author} {\bibinfo {author} {\bibfnamefont {R.}~\bibnamefont
  {Xu}}, \bibinfo {author} {\bibfnamefont {D.}~\bibnamefont {Liang}},\ and\
  \bibinfo {author} {\bibfnamefont {L.}~\bibnamefont {Shao}},\ }\bibfield
  {title} {\bibinfo {title} {{Static spherical vacuum solutions in the
  bumblebee gravity model}},\ }\href
  {https://doi.org/10.1103/PhysRevD.107.024011} {\bibfield  {journal} {\bibinfo
   {journal} {Phys. Rev. D}\ }\textbf {\bibinfo {volume} {107}},\ \bibinfo
  {pages} {024011} (\bibinfo {year} {2023}{\natexlab{a}})},\ \Eprint
  {https://arxiv.org/abs/2209.02209} {arXiv:2209.02209 [gr-qc]} \BibitemShut
  {NoStop}%
\bibitem [{\citenamefont {Ding}\ \emph {et~al.}(2022)\citenamefont {Ding},
  \citenamefont {Chen},\ and\ \citenamefont {Fu}}]{Ding2022}%
  \BibitemOpen
  \bibfield  {author} {\bibinfo {author} {\bibfnamefont {C.}~\bibnamefont
  {Ding}}, \bibinfo {author} {\bibfnamefont {X.}~\bibnamefont {Chen}},\ and\
  \bibinfo {author} {\bibfnamefont {X.}~\bibnamefont {Fu}},\ }\bibfield
  {title} {\bibinfo {title} {{Einstein-Gauss-Bonnet gravity coupled to
  bumblebee field in four dimensional spacetime}},\ }\href
  {https://doi.org/10.1016/j.nuclphysb.2022.115688} {\bibfield  {journal}
  {\bibinfo  {journal} {Nucl. Phys. B}\ }\textbf {\bibinfo {volume} {975}},\
  \bibinfo {pages} {115688} (\bibinfo {year} {2022})},\ \Eprint
  {https://arxiv.org/abs/2102.13335} {arXiv:2102.13335 [gr-qc]} \BibitemShut
  {NoStop}%
\bibitem [{\citenamefont {Poulis}\ and\ \citenamefont
  {Soares}(2022)}]{Poulis:2021nqh}%
  \BibitemOpen
  \bibfield  {author} {\bibinfo {author} {\bibfnamefont {F.~P.}\ \bibnamefont
  {Poulis}}\ and\ \bibinfo {author} {\bibfnamefont {M.~A.~C.}\ \bibnamefont
  {Soares}},\ }\bibfield  {title} {\bibinfo {title} {{Exact modifications on a
  vacuum spacetime due to a gradient bumblebee field at its vacuum expectation
  value}},\ }\href {https://doi.org/10.1140/epjc/s10052-022-10547-y} {\bibfield
   {journal} {\bibinfo  {journal} {Eur. Phys. J. C}\ }\textbf {\bibinfo
  {volume} {82}},\ \bibinfo {pages} {613} (\bibinfo {year} {2022})},\ \Eprint
  {https://arxiv.org/abs/2112.04040} {arXiv:2112.04040 [gr-qc]} \BibitemShut
  {NoStop}%
\bibitem [{\citenamefont {Mai}\ \emph {et~al.}(2023)\citenamefont {Mai},
  \citenamefont {Xu}, \citenamefont {Liang},\ and\ \citenamefont
  {Shao}}]{Mai:2023ggs}%
  \BibitemOpen
  \bibfield  {author} {\bibinfo {author} {\bibfnamefont {Z.-F.}\ \bibnamefont
  {Mai}}, \bibinfo {author} {\bibfnamefont {R.}~\bibnamefont {Xu}}, \bibinfo
  {author} {\bibfnamefont {D.}~\bibnamefont {Liang}},\ and\ \bibinfo {author}
  {\bibfnamefont {L.}~\bibnamefont {Shao}},\ }\bibfield  {title} {\bibinfo
  {title} {{Extended thermodynamics of the bumblebee black holes}},\ }\href
  {https://doi.org/10.1103/PhysRevD.108.024004} {\bibfield  {journal} {\bibinfo
   {journal} {Phys. Rev. D}\ }\textbf {\bibinfo {volume} {108}},\ \bibinfo
  {pages} {024004} (\bibinfo {year} {2023})},\ \Eprint
  {https://arxiv.org/abs/2304.08030} {arXiv:2304.08030 [gr-qc]} \BibitemShut
  {NoStop}%
\bibitem [{\citenamefont {Xu}\ \emph {et~al.}(2023{\natexlab{b}})\citenamefont
  {Xu}, \citenamefont {Liang},\ and\ \citenamefont {Shao}}]{Xu:2023xqh}%
  \BibitemOpen
  \bibfield  {author} {\bibinfo {author} {\bibfnamefont {R.}~\bibnamefont
  {Xu}}, \bibinfo {author} {\bibfnamefont {D.}~\bibnamefont {Liang}},\ and\
  \bibinfo {author} {\bibfnamefont {L.}~\bibnamefont {Shao}},\ }\bibfield
  {title} {\bibinfo {title} {{Bumblebee Black Holes in Light of Event Horizon
  Telescope Observations}},\ }\href {https://doi.org/10.3847/1538-4357/acbdfb}
  {\bibfield  {journal} {\bibinfo  {journal} {Astrophys. J.}\ }\textbf
  {\bibinfo {volume} {945}},\ \bibinfo {pages} {148} (\bibinfo {year}
  {2023}{\natexlab{b}})},\ \Eprint {https://arxiv.org/abs/2302.05671}
  {arXiv:2302.05671 [gr-qc]} \BibitemShut {NoStop}%
\bibitem [{\citenamefont {Zhang}\ \emph {et~al.}(2023)\citenamefont {Zhang},
  \citenamefont {Wang},\ and\ \citenamefont {Jing}}]{Zhang:2023wwk}%
  \BibitemOpen
  \bibfield  {author} {\bibinfo {author} {\bibfnamefont {X.}~\bibnamefont
  {Zhang}}, \bibinfo {author} {\bibfnamefont {M.}~\bibnamefont {Wang}},\ and\
  \bibinfo {author} {\bibfnamefont {J.}~\bibnamefont {Jing}},\ }\bibfield
  {title} {\bibinfo {title} {{Quasinormal modes and late time tails of
  perturbation fields on a Schwarzschild-like black hole with a global monopole
  in the Einstein-bumblebee theory}},\ }\href
  {https://doi.org/10.1007/s11433-023-2153-6} {\bibfield  {journal} {\bibinfo
  {journal} {Sci. China Phys. Mech. Astron.}\ }\textbf {\bibinfo {volume}
  {66}},\ \bibinfo {pages} {100411} (\bibinfo {year} {2023})},\ \Eprint
  {https://arxiv.org/abs/2307.10856} {arXiv:2307.10856 [gr-qc]} \BibitemShut
  {NoStop}%
\bibitem [{\citenamefont {Lin}\ \emph {et~al.}(2023)\citenamefont {Lin},
  \citenamefont {Jiang},\ and\ \citenamefont {Zhai}}]{Lin:2023foj}%
  \BibitemOpen
  \bibfield  {author} {\bibinfo {author} {\bibfnamefont {R.-H.}\ \bibnamefont
  {Lin}}, \bibinfo {author} {\bibfnamefont {R.}~\bibnamefont {Jiang}},\ and\
  \bibinfo {author} {\bibfnamefont {X.-H.}\ \bibnamefont {Zhai}},\ }\bibfield
  {title} {\bibinfo {title} {{Quasinormal modes of the spherical bumblebee
  black holes with a global monopole}},\ }\href
  {https://doi.org/10.1140/epjc/s10052-023-11899-9} {\bibfield  {journal}
  {\bibinfo  {journal} {Eur. Phys. J. C}\ }\textbf {\bibinfo {volume} {83}},\
  \bibinfo {pages} {720} (\bibinfo {year} {2023})},\ \Eprint
  {https://arxiv.org/abs/2308.01575} {arXiv:2308.01575 [gr-qc]} \BibitemShut
  {NoStop}%
\bibitem [{\citenamefont {Chen}\ \emph {et~al.}(2023)\citenamefont {Chen},
  \citenamefont {Pan},\ and\ \citenamefont {Jing}}]{Chen:2023cjd}%
  \BibitemOpen
  \bibfield  {author} {\bibinfo {author} {\bibfnamefont {C.}~\bibnamefont
  {Chen}}, \bibinfo {author} {\bibfnamefont {Q.}~\bibnamefont {Pan}},\ and\
  \bibinfo {author} {\bibfnamefont {J.}~\bibnamefont {Jing}},\ }\bibfield
  {title} {\bibinfo {title} {{Quasinormal modes of a scalar perturbation around
  a rotating BTZ-like black hole in Einstein-bumblebee gravity}},\ }\href
  {https://doi.org/10.1016/j.physletb.2023.138186} {\bibfield  {journal}
  {\bibinfo  {journal} {Phys. Lett. B}\ }\textbf {\bibinfo {volume} {846}},\
  \bibinfo {pages} {138186} (\bibinfo {year} {2023})},\ \Eprint
  {https://arxiv.org/abs/2302.05861} {arXiv:2302.05861 [gr-qc]} \BibitemShut
  {NoStop}%
\bibitem [{\citenamefont {Chen}\ \emph {et~al.}(2020)\citenamefont {Chen},
  \citenamefont {Wang},\ and\ \citenamefont {Jing}}]{Chen2020}%
  \BibitemOpen
  \bibfield  {author} {\bibinfo {author} {\bibfnamefont {S.}~\bibnamefont
  {Chen}}, \bibinfo {author} {\bibfnamefont {M.}~\bibnamefont {Wang}},\ and\
  \bibinfo {author} {\bibfnamefont {J.}~\bibnamefont {Jing}},\ }\bibfield
  {title} {\bibinfo {title} {{Polarization effects in Kerr black hole shadow
  due to the coupling between photon and bumblebee field}},\ }\href
  {https://doi.org/10.1007/JHEP07(2020)054} {\bibfield  {journal} {\bibinfo
  {journal} {JHEP}\ }\textbf {\bibinfo {volume} {07}},\ \bibinfo {pages}
  {054}},\ \Eprint {https://arxiv.org/abs/2004.08857} {arXiv:2004.08857
  [gr-qc]} \BibitemShut {NoStop}%
\bibitem [{\citenamefont {Wang}\ \emph {et~al.}(2022)\citenamefont {Wang},
  \citenamefont {Chen},\ and\ \citenamefont {Jing}}]{Wang:2021gtd}%
  \BibitemOpen
  \bibfield  {author} {\bibinfo {author} {\bibfnamefont {Z.}~\bibnamefont
  {Wang}}, \bibinfo {author} {\bibfnamefont {S.}~\bibnamefont {Chen}},\ and\
  \bibinfo {author} {\bibfnamefont {J.}~\bibnamefont {Jing}},\ }\bibfield
  {title} {\bibinfo {title} {{Constraint on parameters of a rotating black hole
  in Einstein-bumblebee theory by quasi-periodic oscillations}},\ }\href
  {https://doi.org/10.1140/epjc/s10052-022-10475-x} {\bibfield  {journal}
  {\bibinfo  {journal} {Eur. Phys. J. C}\ }\textbf {\bibinfo {volume} {82}},\
  \bibinfo {pages} {528} (\bibinfo {year} {2022})},\ \Eprint
  {https://arxiv.org/abs/2112.02895} {arXiv:2112.02895 [gr-qc]} \BibitemShut
  {NoStop}%
\bibitem [{\citenamefont {Mai}\ \emph {et~al.}(2024)\citenamefont {Mai},
  \citenamefont {Xu}, \citenamefont {Liang},\ and\ \citenamefont
  {Shao}}]{Mai:2024lgk}%
  \BibitemOpen
  \bibfield  {author} {\bibinfo {author} {\bibfnamefont {Z.-F.}\ \bibnamefont
  {Mai}}, \bibinfo {author} {\bibfnamefont {R.}~\bibnamefont {Xu}}, \bibinfo
  {author} {\bibfnamefont {D.}~\bibnamefont {Liang}},\ and\ \bibinfo {author}
  {\bibfnamefont {L.}~\bibnamefont {Shao}},\ }\bibfield  {title} {\bibinfo
  {title} {{Dynamic instability analysis for bumblebee black holes: The odd
  parity}},\ }\href {https://doi.org/10.1103/PhysRevD.109.084076} {\bibfield
  {journal} {\bibinfo  {journal} {Phys. Rev. D}\ }\textbf {\bibinfo {volume}
  {109}},\ \bibinfo {pages} {084076} (\bibinfo {year} {2024})},\ \Eprint
  {https://arxiv.org/abs/2401.07757} {arXiv:2401.07757 [gr-qc]} \BibitemShut
  {NoStop}%
\bibitem [{\citenamefont {Liang}\ \emph {et~al.}(2023)\citenamefont {Liang},
  \citenamefont {Xu}, \citenamefont {Mai},\ and\ \citenamefont
  {Shao}}]{Liang:2022gdk}%
  \BibitemOpen
  \bibfield  {author} {\bibinfo {author} {\bibfnamefont {D.}~\bibnamefont
  {Liang}}, \bibinfo {author} {\bibfnamefont {R.}~\bibnamefont {Xu}}, \bibinfo
  {author} {\bibfnamefont {Z.-F.}\ \bibnamefont {Mai}},\ and\ \bibinfo {author}
  {\bibfnamefont {L.}~\bibnamefont {Shao}},\ }\bibfield  {title} {\bibinfo
  {title} {{Probing vector hair of black holes with extreme-mass-ratio
  inspirals}},\ }\href {https://doi.org/10.1103/PhysRevD.107.044053} {\bibfield
   {journal} {\bibinfo  {journal} {Phys. Rev. D}\ }\textbf {\bibinfo {volume}
  {107}},\ \bibinfo {pages} {044053} (\bibinfo {year} {2023})},\ \Eprint
  {https://arxiv.org/abs/2212.09346} {arXiv:2212.09346 [gr-qc]} \BibitemShut
  {NoStop}%
\bibitem [{\citenamefont {Hosseinifar}\ \emph {et~al.}(2024)\citenamefont
  {Hosseinifar}, \citenamefont {Filho}, \citenamefont {Zhang}, \citenamefont
  {Chen},\ and\ \citenamefont {Hassanabadi}}]{Hosseinifar:2024wwe}%
  \BibitemOpen
  \bibfield  {author} {\bibinfo {author} {\bibfnamefont {F.}~\bibnamefont
  {Hosseinifar}}, \bibinfo {author} {\bibfnamefont {A.~A.~A.}\ \bibnamefont
  {Filho}}, \bibinfo {author} {\bibfnamefont {M.~Y.}\ \bibnamefont {Zhang}},
  \bibinfo {author} {\bibfnamefont {H.}~\bibnamefont {Chen}},\ and\ \bibinfo
  {author} {\bibfnamefont {H.}~\bibnamefont {Hassanabadi}},\ }\bibfield
  {title} {\bibinfo {title} {{Shadows, greybody factors, emission rate,
  topological charge, and phase transitions for a charged black hole with a
  Kalb-Ramond field background}},\ }\href@noop {} {\  (\bibinfo {year}
  {2024})},\ \Eprint {https://arxiv.org/abs/2407.07017} {arXiv:2407.07017
  [gr-qc]} \BibitemShut {NoStop}%
\bibitem [{\citenamefont {Finke}\ and\ \citenamefont
  {Patel}(2024)}]{Finke:2024ada}%
  \BibitemOpen
  \bibfield  {author} {\bibinfo {author} {\bibfnamefont {J.~D.}\ \bibnamefont
  {Finke}}\ and\ \bibinfo {author} {\bibfnamefont {P.}~\bibnamefont {Patel}},\
  }\bibfield  {title} {\bibinfo {title} {{Probing Lorentz Invariance Violation
  with Absorption of Astrophysical \ensuremath{\gamma}-Rays by Solar
  Photons}},\ }\href {https://doi.org/10.3847/1538-4357/ad3212} {\bibfield
  {journal} {\bibinfo  {journal} {Astrophys. J.}\ }\textbf {\bibinfo {volume}
  {965}},\ \bibinfo {pages} {44} (\bibinfo {year} {2024})},\ \Eprint
  {https://arxiv.org/abs/2403.07063} {arXiv:2403.07063 [astro-ph.HE]}
  \BibitemShut {NoStop}%
\bibitem [{\citenamefont {Liu}\ \emph {et~al.}(2024{\natexlab{a}})\citenamefont
  {Liu}, \citenamefont {Guo}, \citenamefont {Wei},\ and\ \citenamefont
  {Liu}}]{Liu:2024axg}%
  \BibitemOpen
  \bibfield  {author} {\bibinfo {author} {\bibfnamefont {J.-Z.}\ \bibnamefont
  {Liu}}, \bibinfo {author} {\bibfnamefont {W.-D.}\ \bibnamefont {Guo}},
  \bibinfo {author} {\bibfnamefont {S.-W.}\ \bibnamefont {Wei}},\ and\ \bibinfo
  {author} {\bibfnamefont {Y.-X.}\ \bibnamefont {Liu}},\ }\bibfield  {title}
  {\bibinfo {title} {{Charged spherically symmetric and slowly rotating charged
  black hole solutions in bumblebee gravity}},\ }\href@noop {} {\  (\bibinfo
  {year} {2024}{\natexlab{a}})},\ \Eprint {https://arxiv.org/abs/2407.08396}
  {arXiv:2407.08396 [gr-qc]} \BibitemShut {NoStop}%
\bibitem [{\citenamefont {Li}\ \emph {et~al.}(2025{\natexlab{a}})\citenamefont
  {Li}, \citenamefont {Liu}, \citenamefont {Guo},\ and\ \citenamefont
  {Liu}}]{Li:2025itp}%
  \BibitemOpen
  \bibfield  {author} {\bibinfo {author} {\bibfnamefont {B.-R.}\ \bibnamefont
  {Li}}, \bibinfo {author} {\bibfnamefont {J.-Z.}\ \bibnamefont {Liu}},
  \bibinfo {author} {\bibfnamefont {W.-D.}\ \bibnamefont {Guo}},\ and\ \bibinfo
  {author} {\bibfnamefont {Y.-X.}\ \bibnamefont {Liu}},\ }\bibfield  {title}
  {\bibinfo {title} {{Quasinormal modes of a charged spherically symmetric
  black hole in bumblebee gravity}},\ }\href@noop {} {\  (\bibinfo {year}
  {2025}{\natexlab{a}})},\ \Eprint {https://arxiv.org/abs/2510.20503}
  {arXiv:2510.20503 [gr-qc]} \BibitemShut {NoStop}%
\bibitem [{\citenamefont {Liu}\ \emph {et~al.}(2025{\natexlab{a}})\citenamefont
  {Liu}, \citenamefont {Wu}, \citenamefont {Wei},\ and\ \citenamefont
  {Liu}}]{Liu:2025oho}%
  \BibitemOpen
  \bibfield  {author} {\bibinfo {author} {\bibfnamefont {J.-Z.}\ \bibnamefont
  {Liu}}, \bibinfo {author} {\bibfnamefont {S.-P.}\ \bibnamefont {Wu}},
  \bibinfo {author} {\bibfnamefont {S.-W.}\ \bibnamefont {Wei}},\ and\ \bibinfo
  {author} {\bibfnamefont {Y.-X.}\ \bibnamefont {Liu}},\ }\bibfield  {title}
  {\bibinfo {title} {{Exact Black Hole Solutions in Bumblebee Gravity with
  Lightlike or Spacelike VEVS}},\ }\href@noop {} {\  (\bibinfo {year}
  {2025}{\natexlab{a}})},\ \Eprint {https://arxiv.org/abs/2510.16731}
  {arXiv:2510.16731 [gr-qc]} \BibitemShut {NoStop}%
\bibitem [{\citenamefont {Gu}\ \emph {et~al.}(2025)\citenamefont {Gu},
  \citenamefont {Guo},\ and\ \citenamefont {Liu}}]{Gu:2025lyz}%
  \BibitemOpen
  \bibfield  {author} {\bibinfo {author} {\bibfnamefont {Y.-T.}\ \bibnamefont
  {Gu}}, \bibinfo {author} {\bibfnamefont {W.-D.}\ \bibnamefont {Guo}},\ and\
  \bibinfo {author} {\bibfnamefont {Y.-X.}\ \bibnamefont {Liu}},\ }\bibfield
  {title} {\bibinfo {title} {{Quasinormal modes of an electrically charged
  Kalb-Ramond black hole}},\ }\href@noop {} {\  (\bibinfo {year} {2025})},\
  \Eprint {https://arxiv.org/abs/2509.23732} {arXiv:2509.23732 [gr-qc]}
  \BibitemShut {NoStop}%
\bibitem [{\citenamefont {Lai}\ \emph {et~al.}(2025)\citenamefont {Lai},
  \citenamefont {Dong}, \citenamefont {Fan},\ and\ \citenamefont
  {Liu}}]{Lai:2025nyo}%
  \BibitemOpen
  \bibfield  {author} {\bibinfo {author} {\bibfnamefont {X.-B.}\ \bibnamefont
  {Lai}}, \bibinfo {author} {\bibfnamefont {Y.-Q.}\ \bibnamefont {Dong}},
  \bibinfo {author} {\bibfnamefont {Y.-Z.}\ \bibnamefont {Fan}},\ and\ \bibinfo
  {author} {\bibfnamefont {Y.-X.}\ \bibnamefont {Liu}},\ }\bibfield  {title}
  {\bibinfo {title} {{Stability Analysis of Cosmological Perturbations in the
  Bumblebee Model: Parameter Constraints and Gravitational Waves}},\
  }\href@noop {} {\  (\bibinfo {year} {2025})},\ \Eprint
  {https://arxiv.org/abs/2509.13958} {arXiv:2509.13958 [gr-qc]} \BibitemShut
  {NoStop}%
\bibitem [{\citenamefont {Li}\ \emph {et~al.}(2025{\natexlab{b}})\citenamefont
  {Li}, \citenamefont {Liang},\ and\ \citenamefont {Ma}}]{Li:2025tcd}%
  \BibitemOpen
  \bibfield  {author} {\bibinfo {author} {\bibfnamefont {S.}~\bibnamefont
  {Li}}, \bibinfo {author} {\bibfnamefont {L.}~\bibnamefont {Liang}},\ and\
  \bibinfo {author} {\bibfnamefont {L.}~\bibnamefont {Ma}},\ }\bibfield
  {title} {\bibinfo {title} {{Dyonic RN-like and Taub-NUT-like black holes in
  Einstein-bumblebee gravity}},\ }\href@noop {} {\  (\bibinfo {year}
  {2025}{\natexlab{b}})},\ \Eprint {https://arxiv.org/abs/2510.04405}
  {arXiv:2510.04405 [gr-qc]} \BibitemShut {NoStop}%
\bibitem [{\citenamefont {Chen}\ and\ \citenamefont
  {Liu}(2025)}]{Chen:2025ypx}%
  \BibitemOpen
  \bibfield  {author} {\bibinfo {author} {\bibfnamefont {Y.-Q.}\ \bibnamefont
  {Chen}}\ and\ \bibinfo {author} {\bibfnamefont {H.-S.}\ \bibnamefont {Liu}},\
  }\bibfield  {title} {\bibinfo {title} {{Taub-NUT-like black holes in
  Einstein-bumblebee gravity}},\ }\href {https://doi.org/10.1103/wmhj-83x3}
  {\bibfield  {journal} {\bibinfo  {journal} {Phys. Rev. D}\ }\textbf {\bibinfo
  {volume} {112}},\ \bibinfo {pages} {084040} (\bibinfo {year} {2025})},\
  \Eprint {https://arxiv.org/abs/2505.23104} {arXiv:2505.23104 [gr-qc]}
  \BibitemShut {NoStop}%
\bibitem [{\citenamefont {Sekhmani}\ \emph {et~al.}(2025)\citenamefont
  {Sekhmani}, \citenamefont {Liu}, \citenamefont {Deng},\ and\ \citenamefont
  {Boshkayev}}]{Sekhmani:2025zen}%
  \BibitemOpen
  \bibfield  {author} {\bibinfo {author} {\bibfnamefont {Y.}~\bibnamefont
  {Sekhmani}}, \bibinfo {author} {\bibfnamefont {W.}~\bibnamefont {Liu}},
  \bibinfo {author} {\bibfnamefont {W.}~\bibnamefont {Deng}},\ and\ \bibinfo
  {author} {\bibfnamefont {K.}~\bibnamefont {Boshkayev}},\ }\bibfield  {title}
  {\bibinfo {title} {{Quasinormal Modes of Massive Scalar Perturbations in
  Slow-Rotation Bumblebee Black Holes with Traceless Conformal
  Electrodynamics}},\ }\href@noop {} {\  (\bibinfo {year} {2025})},\ \Eprint
  {https://arxiv.org/abs/2510.16639} {arXiv:2510.16639 [gr-qc]} \BibitemShut
  {NoStop}%
\bibitem [{\citenamefont {Shi}\ and\ \citenamefont
  {Ara{\'u}jo~Filho}(2025)}]{Shi:2025tvu}%
  \BibitemOpen
  \bibfield  {author} {\bibinfo {author} {\bibfnamefont {Y.}~\bibnamefont
  {Shi}}\ and\ \bibinfo {author} {\bibfnamefont {A.~A.}\ \bibnamefont
  {Ara{\'u}jo~Filho}},\ }\bibfield  {title} {\bibinfo {title} {{Neutrino
  oscillations induced by a new bumblebee black hole}},\ }\href@noop {} {\
  (\bibinfo {year} {2025})},\ \Eprint {https://arxiv.org/abs/2511.16181}
  {arXiv:2511.16181 [gr-qc]} \BibitemShut {NoStop}%
\bibitem [{\citenamefont {Ara{\'u}jo~Filho}\ \emph
  {et~al.}(2025{\natexlab{a}})\citenamefont {Ara{\'u}jo~Filho}, \citenamefont
  {Heidari}, \citenamefont {Lobo},\ and\ \citenamefont
  {Bezerra}}]{AraujoFilho:2025zaj}%
  \BibitemOpen
  \bibfield  {author} {\bibinfo {author} {\bibfnamefont {A.~A.}\ \bibnamefont
  {Ara{\'u}jo~Filho}}, \bibinfo {author} {\bibfnamefont {N.}~\bibnamefont
  {Heidari}}, \bibinfo {author} {\bibfnamefont {I.~P.}\ \bibnamefont {Lobo}},\
  and\ \bibinfo {author} {\bibfnamefont {V.~B.}\ \bibnamefont {Bezerra}},\
  }\bibfield  {title} {\bibinfo {title} {{Gravitational aspects of a new
  bumblebee black hole}},\ }\href@noop {} {\  (\bibinfo {year}
  {2025}{\natexlab{a}})},\ \Eprint {https://arxiv.org/abs/2511.12839}
  {arXiv:2511.12839 [gr-qc]} \BibitemShut {NoStop}%
\bibitem [{\citenamefont {Kalb}\ and\ \citenamefont
  {Ramond}(1974)}]{Kalb:1974yc}%
  \BibitemOpen
  \bibfield  {author} {\bibinfo {author} {\bibfnamefont {M.}~\bibnamefont
  {Kalb}}\ and\ \bibinfo {author} {\bibfnamefont {P.}~\bibnamefont {Ramond}},\
  }\bibfield  {title} {\bibinfo {title} {{Classical direct interstring
  action}},\ }\href {https://doi.org/10.1103/PhysRevD.9.2273} {\bibfield
  {journal} {\bibinfo  {journal} {Phys. Rev. D}\ }\textbf {\bibinfo {volume}
  {9}},\ \bibinfo {pages} {2273} (\bibinfo {year} {1974})}\BibitemShut
  {NoStop}%
\bibitem [{\citenamefont {Kao}\ \emph {et~al.}(1996)\citenamefont {Kao},
  \citenamefont {Dai}, \citenamefont {Wang}, \citenamefont {Chyi},\ and\
  \citenamefont {Lin}}]{Kao:1996ea}%
  \BibitemOpen
  \bibfield  {author} {\bibinfo {author} {\bibfnamefont {W.~F.}\ \bibnamefont
  {Kao}}, \bibinfo {author} {\bibfnamefont {W.~B.}\ \bibnamefont {Dai}},
  \bibinfo {author} {\bibfnamefont {S.-Y.}\ \bibnamefont {Wang}}, \bibinfo
  {author} {\bibfnamefont {T.-K.}\ \bibnamefont {Chyi}},\ and\ \bibinfo
  {author} {\bibfnamefont {S.-Y.}\ \bibnamefont {Lin}},\ }\bibfield  {title}
  {\bibinfo {title} {{Induced Einstein-Kalb-Ramond theory and the black
  hole}},\ }\href {https://doi.org/10.1103/PhysRevD.53.2244} {\bibfield
  {journal} {\bibinfo  {journal} {Phys. Rev. D}\ }\textbf {\bibinfo {volume}
  {53}},\ \bibinfo {pages} {2244} (\bibinfo {year} {1996})}\BibitemShut
  {NoStop}%
\bibitem [{\citenamefont {Chakraborty}\ and\ \citenamefont
  {SenGupta}(2016)}]{Chakraborty:2014fva}%
  \BibitemOpen
  \bibfield  {author} {\bibinfo {author} {\bibfnamefont {S.}~\bibnamefont
  {Chakraborty}}\ and\ \bibinfo {author} {\bibfnamefont {S.}~\bibnamefont
  {SenGupta}},\ }\bibfield  {title} {\bibinfo {title} {{Solutions on a brane in
  a bulk spacetime with Kalb-Ramond field}},\ }\href
  {https://doi.org/10.1016/j.aop.2016.01.023} {\bibfield  {journal} {\bibinfo
  {journal} {Annals Phys.}\ }\textbf {\bibinfo {volume} {367}},\ \bibinfo
  {pages} {258} (\bibinfo {year} {2016})},\ \Eprint
  {https://arxiv.org/abs/1412.7783} {arXiv:1412.7783 [gr-qc]} \BibitemShut
  {NoStop}%
\bibitem [{\citenamefont {Maluf}\ and\ \citenamefont
  {Muniz}(2022)}]{Maluf:2021ywn}%
  \BibitemOpen
  \bibfield  {author} {\bibinfo {author} {\bibfnamefont {R.~V.}\ \bibnamefont
  {Maluf}}\ and\ \bibinfo {author} {\bibfnamefont {C.~R.}\ \bibnamefont
  {Muniz}},\ }\bibfield  {title} {\bibinfo {title} {{Exact solution for a
  traversable wormhole in a curvature-coupled antisymmetric background
  field}},\ }\href {https://doi.org/10.1140/epjc/s10052-022-10409-7} {\bibfield
   {journal} {\bibinfo  {journal} {Eur. Phys. J. C}\ }\textbf {\bibinfo
  {volume} {82}},\ \bibinfo {pages} {445} (\bibinfo {year} {2022})},\ \Eprint
  {https://arxiv.org/abs/2110.12202} {arXiv:2110.12202 [gr-qc]} \BibitemShut
  {NoStop}%
\bibitem [{\citenamefont {Duan}\ \emph {et~al.}(2023)\citenamefont {Duan},
  \citenamefont {Zhao},\ and\ \citenamefont {Yang}}]{Duan:2023gng}%
  \BibitemOpen
  \bibfield  {author} {\bibinfo {author} {\bibfnamefont {Z.-Q.}\ \bibnamefont
  {Duan}}, \bibinfo {author} {\bibfnamefont {J.-Y.}\ \bibnamefont {Zhao}},\
  and\ \bibinfo {author} {\bibfnamefont {K.}~\bibnamefont {Yang}},\ }\bibfield
  {title} {\bibinfo {title} {{Electrically charged black holes in gravity with
  a background Kalb-Ramond field}},\ }\href@noop {} {\  (\bibinfo {year}
  {2023})},\ \Eprint {https://arxiv.org/abs/2310.13555} {arXiv:2310.13555
  [gr-qc]} \BibitemShut {NoStop}%
\bibitem [{\citenamefont {Yang}\ \emph {et~al.}(2023)\citenamefont {Yang},
  \citenamefont {Chen}, \citenamefont {Duan},\ and\ \citenamefont
  {Zhao}}]{Yang:2023wtu}%
  \BibitemOpen
  \bibfield  {author} {\bibinfo {author} {\bibfnamefont {K.}~\bibnamefont
  {Yang}}, \bibinfo {author} {\bibfnamefont {Y.-Z.}\ \bibnamefont {Chen}},
  \bibinfo {author} {\bibfnamefont {Z.-Q.}\ \bibnamefont {Duan}},\ and\
  \bibinfo {author} {\bibfnamefont {J.-Y.}\ \bibnamefont {Zhao}},\ }\bibfield
  {title} {\bibinfo {title} {{Static and spherically symmetric black holes in
  gravity with a background Kalb-Ramond field}},\ }\href
  {https://doi.org/10.1103/PhysRevD.108.124004} {\bibfield  {journal} {\bibinfo
   {journal} {Phys. Rev. D}\ }\textbf {\bibinfo {volume} {108}},\ \bibinfo
  {pages} {124004} (\bibinfo {year} {2023})},\ \Eprint
  {https://arxiv.org/abs/2308.06613} {arXiv:2308.06613 [gr-qc]} \BibitemShut
  {NoStop}%
\bibitem [{\citenamefont {Liu}\ \emph {et~al.}(2024{\natexlab{b}})\citenamefont
  {Liu}, \citenamefont {Wu},\ and\ \citenamefont {Wang}}]{Liu:2024oas}%
  \BibitemOpen
  \bibfield  {author} {\bibinfo {author} {\bibfnamefont {W.}~\bibnamefont
  {Liu}}, \bibinfo {author} {\bibfnamefont {D.}~\bibnamefont {Wu}},\ and\
  \bibinfo {author} {\bibfnamefont {J.}~\bibnamefont {Wang}},\ }\bibfield
  {title} {\bibinfo {title} {{Static neutral black holes in Kalb-Ramond
  gravity}},\ }\href {https://doi.org/10.1088/1475-7516/2024/09/017} {\bibfield
   {journal} {\bibinfo  {journal} {JCAP}\ }\textbf {\bibinfo {volume} {09}},\
  \bibinfo {pages} {017}},\ \Eprint {https://arxiv.org/abs/2406.13461}
  {arXiv:2406.13461 [hep-th]} \BibitemShut {NoStop}%
\bibitem [{\citenamefont {Liu}\ \emph {et~al.}(2025{\natexlab{b}})\citenamefont
  {Liu}, \citenamefont {Wu}, \citenamefont {Wei},\ and\ \citenamefont
  {Liu}}]{Liu:2025fxj}%
  \BibitemOpen
  \bibfield  {author} {\bibinfo {author} {\bibfnamefont {J.-Z.}\ \bibnamefont
  {Liu}}, \bibinfo {author} {\bibfnamefont {S.-P.}\ \bibnamefont {Wu}},
  \bibinfo {author} {\bibfnamefont {S.-W.}\ \bibnamefont {Wei}},\ and\ \bibinfo
  {author} {\bibfnamefont {Y.-X.}\ \bibnamefont {Liu}},\ }\bibfield  {title}
  {\bibinfo {title} {{Exact black hole solutions in gravity with a background
  Kalb-Ramond field}},\ }\href {https://doi.org/10.1088/1475-7516/2025/11/056}
  {\bibfield  {journal} {\bibinfo  {journal} {JCAP}\ }\textbf {\bibinfo
  {volume} {11}},\ \bibinfo {pages} {056}},\ \Eprint
  {https://arxiv.org/abs/2505.07404} {arXiv:2505.07404 [gr-qc]} \BibitemShut
  {NoStop}%
\bibitem [{\citenamefont {Ara{\'u}jo~Filho}\ \emph
  {et~al.}(2025{\natexlab{b}})\citenamefont {Ara{\'u}jo~Filho}, \citenamefont
  {Heidari},\ and\ \citenamefont {Lobo}}]{AraujoFilho:2025jcu}%
  \BibitemOpen
  \bibfield  {author} {\bibinfo {author} {\bibfnamefont {A.~A.}\ \bibnamefont
  {Ara{\'u}jo~Filho}}, \bibinfo {author} {\bibfnamefont {N.}~\bibnamefont
  {Heidari}},\ and\ \bibinfo {author} {\bibfnamefont {I.~P.}\ \bibnamefont
  {Lobo}},\ }\bibfield  {title} {\bibinfo {title} {{A non-commutative
  Kalb-Ramond black hole}},\ }\href
  {https://doi.org/10.1088/1475-7516/2025/09/076} {\bibfield  {journal}
  {\bibinfo  {journal} {JCAP}\ }\textbf {\bibinfo {volume} {09}},\ \bibinfo
  {pages} {076}},\ \Eprint {https://arxiv.org/abs/2507.17390} {arXiv:2507.17390
  [gr-qc]} \BibitemShut {NoStop}%
\bibitem [{\citenamefont {Guo}\ \emph {et~al.}(2023{\natexlab{a}})\citenamefont
  {Guo}, \citenamefont {Tan},\ and\ \citenamefont {Liu}}]{Guo:2023nkd}%
  \BibitemOpen
  \bibfield  {author} {\bibinfo {author} {\bibfnamefont {W.-D.}\ \bibnamefont
  {Guo}}, \bibinfo {author} {\bibfnamefont {Q.}~\bibnamefont {Tan}},\ and\
  \bibinfo {author} {\bibfnamefont {Y.-X.}\ \bibnamefont {Liu}},\ }\bibfield
  {title} {\bibinfo {title} {{Quasinormal modes and greybody factor of a
  Lorentz-violating black hole}},\ }\href@noop {} {\  (\bibinfo {year}
  {2023}{\natexlab{a}})},\ \Eprint {https://arxiv.org/abs/2312.16605}
  {arXiv:2312.16605 [gr-qc]} \BibitemShut {NoStop}%
\bibitem [{\citenamefont {Filho}\ \emph
  {et~al.}(2024{\natexlab{a}})\citenamefont {Filho}, \citenamefont {Reis},\
  and\ \citenamefont {Hassanabadi}}]{Filho:2023ycx}%
  \BibitemOpen
  \bibfield  {author} {\bibinfo {author} {\bibfnamefont {A.~A.~A.}\
  \bibnamefont {Filho}}, \bibinfo {author} {\bibfnamefont {J.~A. A.~S.}\
  \bibnamefont {Reis}},\ and\ \bibinfo {author} {\bibfnamefont
  {H.}~\bibnamefont {Hassanabadi}},\ }\bibfield  {title} {\bibinfo {title}
  {{Exploring antisymmetric tensor effects on black hole shadows and
  quasinormal frequencies}},\ }\href
  {https://doi.org/10.1088/1475-7516/2024/05/029} {\bibfield  {journal}
  {\bibinfo  {journal} {JCAP}\ }\textbf {\bibinfo {volume} {05}},\ \bibinfo
  {pages} {029}},\ \Eprint {https://arxiv.org/abs/2309.15778} {arXiv:2309.15778
  [gr-qc]} \BibitemShut {NoStop}%
\bibitem [{\citenamefont {Ara\'ujo~Filho}\ \emph {et~al.}(2024)\citenamefont
  {Ara\'ujo~Filho}, \citenamefont {Heidari}, \citenamefont {Reis},\ and\
  \citenamefont {Hassanabadi}}]{AraujoFilho:2024rcr}%
  \BibitemOpen
  \bibfield  {author} {\bibinfo {author} {\bibfnamefont {A.~A.}\ \bibnamefont
  {Ara\'ujo~Filho}}, \bibinfo {author} {\bibfnamefont {N.}~\bibnamefont
  {Heidari}}, \bibinfo {author} {\bibfnamefont {J.~A. A.~S.}\ \bibnamefont
  {Reis}},\ and\ \bibinfo {author} {\bibfnamefont {H.}~\bibnamefont
  {Hassanabadi}},\ }\bibfield  {title} {\bibinfo {title} {{The impact of an
  antisymmetric tensor on charged black holes: evaporation process, geodesics,
  deflection angle, scattering effects and quasinormal modes}},\ }\href@noop {}
  {\  (\bibinfo {year} {2024})},\ \Eprint {https://arxiv.org/abs/2404.10721}
  {arXiv:2404.10721 [gr-qc]} \BibitemShut {NoStop}%
\bibitem [{\citenamefont {Jha}(2024)}]{Jha:2024xtr}%
  \BibitemOpen
  \bibfield  {author} {\bibinfo {author} {\bibfnamefont {S.~K.}\ \bibnamefont
  {Jha}},\ }\bibfield  {title} {\bibinfo {title} {{Observational signature of
  Lorentz violation in Kalb-Ramond field model and Bumblebee model: A
  comprehensive comparative study}},\ }\href@noop {} {\  (\bibinfo {year}
  {2024})},\ \Eprint {https://arxiv.org/abs/2404.15808} {arXiv:2404.15808
  [gr-qc]} \BibitemShut {NoStop}%
\bibitem [{\citenamefont {Junior}\ \emph
  {et~al.}(2024{\natexlab{a}})\citenamefont {Junior}, \citenamefont {Junior},
  \citenamefont {Lobo}, \citenamefont {Rodrigues}, \citenamefont
  {Rubiera-Garcia}, \citenamefont {da~Silva},\ and\ \citenamefont
  {Vieira}}]{Junior:2024ety}%
  \BibitemOpen
  \bibfield  {author} {\bibinfo {author} {\bibfnamefont {E.~L.~B.}\
  \bibnamefont {Junior}}, \bibinfo {author} {\bibfnamefont {J.~T. S.~S.}\
  \bibnamefont {Junior}}, \bibinfo {author} {\bibfnamefont {F.~S.~N.}\
  \bibnamefont {Lobo}}, \bibinfo {author} {\bibfnamefont {M.~E.}\ \bibnamefont
  {Rodrigues}}, \bibinfo {author} {\bibfnamefont {D.}~\bibnamefont
  {Rubiera-Garcia}}, \bibinfo {author} {\bibfnamefont {L.~F.~D.}\ \bibnamefont
  {da~Silva}},\ and\ \bibinfo {author} {\bibfnamefont {H.~A.}\ \bibnamefont
  {Vieira}},\ }\bibfield  {title} {\bibinfo {title} {{Spontaneous Lorentz
  symmetry-breaking constraints in Kalb-Ramond gravity}},\ }\href@noop {} {\
  (\bibinfo {year} {2024}{\natexlab{a}})},\ \Eprint
  {https://arxiv.org/abs/2405.03291} {arXiv:2405.03291 [gr-qc]} \BibitemShut
  {NoStop}%
\bibitem [{\citenamefont {Junior}\ \emph
  {et~al.}(2024{\natexlab{b}})\citenamefont {Junior}, \citenamefont {Junior},
  \citenamefont {Lobo}, \citenamefont {Rodrigues}, \citenamefont
  {Rubiera-Garcia}, \citenamefont {da~Silva},\ and\ \citenamefont
  {Vieira}}]{Junior:2024vdk}%
  \BibitemOpen
  \bibfield  {author} {\bibinfo {author} {\bibfnamefont {E.~L.~B.}\
  \bibnamefont {Junior}}, \bibinfo {author} {\bibfnamefont {J.~T. S.~S.}\
  \bibnamefont {Junior}}, \bibinfo {author} {\bibfnamefont {F.~S.~N.}\
  \bibnamefont {Lobo}}, \bibinfo {author} {\bibfnamefont {M.~E.}\ \bibnamefont
  {Rodrigues}}, \bibinfo {author} {\bibfnamefont {D.}~\bibnamefont
  {Rubiera-Garcia}}, \bibinfo {author} {\bibfnamefont {L.~F.~D.}\ \bibnamefont
  {da~Silva}},\ and\ \bibinfo {author} {\bibfnamefont {H.~A.}\ \bibnamefont
  {Vieira}},\ }\bibfield  {title} {\bibinfo {title} {{Gravitational lensing of
  a Schwarzschild-like black hole in Kalb-Ramond gravity}},\ }\href@noop {} {\
  (\bibinfo {year} {2024}{\natexlab{b}})},\ \Eprint
  {https://arxiv.org/abs/2405.03284} {arXiv:2405.03284 [gr-qc]} \BibitemShut
  {NoStop}%
\bibitem [{\citenamefont {Filho}\ \emph
  {et~al.}(2024{\natexlab{b}})\citenamefont {Filho}, \citenamefont {Heidari},
  \citenamefont {Reis},\ and\ \citenamefont {Hassanabadi}}]{Filho:2024kbq}%
  \BibitemOpen
  \bibfield  {author} {\bibinfo {author} {\bibfnamefont {A.~A.~A.}\
  \bibnamefont {Filho}}, \bibinfo {author} {\bibfnamefont {N.}~\bibnamefont
  {Heidari}}, \bibinfo {author} {\bibfnamefont {J.~A. A.~S.}\ \bibnamefont
  {Reis}},\ and\ \bibinfo {author} {\bibfnamefont {H.}~\bibnamefont
  {Hassanabadi}},\ }\bibfield  {title} {\bibinfo {title} {{The impact of an
  antisymmetric tensor on charged black holes: evaporation process, geodesics,
  deflection angle, scattering effects and quasinormal modes}},\ }\href@noop {}
  {\  (\bibinfo {year} {2024}{\natexlab{b}})},\ \Eprint
  {https://arxiv.org/abs/2404.10721} {arXiv:2404.10721 [gr-qc]} \BibitemShut
  {NoStop}%
\bibitem [{\citenamefont {Du}\ \emph {et~al.}(2024)\citenamefont {Du},
  \citenamefont {Li}, \citenamefont {Ma},\ and\ \citenamefont
  {Gu}}]{Du:2024uhd}%
  \BibitemOpen
  \bibfield  {author} {\bibinfo {author} {\bibfnamefont {Y.-Z.}\ \bibnamefont
  {Du}}, \bibinfo {author} {\bibfnamefont {H.-F.}\ \bibnamefont {Li}}, \bibinfo
  {author} {\bibfnamefont {Y.-B.}\ \bibnamefont {Ma}},\ and\ \bibinfo {author}
  {\bibfnamefont {Q.}~\bibnamefont {Gu}},\ }\bibfield  {title} {\bibinfo
  {title} {{Phase structure of the de Sitter Spacetime with KR field based on
  the Lyapunov exponent}},\ }\href@noop {} {\  (\bibinfo {year} {2024})},\
  \Eprint {https://arxiv.org/abs/2403.20083} {arXiv:2403.20083 [hep-th]}
  \BibitemShut {NoStop}%
\bibitem [{\citenamefont {Filho}(2024)}]{Filho:2024tgy}%
  \BibitemOpen
  \bibfield  {author} {\bibinfo {author} {\bibfnamefont {A.~A.~A.}\
  \bibnamefont {Filho}},\ }\bibfield  {title} {\bibinfo {title} {{Antisymmetric
  tensor influence on charged black hole lensing phenomena and time delay}},\
  }\href@noop {} {\  (\bibinfo {year} {2024})},\ \Eprint
  {https://arxiv.org/abs/2406.11582} {arXiv:2406.11582 [gr-qc]} \BibitemShut
  {NoStop}%
\bibitem [{\citenamefont {Liu}\ \emph {et~al.}(2025{\natexlab{c}})\citenamefont
  {Liu}, \citenamefont {Wu},\ and\ \citenamefont {Wang}}]{Liu:2024lve}%
  \BibitemOpen
  \bibfield  {author} {\bibinfo {author} {\bibfnamefont {W.}~\bibnamefont
  {Liu}}, \bibinfo {author} {\bibfnamefont {D.}~\bibnamefont {Wu}},\ and\
  \bibinfo {author} {\bibfnamefont {J.}~\bibnamefont {Wang}},\ }\bibfield
  {title} {\bibinfo {title} {{Shadow of slowly rotating Kalb-Ramond black
  holes}},\ }\href {https://doi.org/10.1088/1475-7516/2025/05/017} {\bibfield
  {journal} {\bibinfo  {journal} {JCAP}\ }\textbf {\bibinfo {volume} {05}},\
  \bibinfo {pages} {017}},\ \Eprint {https://arxiv.org/abs/2407.07416}
  {arXiv:2407.07416 [gr-qc]} \BibitemShut {NoStop}%
\bibitem [{\citenamefont {Roulet}\ and\ \citenamefont
  {Zaldarriaga}(2019)}]{Roulet:2018jbe}%
  \BibitemOpen
  \bibfield  {author} {\bibinfo {author} {\bibfnamefont {J.}~\bibnamefont
  {Roulet}}\ and\ \bibinfo {author} {\bibfnamefont {M.}~\bibnamefont
  {Zaldarriaga}},\ }\bibfield  {title} {\bibinfo {title} {{Constraints on
  binary black hole populations from LIGO\textendash{}Virgo detections}},\
  }\href {https://doi.org/10.1093/mnras/stz226} {\bibfield  {journal} {\bibinfo
   {journal} {Mon. Not. Roy. Astron. Soc.}\ }\textbf {\bibinfo {volume}
  {484}},\ \bibinfo {pages} {4216} (\bibinfo {year} {2019})},\ \Eprint
  {https://arxiv.org/abs/1806.10610} {arXiv:1806.10610 [astro-ph.HE]}
  \BibitemShut {NoStop}%
\bibitem [{\citenamefont {Abbott}\ \emph {et~al.}(2020)\citenamefont {Abbott}
  \emph {et~al.}}]{LIGOScientific:2020zkf}%
  \BibitemOpen
  \bibfield  {author} {\bibinfo {author} {\bibfnamefont {R.}~\bibnamefont
  {Abbott}} \emph {et~al.} (\bibinfo {collaboration} {LIGO Scientific,
  Virgo}),\ }\bibfield  {title} {\bibinfo {title} {{GW190814: Gravitational
  Waves from the Coalescence of a 23 Solar Mass Black Hole with a 2.6 Solar
  Mass Compact Object}},\ }\href {https://doi.org/10.3847/2041-8213/ab960f}
  {\bibfield  {journal} {\bibinfo  {journal} {Astrophys. J. Lett.}\ }\textbf
  {\bibinfo {volume} {896}},\ \bibinfo {pages} {L44} (\bibinfo {year}
  {2020})},\ \Eprint {https://arxiv.org/abs/2006.12611} {arXiv:2006.12611
  [astro-ph.HE]} \BibitemShut {NoStop}%
\bibitem [{\citenamefont {Pani}(2013)}]{Pani2013IJMPA}%
  \BibitemOpen
  \bibfield  {author} {\bibinfo {author} {\bibfnamefont {P.}~\bibnamefont
  {Pani}},\ }\bibfield  {title} {\bibinfo {title} {{Advanced Methods in
  Black-Hole Perturbation Theory}},\ }\href
  {https://doi.org/10.1142/S0217751X13400186} {\bibfield  {journal} {\bibinfo
  {journal} {Int. J. Mod. Phys. A}\ }\textbf {\bibinfo {volume} {28}},\
  \bibinfo {pages} {1340018} (\bibinfo {year} {2013})},\ \Eprint
  {https://arxiv.org/abs/1305.6759} {arXiv:1305.6759 [gr-qc]} \BibitemShut
  {NoStop}%
\bibitem [{\citenamefont {Pani}\ \emph {et~al.}(2011)\citenamefont {Pani},
  \citenamefont {Macedo}, \citenamefont {Crispino},\ and\ \citenamefont
  {Cardoso}}]{Pani2011}%
  \BibitemOpen
  \bibfield  {author} {\bibinfo {author} {\bibfnamefont {P.}~\bibnamefont
  {Pani}}, \bibinfo {author} {\bibfnamefont {C.~F.~B.}\ \bibnamefont {Macedo}},
  \bibinfo {author} {\bibfnamefont {L.~C.~B.}\ \bibnamefont {Crispino}},\ and\
  \bibinfo {author} {\bibfnamefont {V.}~\bibnamefont {Cardoso}},\ }\bibfield
  {title} {\bibinfo {title} {{Slowly rotating black holes in alternative
  theories of gravity}},\ }\href {https://doi.org/10.1103/PhysRevD.84.087501}
  {\bibfield  {journal} {\bibinfo  {journal} {Phys. Rev. D}\ }\textbf {\bibinfo
  {volume} {84}},\ \bibinfo {pages} {087501} (\bibinfo {year} {2011})},\
  \Eprint {https://arxiv.org/abs/1109.3996} {arXiv:1109.3996 [gr-qc]}
  \BibitemShut {NoStop}%
\bibitem [{\citenamefont {Pani}\ \emph
  {et~al.}(2012{\natexlab{a}})\citenamefont {Pani}, \citenamefont {Cardoso},
  \citenamefont {Gualtieri}, \citenamefont {Berti},\ and\ \citenamefont
  {Ishibashi}}]{Pani2012}%
  \BibitemOpen
  \bibfield  {author} {\bibinfo {author} {\bibfnamefont {P.}~\bibnamefont
  {Pani}}, \bibinfo {author} {\bibfnamefont {V.}~\bibnamefont {Cardoso}},
  \bibinfo {author} {\bibfnamefont {L.}~\bibnamefont {Gualtieri}}, \bibinfo
  {author} {\bibfnamefont {E.}~\bibnamefont {Berti}},\ and\ \bibinfo {author}
  {\bibfnamefont {A.}~\bibnamefont {Ishibashi}},\ }\bibfield  {title} {\bibinfo
  {title} {{Perturbations of slowly rotating black holes: massive vector fields
  in the Kerr metric}},\ }\href {https://doi.org/10.1103/PhysRevD.86.104017}
  {\bibfield  {journal} {\bibinfo  {journal} {Phys. Rev. D}\ }\textbf {\bibinfo
  {volume} {86}},\ \bibinfo {pages} {104017} (\bibinfo {year}
  {2012}{\natexlab{a}})},\ \Eprint {https://arxiv.org/abs/1209.0773}
  {arXiv:1209.0773 [gr-qc]} \BibitemShut {NoStop}%
\bibitem [{\citenamefont {Pani}\ \emph
  {et~al.}(2013{\natexlab{a}})\citenamefont {Pani}, \citenamefont {Berti},\
  and\ \citenamefont {Gualtieri}}]{Pani2013prd}%
  \BibitemOpen
  \bibfield  {author} {\bibinfo {author} {\bibfnamefont {P.}~\bibnamefont
  {Pani}}, \bibinfo {author} {\bibfnamefont {E.}~\bibnamefont {Berti}},\ and\
  \bibinfo {author} {\bibfnamefont {L.}~\bibnamefont {Gualtieri}},\ }\bibfield
  {title} {\bibinfo {title} {{Scalar, Electromagnetic and Gravitational
  Perturbations of Kerr-Newman Black Holes in the Slow-Rotation Limit}},\
  }\href {https://doi.org/10.1103/PhysRevD.88.064048} {\bibfield  {journal}
  {\bibinfo  {journal} {Phys. Rev. D}\ }\textbf {\bibinfo {volume} {88}},\
  \bibinfo {pages} {064048} (\bibinfo {year} {2013}{\natexlab{a}})},\ \Eprint
  {https://arxiv.org/abs/1307.7315} {arXiv:1307.7315 [gr-qc]} \BibitemShut
  {NoStop}%
\bibitem [{\citenamefont {Pani}\ \emph
  {et~al.}(2012{\natexlab{b}})\citenamefont {Pani}, \citenamefont {Cardoso},
  \citenamefont {Gualtieri}, \citenamefont {Berti},\ and\ \citenamefont
  {Ishibashi}}]{Pani2012prl}%
  \BibitemOpen
  \bibfield  {author} {\bibinfo {author} {\bibfnamefont {P.}~\bibnamefont
  {Pani}}, \bibinfo {author} {\bibfnamefont {V.}~\bibnamefont {Cardoso}},
  \bibinfo {author} {\bibfnamefont {L.}~\bibnamefont {Gualtieri}}, \bibinfo
  {author} {\bibfnamefont {E.}~\bibnamefont {Berti}},\ and\ \bibinfo {author}
  {\bibfnamefont {A.}~\bibnamefont {Ishibashi}},\ }\bibfield  {title} {\bibinfo
  {title} {{Black hole bombs and photon mass bounds}},\ }\href
  {https://doi.org/10.1103/PhysRevLett.109.131102} {\bibfield  {journal}
  {\bibinfo  {journal} {Phys. Rev. Lett.}\ }\textbf {\bibinfo {volume} {109}},\
  \bibinfo {pages} {131102} (\bibinfo {year} {2012}{\natexlab{b}})},\ \Eprint
  {https://arxiv.org/abs/1209.0465} {arXiv:1209.0465 [gr-qc]} \BibitemShut
  {NoStop}%
\bibitem [{\citenamefont {Pani}\ \emph
  {et~al.}(2013{\natexlab{b}})\citenamefont {Pani}, \citenamefont {Berti},\
  and\ \citenamefont {Gualtieri}}]{Pani2013prl}%
  \BibitemOpen
  \bibfield  {author} {\bibinfo {author} {\bibfnamefont {P.}~\bibnamefont
  {Pani}}, \bibinfo {author} {\bibfnamefont {E.}~\bibnamefont {Berti}},\ and\
  \bibinfo {author} {\bibfnamefont {L.}~\bibnamefont {Gualtieri}},\ }\bibfield
  {title} {\bibinfo {title} {{Gravitoelectromagnetic Perturbations of
  Kerr-Newman Black Holes: Stability and Isospectrality in the Slow-Rotation
  Limit}},\ }\href {https://doi.org/10.1103/PhysRevLett.110.241103} {\bibfield
  {journal} {\bibinfo  {journal} {Phys. Rev. Lett.}\ }\textbf {\bibinfo
  {volume} {110}},\ \bibinfo {pages} {241103} (\bibinfo {year}
  {2013}{\natexlab{b}})},\ \Eprint {https://arxiv.org/abs/1304.1160}
  {arXiv:1304.1160 [gr-qc]} \BibitemShut {NoStop}%
\bibitem [{\citenamefont {Liu}\ \emph {et~al.}(2023{\natexlab{a}})\citenamefont
  {Liu}, \citenamefont {Fang}, \citenamefont {Jing},\ and\ \citenamefont
  {Wang}}]{Liu:2022dcn}%
  \BibitemOpen
  \bibfield  {author} {\bibinfo {author} {\bibfnamefont {W.}~\bibnamefont
  {Liu}}, \bibinfo {author} {\bibfnamefont {X.}~\bibnamefont {Fang}}, \bibinfo
  {author} {\bibfnamefont {J.}~\bibnamefont {Jing}},\ and\ \bibinfo {author}
  {\bibfnamefont {J.}~\bibnamefont {Wang}},\ }\bibfield  {title} {\bibinfo
  {title} {{QNMs of slowly rotating Einstein{\textendash}Bumblebee black
  hole}},\ }\href {https://doi.org/10.1140/epjc/s10052-023-11231-5} {\bibfield
  {journal} {\bibinfo  {journal} {Eur. Phys. J. C}\ }\textbf {\bibinfo {volume}
  {83}},\ \bibinfo {pages} {83} (\bibinfo {year} {2023}{\natexlab{a}})},\
  \Eprint {https://arxiv.org/abs/2211.03156} {arXiv:2211.03156 [gr-qc]}
  \BibitemShut {NoStop}%
\bibitem [{\citenamefont {Altschul}\ \emph {et~al.}(2010)\citenamefont
  {Altschul}, \citenamefont {Bailey},\ and\ \citenamefont
  {Kostelecky}}]{Altschul:2009ae}%
  \BibitemOpen
  \bibfield  {author} {\bibinfo {author} {\bibfnamefont {B.}~\bibnamefont
  {Altschul}}, \bibinfo {author} {\bibfnamefont {Q.~G.}\ \bibnamefont
  {Bailey}},\ and\ \bibinfo {author} {\bibfnamefont {V.~A.}\ \bibnamefont
  {Kostelecky}},\ }\bibfield  {title} {\bibinfo {title} {{Lorentz violation
  with an antisymmetric tensor}},\ }\href
  {https://doi.org/10.1103/PhysRevD.81.065028} {\bibfield  {journal} {\bibinfo
  {journal} {Phys. Rev. D}\ }\textbf {\bibinfo {volume} {81}},\ \bibinfo
  {pages} {065028} (\bibinfo {year} {2010})},\ \Eprint
  {https://arxiv.org/abs/0912.4852} {arXiv:0912.4852 [gr-qc]} \BibitemShut
  {NoStop}%
\bibitem [{\citenamefont {Lessa}\ \emph {et~al.}(2020)\citenamefont {Lessa},
  \citenamefont {Silva}, \citenamefont {Maluf},\ and\ \citenamefont
  {Almeida}}]{Lessa:2019bgi}%
  \BibitemOpen
  \bibfield  {author} {\bibinfo {author} {\bibfnamefont {L.~A.}\ \bibnamefont
  {Lessa}}, \bibinfo {author} {\bibfnamefont {J.~E.~G.}\ \bibnamefont {Silva}},
  \bibinfo {author} {\bibfnamefont {R.~V.}\ \bibnamefont {Maluf}},\ and\
  \bibinfo {author} {\bibfnamefont {C.~A.~S.}\ \bibnamefont {Almeida}},\
  }\bibfield  {title} {\bibinfo {title} {{Modified black hole solution with a
  background Kalb\textendash{}Ramond field}},\ }\href
  {https://doi.org/10.1140/epjc/s10052-020-7902-1} {\bibfield  {journal}
  {\bibinfo  {journal} {Eur. Phys. J. C}\ }\textbf {\bibinfo {volume} {80}},\
  \bibinfo {pages} {335} (\bibinfo {year} {2020})},\ \Eprint
  {https://arxiv.org/abs/1911.10296} {arXiv:1911.10296 [gr-qc]} \BibitemShut
  {NoStop}%
\bibitem [{\citenamefont {Bluhm}\ \emph {et~al.}(2008)\citenamefont {Bluhm},
  \citenamefont {Fung},\ and\ \citenamefont {Kostelecky}}]{Bluhm:2007bd}%
  \BibitemOpen
  \bibfield  {author} {\bibinfo {author} {\bibfnamefont {R.}~\bibnamefont
  {Bluhm}}, \bibinfo {author} {\bibfnamefont {S.-H.}\ \bibnamefont {Fung}},\
  and\ \bibinfo {author} {\bibfnamefont {V.~A.}\ \bibnamefont {Kostelecky}},\
  }\bibfield  {title} {\bibinfo {title} {{Spontaneous Lorentz and
  Diffeomorphism Violation, Massive Modes, and Gravity}},\ }\href
  {https://doi.org/10.1103/PhysRevD.77.065020} {\bibfield  {journal} {\bibinfo
  {journal} {Phys. Rev. D}\ }\textbf {\bibinfo {volume} {77}},\ \bibinfo
  {pages} {065020} (\bibinfo {year} {2008})},\ \Eprint
  {https://arxiv.org/abs/0712.4119} {arXiv:0712.4119 [hep-th]} \BibitemShut
  {NoStop}%
\bibitem [{\citenamefont {Regge}\ and\ \citenamefont
  {Wheeler}(1957)}]{ReggeWheeler1957}%
  \BibitemOpen
  \bibfield  {author} {\bibinfo {author} {\bibfnamefont {T.}~\bibnamefont
  {Regge}}\ and\ \bibinfo {author} {\bibfnamefont {J.~A.}\ \bibnamefont
  {Wheeler}},\ }\bibfield  {title} {\bibinfo {title} {{Stability of a
  Schwarzschild singularity}},\ }\href
  {https://doi.org/10.1103/PhysRev.108.1063} {\bibfield  {journal} {\bibinfo
  {journal} {Phys. Rev.}\ }\textbf {\bibinfo {volume} {108}},\ \bibinfo {pages}
  {1063} (\bibinfo {year} {1957})}\BibitemShut {NoStop}%
\bibitem [{\citenamefont {Zerilli}(1970)}]{Zerilli1970PRD}%
  \BibitemOpen
  \bibfield  {author} {\bibinfo {author} {\bibfnamefont {F.~J.}\ \bibnamefont
  {Zerilli}},\ }\bibfield  {title} {\bibinfo {title} {{Gravitational field of a
  particle falling in a schwarzschild geometry analyzed in tensor harmonics}},\
  }\href {https://doi.org/10.1103/PhysRevD.2.2141} {\bibfield  {journal}
  {\bibinfo  {journal} {Phys. Rev. D}\ }\textbf {\bibinfo {volume} {2}},\
  \bibinfo {pages} {2141} (\bibinfo {year} {1970})}\BibitemShut {NoStop}%
\bibitem [{\citenamefont {Liu}\ \emph {et~al.}(2023{\natexlab{b}})\citenamefont
  {Liu}, \citenamefont {Fang}, \citenamefont {Jing},\ and\ \citenamefont
  {Wang}}]{Liu:2022csl}%
  \BibitemOpen
  \bibfield  {author} {\bibinfo {author} {\bibfnamefont {W.}~\bibnamefont
  {Liu}}, \bibinfo {author} {\bibfnamefont {X.}~\bibnamefont {Fang}}, \bibinfo
  {author} {\bibfnamefont {J.}~\bibnamefont {Jing}},\ and\ \bibinfo {author}
  {\bibfnamefont {A.}~\bibnamefont {Wang}},\ }\bibfield  {title} {\bibinfo
  {title} {{Gauge invariant perturbations of general spherically symmetric
  spacetimes}},\ }\href {https://doi.org/10.1007/s11433-022-1956-4} {\bibfield
  {journal} {\bibinfo  {journal} {Sci. China Phys. Mech. Astron.}\ }\textbf
  {\bibinfo {volume} {66}},\ \bibinfo {pages} {210411} (\bibinfo {year}
  {2023}{\natexlab{b}})},\ \Eprint {https://arxiv.org/abs/2201.01259}
  {arXiv:2201.01259 [gr-qc]} \BibitemShut {NoStop}%
\bibitem [{\citenamefont {Kojima}(1992)}]{Kojima}%
  \BibitemOpen
  \bibfield  {author} {\bibinfo {author} {\bibfnamefont {Y.}~\bibnamefont
  {Kojima}},\ }\bibfield  {title} {\bibinfo {title} {{Equations governing the
  nonradial oscillations of a slowly rotating relativistic star}},\ }\href
  {https://doi.org/10.1103/PhysRevD.46.4289} {\bibfield  {journal} {\bibinfo
  {journal} {Phys. Rev. D}\ }\textbf {\bibinfo {volume} {46}},\ \bibinfo
  {pages} {4289} (\bibinfo {year} {1992})}\BibitemShut {NoStop}%
\bibitem [{\citenamefont {Lin}\ and\ \citenamefont {Qian}(2017)}]{Lin:2016sch}%
  \BibitemOpen
  \bibfield  {author} {\bibinfo {author} {\bibfnamefont {K.}~\bibnamefont
  {Lin}}\ and\ \bibinfo {author} {\bibfnamefont {W.-L.}\ \bibnamefont {Qian}},\
  }\bibfield  {title} {\bibinfo {title} {{A Matrix Method for Quasinormal
  Modes: Schwarzschild Black Holes in Asymptotically Flat and (Anti-) de Sitter
  Spacetimes}},\ }\href {https://doi.org/10.1088/1361-6382/aa6643} {\bibfield
  {journal} {\bibinfo  {journal} {Class. Quant. Grav.}\ }\textbf {\bibinfo
  {volume} {34}},\ \bibinfo {pages} {095004} (\bibinfo {year} {2017})},\
  \Eprint {https://arxiv.org/abs/1610.08135} {arXiv:1610.08135 [gr-qc]}
  \BibitemShut {NoStop}%
\bibitem [{\citenamefont {Lin}\ \emph {et~al.}(2017)\citenamefont {Lin},
  \citenamefont {Qian}, \citenamefont {Pavan},\ and\ \citenamefont
  {Abdalla}}]{Lin:2017oag}%
  \BibitemOpen
  \bibfield  {author} {\bibinfo {author} {\bibfnamefont {K.}~\bibnamefont
  {Lin}}, \bibinfo {author} {\bibfnamefont {W.-L.}\ \bibnamefont {Qian}},
  \bibinfo {author} {\bibfnamefont {A.~B.}\ \bibnamefont {Pavan}},\ and\
  \bibinfo {author} {\bibfnamefont {E.}~\bibnamefont {Abdalla}},\ }\bibfield
  {title} {\bibinfo {title} {{A matrix method for quasinormal modes: Kerr and
  Kerr{\textendash}Sen black holes}},\ }\href
  {https://doi.org/10.1142/S0217732317501346} {\bibfield  {journal} {\bibinfo
  {journal} {Mod. Phys. Lett. A}\ }\textbf {\bibinfo {volume} {32}},\ \bibinfo
  {pages} {1750134} (\bibinfo {year} {2017})},\ \Eprint
  {https://arxiv.org/abs/1703.06439} {arXiv:1703.06439 [gr-qc]} \BibitemShut
  {NoStop}%
\bibitem [{\citenamefont {Lin}\ and\ \citenamefont {Qian}(2019)}]{Lin:2019mmf}%
  \BibitemOpen
  \bibfield  {author} {\bibinfo {author} {\bibfnamefont {K.}~\bibnamefont
  {Lin}}\ and\ \bibinfo {author} {\bibfnamefont {W.-L.}\ \bibnamefont {Qian}},\
  }\bibfield  {title} {\bibinfo {title} {{On matrix method for black hole
  quasinormal modes}},\ }\href {https://doi.org/10.1088/1674-1137/43/3/035105}
  {\bibfield  {journal} {\bibinfo  {journal} {Chin. Phys. C}\ }\textbf
  {\bibinfo {volume} {43}},\ \bibinfo {pages} {035105} (\bibinfo {year}
  {2019})},\ \Eprint {https://arxiv.org/abs/1902.08352} {arXiv:1902.08352
  [gr-qc]} \BibitemShut {NoStop}%
\bibitem [{\citenamefont {Lin}\ and\ \citenamefont {Qian}(2023)}]{Lin:2022ynv}%
  \BibitemOpen
  \bibfield  {author} {\bibinfo {author} {\bibfnamefont {K.}~\bibnamefont
  {Lin}}\ and\ \bibinfo {author} {\bibfnamefont {W.-L.}\ \bibnamefont {Qian}},\
  }\bibfield  {title} {\bibinfo {title} {{High-order matrix method with
  delimited expansion domain}},\ }\href
  {https://doi.org/10.1088/1361-6382/acc50f} {\bibfield  {journal} {\bibinfo
  {journal} {Class. Quant. Grav.}\ }\textbf {\bibinfo {volume} {40}},\ \bibinfo
  {pages} {085019} (\bibinfo {year} {2023})},\ \Eprint
  {https://arxiv.org/abs/2209.11612} {arXiv:2209.11612 [gr-qc]} \BibitemShut
  {NoStop}%
\bibitem [{\citenamefont {Shen}\ \emph {et~al.}(2022)\citenamefont {Shen},
  \citenamefont {Qian}, \citenamefont {Lin}, \citenamefont {Shao},\ and\
  \citenamefont {Pan}}]{Shen:2022xdp}%
  \BibitemOpen
  \bibfield  {author} {\bibinfo {author} {\bibfnamefont {S.-F.}\ \bibnamefont
  {Shen}}, \bibinfo {author} {\bibfnamefont {W.-L.}\ \bibnamefont {Qian}},
  \bibinfo {author} {\bibfnamefont {K.}~\bibnamefont {Lin}}, \bibinfo {author}
  {\bibfnamefont {C.-G.}\ \bibnamefont {Shao}},\ and\ \bibinfo {author}
  {\bibfnamefont {Y.}~\bibnamefont {Pan}},\ }\bibfield  {title} {\bibinfo
  {title} {{Matrix method for perturbed black hole metric with
  discontinuity}},\ }\href {https://doi.org/10.1088/1361-6382/ac95f1}
  {\bibfield  {journal} {\bibinfo  {journal} {Class. Quant. Grav.}\ }\textbf
  {\bibinfo {volume} {39}},\ \bibinfo {pages} {225004} (\bibinfo {year}
  {2022})},\ \Eprint {https://arxiv.org/abs/2203.14320} {arXiv:2203.14320
  [gr-qc]} \BibitemShut {NoStop}%
\bibitem [{\citenamefont {Lei}\ \emph {et~al.}(2021)\citenamefont {Lei},
  \citenamefont {Wang},\ and\ \citenamefont {Jing}}]{Lei2021}%
  \BibitemOpen
  \bibfield  {author} {\bibinfo {author} {\bibfnamefont {Y.}~\bibnamefont
  {Lei}}, \bibinfo {author} {\bibfnamefont {M.}~\bibnamefont {Wang}},\ and\
  \bibinfo {author} {\bibfnamefont {J.}~\bibnamefont {Jing}},\ }\bibfield
  {title} {\bibinfo {title} {{Maxwell perturbations in a cavity with Robin
  boundary conditions: two branches of modes with spectrum bifurcation on
  Schwarzschild black holes}},\ }\href
  {https://doi.org/10.1140/epjc/s10052-021-09942-8} {\bibfield  {journal}
  {\bibinfo  {journal} {Eur. Phys. J. C}\ }\textbf {\bibinfo {volume} {81}},\
  \bibinfo {pages} {1129} (\bibinfo {year} {2021})},\ \Eprint
  {https://arxiv.org/abs/2108.04146} {arXiv:2108.04146 [gr-qc]} \BibitemShut
  {NoStop}%
\bibitem [{\citenamefont {Liu}\ \emph {et~al.}(2024{\natexlab{c}})\citenamefont
  {Liu}, \citenamefont {Fang}, \citenamefont {Jing},\ and\ \citenamefont
  {Wang}}]{Liu:2024oeq}%
  \BibitemOpen
  \bibfield  {author} {\bibinfo {author} {\bibfnamefont {W.}~\bibnamefont
  {Liu}}, \bibinfo {author} {\bibfnamefont {X.}~\bibnamefont {Fang}}, \bibinfo
  {author} {\bibfnamefont {J.}~\bibnamefont {Jing}},\ and\ \bibinfo {author}
  {\bibfnamefont {J.}~\bibnamefont {Wang}},\ }\bibfield  {title} {\bibinfo
  {title} {{Lorentz violation induces isospectrality breaking in
  Einstein-bumblebee gravity theory}},\ }\href
  {https://doi.org/10.1007/s11433-024-2405-y} {\bibfield  {journal} {\bibinfo
  {journal} {Sci. China Phys. Mech. Astron.}\ }\textbf {\bibinfo {volume}
  {67}},\ \bibinfo {pages} {280413} (\bibinfo {year} {2024}{\natexlab{c}})},\
  \Eprint {https://arxiv.org/abs/2402.09686} {arXiv:2402.09686 [gr-qc]}
  \BibitemShut {NoStop}%
\bibitem [{\citenamefont {Leaver}(1985)}]{Leaver:1985ax}%
  \BibitemOpen
  \bibfield  {author} {\bibinfo {author} {\bibfnamefont {E.~W.}\ \bibnamefont
  {Leaver}},\ }\bibfield  {title} {\bibinfo {title} {{An Analytic
  representation for the quasi normal modes of Kerr black holes}},\ }\href
  {https://doi.org/10.1098/rspa.1985.0119} {\bibfield  {journal} {\bibinfo
  {journal} {Proc. Roy. Soc. Lond. A}\ }\textbf {\bibinfo {volume} {402}},\
  \bibinfo {pages} {285} (\bibinfo {year} {1985})}\BibitemShut {NoStop}%
\bibitem [{\citenamefont {Leaver}(1990)}]{Leaver:1990zz}%
  \BibitemOpen
  \bibfield  {author} {\bibinfo {author} {\bibfnamefont {E.~W.}\ \bibnamefont
  {Leaver}},\ }\bibfield  {title} {\bibinfo {title} {{Quasinormal modes of
  Reissner-Nordstrom black holes}},\ }\href
  {https://doi.org/10.1103/PhysRevD.41.2986} {\bibfield  {journal} {\bibinfo
  {journal} {Phys. Rev. D}\ }\textbf {\bibinfo {volume} {41}},\ \bibinfo
  {pages} {2986} (\bibinfo {year} {1990})}\BibitemShut {NoStop}%
\bibitem [{\citenamefont {Percival}\ and\ \citenamefont
  {Dolan}(2020)}]{Percival:2020skc}%
  \BibitemOpen
  \bibfield  {author} {\bibinfo {author} {\bibfnamefont {J.}~\bibnamefont
  {Percival}}\ and\ \bibinfo {author} {\bibfnamefont {S.~R.}\ \bibnamefont
  {Dolan}},\ }\bibfield  {title} {\bibinfo {title} {{Quasinormal modes of
  massive vector fields on the Kerr spacetime}},\ }\href
  {https://doi.org/10.1103/PhysRevD.102.104055} {\bibfield  {journal} {\bibinfo
   {journal} {Phys. Rev. D}\ }\textbf {\bibinfo {volume} {102}},\ \bibinfo
  {pages} {104055} (\bibinfo {year} {2020})},\ \Eprint
  {https://arxiv.org/abs/2008.10621} {arXiv:2008.10621 [gr-qc]} \BibitemShut
  {NoStop}%
\bibitem [{\citenamefont {Guo}\ \emph {et~al.}(2023{\natexlab{b}})\citenamefont
  {Guo}, \citenamefont {Tan},\ and\ \citenamefont {Liu}}]{Guo:2022rms}%
  \BibitemOpen
  \bibfield  {author} {\bibinfo {author} {\bibfnamefont {W.-D.}\ \bibnamefont
  {Guo}}, \bibinfo {author} {\bibfnamefont {Q.}~\bibnamefont {Tan}},\ and\
  \bibinfo {author} {\bibfnamefont {Y.-X.}\ \bibnamefont {Liu}},\ }\bibfield
  {title} {\bibinfo {title} {{Gravitoelectromagnetic coupled perturbations and
  quasinormal modes of a charged black hole with scalar hair}},\ }\href
  {https://doi.org/10.1103/PhysRevD.107.124046} {\bibfield  {journal} {\bibinfo
   {journal} {Phys. Rev. D}\ }\textbf {\bibinfo {volume} {107}},\ \bibinfo
  {pages} {124046} (\bibinfo {year} {2023}{\natexlab{b}})},\ \Eprint
  {https://arxiv.org/abs/2212.08784} {arXiv:2212.08784 [gr-qc]} \BibitemShut
  {NoStop}%
\bibitem [{\citenamefont {Liu}\ \emph {et~al.}(2023{\natexlab{c}})\citenamefont
  {Liu}, \citenamefont {Fang}, \citenamefont {Jing},\ and\ \citenamefont
  {Wang}}]{Liu:2023uft}%
  \BibitemOpen
  \bibfield  {author} {\bibinfo {author} {\bibfnamefont {W.}~\bibnamefont
  {Liu}}, \bibinfo {author} {\bibfnamefont {X.}~\bibnamefont {Fang}}, \bibinfo
  {author} {\bibfnamefont {J.}~\bibnamefont {Jing}},\ and\ \bibinfo {author}
  {\bibfnamefont {J.}~\bibnamefont {Wang}},\ }\bibfield  {title} {\bibinfo
  {title} {{Gravito-electromagnetic perturbations of MOG black holes with a
  cosmological constant: quasinormal modes and ringdown waveforms}},\ }\href
  {https://doi.org/10.1088/1475-7516/2023/11/057} {\bibfield  {journal}
  {\bibinfo  {journal} {JCAP}\ }\textbf {\bibinfo {volume} {11}},\ \bibinfo
  {pages} {057}},\ \Eprint {https://arxiv.org/abs/2306.03599} {arXiv:2306.03599
  [gr-qc]} \BibitemShut {NoStop}%
\bibitem [{\citenamefont {Liu}\ \emph {et~al.}(2025{\natexlab{d}})\citenamefont
  {Liu}, \citenamefont {Liu}, \citenamefont {Wu},\ and\ \citenamefont
  {Liu}}]{Liu:2025wwq}%
  \BibitemOpen
  \bibfield  {author} {\bibinfo {author} {\bibfnamefont {W.}~\bibnamefont
  {Liu}}, \bibinfo {author} {\bibfnamefont {Y.}~\bibnamefont {Liu}}, \bibinfo
  {author} {\bibfnamefont {D.}~\bibnamefont {Wu}},\ and\ \bibinfo {author}
  {\bibfnamefont {Y.-X.}\ \bibnamefont {Liu}},\ }\bibfield  {title} {\bibinfo
  {title} {{A Universal Framework for Horizon-Scale Tests of Gravity with Black
  Hole Shadows}},\ }\href@noop {} {\  (\bibinfo {year} {2025}{\natexlab{d}})},\
  \Eprint {https://arxiv.org/abs/2511.06017} {arXiv:2511.06017 [gr-qc]}
  \BibitemShut {NoStop}%
\bibitem [{\citenamefont {Liu}\ \emph {et~al.}(2024{\natexlab{d}})\citenamefont
  {Liu}, \citenamefont {Wu},\ and\ \citenamefont {Wang}}]{Liu:2024soc}%
  \BibitemOpen
  \bibfield  {author} {\bibinfo {author} {\bibfnamefont {W.}~\bibnamefont
  {Liu}}, \bibinfo {author} {\bibfnamefont {D.}~\bibnamefont {Wu}},\ and\
  \bibinfo {author} {\bibfnamefont {J.}~\bibnamefont {Wang}},\ }\bibfield
  {title} {\bibinfo {title} {{Light rings and shadows of static black holes in
  effective quantum gravity}},\ }\href
  {https://doi.org/10.1016/j.physletb.2024.139052} {\bibfield  {journal}
  {\bibinfo  {journal} {Phys. Lett. B}\ }\textbf {\bibinfo {volume} {858}},\
  \bibinfo {pages} {139052} (\bibinfo {year} {2024}{\natexlab{d}})},\ \Eprint
  {https://arxiv.org/abs/2408.05569} {arXiv:2408.05569 [gr-qc]} \BibitemShut
  {NoStop}%
\bibitem [{\citenamefont {Zhang}\ \emph {et~al.}(2024)\citenamefont {Zhang},
  \citenamefont {Chen},\ and\ \citenamefont {Jing}}]{Zhang:2024jrw}%
  \BibitemOpen
  \bibfield  {author} {\bibinfo {author} {\bibfnamefont {Z.}~\bibnamefont
  {Zhang}}, \bibinfo {author} {\bibfnamefont {S.}~\bibnamefont {Chen}},\ and\
  \bibinfo {author} {\bibfnamefont {J.}~\bibnamefont {Jing}},\ }\bibfield
  {title} {\bibinfo {title} {{Images of Kerr-MOG black holes surrounded by
  geometrically thick magnetized equilibrium tori}},\ }\href@noop {} {\
  (\bibinfo {year} {2024})},\ \Eprint {https://arxiv.org/abs/2404.12223}
  {arXiv:2404.12223 [gr-qc]} \BibitemShut {NoStop}%
\bibitem [{\citenamefont {Zhao}\ \emph {et~al.}(2025)\citenamefont {Zhao},
  \citenamefont {Fan}, \citenamefont {Wang}, \citenamefont {Guo},\ and\
  \citenamefont {Chen}}]{Zhao:2025ouq}%
  \BibitemOpen
  \bibfield  {author} {\bibinfo {author} {\bibfnamefont {Z.}~\bibnamefont
  {Zhao}}, \bibinfo {author} {\bibfnamefont {Z.-Y.}\ \bibnamefont {Fan}},
  \bibinfo {author} {\bibfnamefont {X.}~\bibnamefont {Wang}}, \bibinfo {author}
  {\bibfnamefont {M.}~\bibnamefont {Guo}},\ and\ \bibinfo {author}
  {\bibfnamefont {B.}~\bibnamefont {Chen}},\ }\bibfield  {title} {\bibinfo
  {title} {{Probing the Penrose Process: Images of Split Hotspots and Their
  Observational Signatures}},\ }\href@noop {} {\  (\bibinfo {year} {2025})},\
  \Eprint {https://arxiv.org/abs/2510.27409} {arXiv:2510.27409 [astro-ph.HE]}
  \BibitemShut {NoStop}%
\bibitem [{\citenamefont {Wang}\ \emph {et~al.}(2025)\citenamefont {Wang},
  \citenamefont {Chen}, \citenamefont {Guo},\ and\ \citenamefont
  {Chen}}]{Wang:2025btn}%
  \BibitemOpen
  \bibfield  {author} {\bibinfo {author} {\bibfnamefont {X.}~\bibnamefont
  {Wang}}, \bibinfo {author} {\bibfnamefont {S.}~\bibnamefont {Chen}}, \bibinfo
  {author} {\bibfnamefont {M.}~\bibnamefont {Guo}},\ and\ \bibinfo {author}
  {\bibfnamefont {B.}~\bibnamefont {Chen}},\ }\bibfield  {title} {\bibinfo
  {title} {{Semi-analytical Study on the Polarized Images of Black Hole due to
  Frame Dragging}},\ }\href@noop {} {\  (\bibinfo {year} {2025})},\ \Eprint
  {https://arxiv.org/abs/2508.15178} {arXiv:2508.15178 [gr-qc]} \BibitemShut
  {NoStop}%
\bibitem [{\citenamefont {Liu}\ \emph {et~al.}(2025{\natexlab{e}})\citenamefont
  {Liu}, \citenamefont {Huang}, \citenamefont {Wu},\ and\ \citenamefont
  {Wang}}]{Liu:2025lwj}%
  \BibitemOpen
  \bibfield  {author} {\bibinfo {author} {\bibfnamefont {W.}~\bibnamefont
  {Liu}}, \bibinfo {author} {\bibfnamefont {H.}~\bibnamefont {Huang}}, \bibinfo
  {author} {\bibfnamefont {D.}~\bibnamefont {Wu}},\ and\ \bibinfo {author}
  {\bibfnamefont {J.}~\bibnamefont {Wang}},\ }\bibfield  {title} {\bibinfo
  {title} {{Lorentz violation signatures in the low-energy sector of
  Ho{\v{r}}ava gravity from black hole shadow observations}},\ }\href
  {https://doi.org/10.1016/j.physletb.2025.139812} {\bibfield  {journal}
  {\bibinfo  {journal} {Phys. Lett. B}\ }\textbf {\bibinfo {volume} {868}},\
  \bibinfo {pages} {139812} (\bibinfo {year} {2025}{\natexlab{e}})},\ \Eprint
  {https://arxiv.org/abs/2506.13504} {arXiv:2506.13504 [gr-qc]} \BibitemShut
  {NoStop}%
\bibitem [{\citenamefont {Wei}\ \emph {et~al.}(2022)\citenamefont {Wei},
  \citenamefont {Liu},\ and\ \citenamefont {Mann}}]{Wei:2022dzw}%
  \BibitemOpen
  \bibfield  {author} {\bibinfo {author} {\bibfnamefont {S.-W.}\ \bibnamefont
  {Wei}}, \bibinfo {author} {\bibfnamefont {Y.-X.}\ \bibnamefont {Liu}},\ and\
  \bibinfo {author} {\bibfnamefont {R.~B.}\ \bibnamefont {Mann}},\ }\bibfield
  {title} {\bibinfo {title} {{Black Hole Solutions as Topological Thermodynamic
  Defects}},\ }\href {https://doi.org/10.1103/PhysRevLett.129.191101}
  {\bibfield  {journal} {\bibinfo  {journal} {Phys. Rev. Lett.}\ }\textbf
  {\bibinfo {volume} {129}},\ \bibinfo {pages} {191101} (\bibinfo {year}
  {2022})},\ \Eprint {https://arxiv.org/abs/2208.01932} {arXiv:2208.01932
  [gr-qc]} \BibitemShut {NoStop}%
\bibitem [{\citenamefont {Wei}\ and\ \citenamefont {Liu}(2022)}]{Wei:2021vdx}%
  \BibitemOpen
  \bibfield  {author} {\bibinfo {author} {\bibfnamefont {S.-W.}\ \bibnamefont
  {Wei}}\ and\ \bibinfo {author} {\bibfnamefont {Y.-X.}\ \bibnamefont {Liu}},\
  }\bibfield  {title} {\bibinfo {title} {{Topology of black hole
  thermodynamics}},\ }\href {https://doi.org/10.1103/PhysRevD.105.104003}
  {\bibfield  {journal} {\bibinfo  {journal} {Phys. Rev. D}\ }\textbf {\bibinfo
  {volume} {105}},\ \bibinfo {pages} {104003} (\bibinfo {year} {2022})},\
  \Eprint {https://arxiv.org/abs/2112.01706} {arXiv:2112.01706 [gr-qc]}
  \BibitemShut {NoStop}%
\bibitem [{\citenamefont {Wu}(2023{\natexlab{a}})}]{Wu:2023fcw}%
  \BibitemOpen
  \bibfield  {author} {\bibinfo {author} {\bibfnamefont {D.}~\bibnamefont
  {Wu}},\ }\bibfield  {title} {\bibinfo {title} {{Consistent thermodynamics and
  topological classes for the four-dimensional Lorentzian charged Taub-NUT
  spacetimes}},\ }\href {https://doi.org/10.1140/epjc/s10052-023-11782-7}
  {\bibfield  {journal} {\bibinfo  {journal} {Eur. Phys. J. C}\ }\textbf
  {\bibinfo {volume} {83}},\ \bibinfo {pages} {589} (\bibinfo {year}
  {2023}{\natexlab{a}})},\ \Eprint {https://arxiv.org/abs/2306.02324}
  {arXiv:2306.02324 [gr-qc]} \BibitemShut {NoStop}%
\bibitem [{\citenamefont {Wu}(2023{\natexlab{b}})}]{Wu:2023meo}%
  \BibitemOpen
  \bibfield  {author} {\bibinfo {author} {\bibfnamefont {D.}~\bibnamefont
  {Wu}},\ }\bibfield  {title} {\bibinfo {title} {{Topological classes of
  thermodynamics of the four-dimensional static accelerating black holes}},\
  }\href {https://doi.org/10.1103/PhysRevD.108.084041} {\bibfield  {journal}
  {\bibinfo  {journal} {Phys. Rev. D}\ }\textbf {\bibinfo {volume} {108}},\
  \bibinfo {pages} {084041} (\bibinfo {year} {2023}{\natexlab{b}})},\ \Eprint
  {https://arxiv.org/abs/2307.02030} {arXiv:2307.02030 [hep-th]} \BibitemShut
  {NoStop}%
\bibitem [{\citenamefont {Wu}\ \emph {et~al.}(2024{\natexlab{a}})\citenamefont
  {Wu}, \citenamefont {Gu}, \citenamefont {Zhu}, \citenamefont {Jiang},\ and\
  \citenamefont {Yang}}]{Wu:2024rmv}%
  \BibitemOpen
  \bibfield  {author} {\bibinfo {author} {\bibfnamefont {D.}~\bibnamefont
  {Wu}}, \bibinfo {author} {\bibfnamefont {S.-Y.}\ \bibnamefont {Gu}}, \bibinfo
  {author} {\bibfnamefont {X.-D.}\ \bibnamefont {Zhu}}, \bibinfo {author}
  {\bibfnamefont {Q.-Q.}\ \bibnamefont {Jiang}},\ and\ \bibinfo {author}
  {\bibfnamefont {S.-Z.}\ \bibnamefont {Yang}},\ }\bibfield  {title} {\bibinfo
  {title} {{Topological classes of thermodynamics of the static multi-charge
  AdS black holes in gauged supergravities}},\ }\href@noop {} {\  (\bibinfo
  {year} {2024}{\natexlab{a}})},\ \Eprint {https://arxiv.org/abs/2402.00106}
  {arXiv:2402.00106 [hep-th]} \BibitemShut {NoStop}%
\bibitem [{\citenamefont {Wu}\ \emph {et~al.}(2025{\natexlab{a}})\citenamefont
  {Wu}, \citenamefont {Liu}, \citenamefont {Wu},\ and\ \citenamefont
  {Mann}}]{Wu:2024asq}%
  \BibitemOpen
  \bibfield  {author} {\bibinfo {author} {\bibfnamefont {D.}~\bibnamefont
  {Wu}}, \bibinfo {author} {\bibfnamefont {W.}~\bibnamefont {Liu}}, \bibinfo
  {author} {\bibfnamefont {S.-Q.}\ \bibnamefont {Wu}},\ and\ \bibinfo {author}
  {\bibfnamefont {R.~B.}\ \bibnamefont {Mann}},\ }\bibfield  {title} {\bibinfo
  {title} {{Novel topological classes in black hole thermodynamics}},\ }\href
  {https://doi.org/10.1103/PhysRevD.111.L061501} {\bibfield  {journal}
  {\bibinfo  {journal} {Phys. Rev. D}\ }\textbf {\bibinfo {volume} {111}},\
  \bibinfo {pages} {L061501} (\bibinfo {year} {2025}{\natexlab{a}})},\ \Eprint
  {https://arxiv.org/abs/2411.10102} {arXiv:2411.10102 [hep-th]} \BibitemShut
  {NoStop}%
\bibitem [{\citenamefont {Zhu}\ \emph {et~al.}(2025)\citenamefont {Zhu},
  \citenamefont {Liu},\ and\ \citenamefont {Wu}}]{Zhu:2024zcl}%
  \BibitemOpen
  \bibfield  {author} {\bibinfo {author} {\bibfnamefont {X.-D.}\ \bibnamefont
  {Zhu}}, \bibinfo {author} {\bibfnamefont {W.}~\bibnamefont {Liu}},\ and\
  \bibinfo {author} {\bibfnamefont {D.}~\bibnamefont {Wu}},\ }\bibfield
  {title} {\bibinfo {title} {{Universal thermodynamic topological classes of
  rotating black holes}},\ }\href
  {https://doi.org/10.1016/j.physletb.2024.139163} {\bibfield  {journal}
  {\bibinfo  {journal} {Phys. Lett. B}\ }\textbf {\bibinfo {volume} {860}},\
  \bibinfo {pages} {139163} (\bibinfo {year} {2025})},\ \Eprint
  {https://arxiv.org/abs/2409.12747} {arXiv:2409.12747 [hep-th]} \BibitemShut
  {NoStop}%
\bibitem [{\citenamefont {Liu}\ \emph {et~al.}(2025{\natexlab{f}})\citenamefont
  {Liu}, \citenamefont {Zhang}, \citenamefont {Wu},\ and\ \citenamefont
  {Wang}}]{Liu:2025iyl}%
  \BibitemOpen
  \bibfield  {author} {\bibinfo {author} {\bibfnamefont {W.}~\bibnamefont
  {Liu}}, \bibinfo {author} {\bibfnamefont {L.}~\bibnamefont {Zhang}}, \bibinfo
  {author} {\bibfnamefont {D.}~\bibnamefont {Wu}},\ and\ \bibinfo {author}
  {\bibfnamefont {J.}~\bibnamefont {Wang}},\ }\bibfield  {title} {\bibinfo
  {title} {{Thermodynamic topological classes of the rotating, accelerating
  black holes}},\ }\href {https://doi.org/10.1088/1361-6382/ade35b} {\bibfield
  {journal} {\bibinfo  {journal} {Class. Quant. Grav.}\ }\textbf {\bibinfo
  {volume} {42}},\ \bibinfo {pages} {125007} (\bibinfo {year}
  {2025}{\natexlab{f}})},\ \Eprint {https://arxiv.org/abs/2409.11666}
  {arXiv:2409.11666 [hep-th]} \BibitemShut {NoStop}%
\bibitem [{\citenamefont {Fuentes-Schuller}\ and\ \citenamefont
  {Mann}(2005)}]{Fuentes-Schuller:2004iaz}%
  \BibitemOpen
  \bibfield  {author} {\bibinfo {author} {\bibfnamefont {I.}~\bibnamefont
  {Fuentes-Schuller}}\ and\ \bibinfo {author} {\bibfnamefont {R.~B.}\
  \bibnamefont {Mann}},\ }\bibfield  {title} {\bibinfo {title} {{Alice falls
  into a black hole: Entanglement in non-inertial frames}},\ }\href
  {https://doi.org/10.1103/PhysRevLett.95.120404} {\bibfield  {journal}
  {\bibinfo  {journal} {Phys. Rev. Lett.}\ }\textbf {\bibinfo {volume} {95}},\
  \bibinfo {pages} {120404} (\bibinfo {year} {2005})},\ \Eprint
  {https://arxiv.org/abs/0410172} {arXiv:0410172 [quant-ph]} \BibitemShut
  {NoStop}%
\bibitem [{\citenamefont {Liu}\ \emph {et~al.}(2025{\natexlab{g}})\citenamefont
  {Liu}, \citenamefont {Wen},\ and\ \citenamefont {Wang}}]{Liu:2024wpa}%
  \BibitemOpen
  \bibfield  {author} {\bibinfo {author} {\bibfnamefont {W.}~\bibnamefont
  {Liu}}, \bibinfo {author} {\bibfnamefont {C.}~\bibnamefont {Wen}},\ and\
  \bibinfo {author} {\bibfnamefont {J.}~\bibnamefont {Wang}},\ }\bibfield
  {title} {\bibinfo {title} {{Lorentz violation alleviates gravitationally
  induced entanglement degradation}},\ }\href
  {https://doi.org/10.1007/JHEP01(2025)184} {\bibfield  {journal} {\bibinfo
  {journal} {JHEP}\ }\textbf {\bibinfo {volume} {01}},\ \bibinfo {pages}
  {184}},\ \Eprint {https://arxiv.org/abs/2410.21681} {arXiv:2410.21681
  [gr-qc]} \BibitemShut {NoStop}%
\bibitem [{\citenamefont {Liu}\ \emph {et~al.}(2025{\natexlab{h}})\citenamefont
  {Liu}, \citenamefont {Liu}, \citenamefont {Liu},\ and\ \citenamefont
  {Wang}}]{Liu:2025bpp}%
  \BibitemOpen
  \bibfield  {author} {\bibinfo {author} {\bibfnamefont {X.}~\bibnamefont
  {Liu}}, \bibinfo {author} {\bibfnamefont {W.}~\bibnamefont {Liu}}, \bibinfo
  {author} {\bibfnamefont {Z.}~\bibnamefont {Liu}},\ and\ \bibinfo {author}
  {\bibfnamefont {J.}~\bibnamefont {Wang}},\ }\bibfield  {title} {\bibinfo
  {title} {{Harvesting correlations from BTZ black hole coupled to a
  Lorentz-violating vector field}},\ }\href
  {https://doi.org/10.1007/JHEP08(2025)094} {\bibfield  {journal} {\bibinfo
  {journal} {JHEP}\ }\textbf {\bibinfo {volume} {08}},\ \bibinfo {pages}
  {094}},\ \Eprint {https://arxiv.org/abs/2503.06404} {arXiv:2503.06404
  [gr-qc]} \BibitemShut {NoStop}%
\bibitem [{\citenamefont {Wu}\ and\ \citenamefont {Zeng}(2022)}]{Wu:2022xwy}%
  \BibitemOpen
  \bibfield  {author} {\bibinfo {author} {\bibfnamefont {S.~M.}\ \bibnamefont
  {Wu}}\ and\ \bibinfo {author} {\bibfnamefont {H.~S.}\ \bibnamefont {Zeng}},\
  }\bibfield  {title} {\bibinfo {title} {{Genuine tripartite nonlocality and
  entanglement in curved spacetime}},\ }\href
  {https://doi.org/10.1140/epjc/s10052-021-09954-4} {\bibfield  {journal}
  {\bibinfo  {journal} {Eur. Phys. J. C}\ }\textbf {\bibinfo {volume} {82}},\
  \bibinfo {pages} {4} (\bibinfo {year} {2022})},\ \Eprint
  {https://arxiv.org/abs/2201.02333} {arXiv:2201.02333 [quant-ph]} \BibitemShut
  {NoStop}%
\bibitem [{\citenamefont {Wu}\ \emph {et~al.}(2024{\natexlab{b}})\citenamefont
  {Wu}, \citenamefont {Teng}, \citenamefont {Li}, \citenamefont {Li},
  \citenamefont {Liu},\ and\ \citenamefont {Wang}}]{Wu:2023spa}%
  \BibitemOpen
  \bibfield  {author} {\bibinfo {author} {\bibfnamefont {S.~M.}\ \bibnamefont
  {Wu}}, \bibinfo {author} {\bibfnamefont {X.~W.}\ \bibnamefont {Teng}},
  \bibinfo {author} {\bibfnamefont {J.~X.}\ \bibnamefont {Li}}, \bibinfo
  {author} {\bibfnamefont {S.~H.}\ \bibnamefont {Li}}, \bibinfo {author}
  {\bibfnamefont {T.~H.}\ \bibnamefont {Liu}},\ and\ \bibinfo {author}
  {\bibfnamefont {J.}~\bibnamefont {Wang}},\ }\bibfield  {title} {\bibinfo
  {title} {{Genuinely accessible and inaccessible entanglement in Schwarzschild
  black hole}},\ }\href {https://doi.org/10.1016/j.physletb.2023.138334}
  {\bibfield  {journal} {\bibinfo  {journal} {Phys. Lett. B}\ }\textbf
  {\bibinfo {volume} {848}},\ \bibinfo {pages} {138334} (\bibinfo {year}
  {2024}{\natexlab{b}})},\ \Eprint {https://arxiv.org/abs/2311.12362}
  {arXiv:2311.12362 [gr-qc]} \BibitemShut {NoStop}%
\bibitem [{\citenamefont {Wu}\ \emph {et~al.}(2023)\citenamefont {Wu},
  \citenamefont {Fan}, \citenamefont {Wang}, \citenamefont {Wu}, \citenamefont
  {Huang},\ and\ \citenamefont {Zeng}}]{Wu:2023sye}%
  \BibitemOpen
  \bibfield  {author} {\bibinfo {author} {\bibfnamefont {S.~M.}\ \bibnamefont
  {Wu}}, \bibinfo {author} {\bibfnamefont {X.~W.}\ \bibnamefont {Fan}},
  \bibinfo {author} {\bibfnamefont {R.~D.}\ \bibnamefont {Wang}}, \bibinfo
  {author} {\bibfnamefont {H.~Y.}\ \bibnamefont {Wu}}, \bibinfo {author}
  {\bibfnamefont {X.~L.}\ \bibnamefont {Huang}},\ and\ \bibinfo {author}
  {\bibfnamefont {H.~S.}\ \bibnamefont {Zeng}},\ }\bibfield  {title} {\bibinfo
  {title} {{Does Hawking effect always degrade fidelity of quantum
  teleportation in Schwarzschild spacetime?}},\ }\href
  {https://doi.org/10.1007/JHEP11(2023)232} {\bibfield  {journal} {\bibinfo
  {journal} {JHEP}\ }\textbf {\bibinfo {volume} {11}},\ \bibinfo {pages}
  {232}},\ \Eprint {https://arxiv.org/abs/2304.00984} {arXiv:2304.00984
  [gr-qc]} \BibitemShut {NoStop}%
\bibitem [{\citenamefont {Wu}\ \emph {et~al.}(2025{\natexlab{b}})\citenamefont
  {Wu}, \citenamefont {Teng}, \citenamefont {Li}, \citenamefont {Wang},\ and\
  \citenamefont {Lu}}]{Wu:2025euf}%
  \BibitemOpen
  \bibfield  {author} {\bibinfo {author} {\bibfnamefont {S.-M.}\ \bibnamefont
  {Wu}}, \bibinfo {author} {\bibfnamefont {X.-W.}\ \bibnamefont {Teng}},
  \bibinfo {author} {\bibfnamefont {W.-M.}\ \bibnamefont {Li}}, \bibinfo
  {author} {\bibfnamefont {Y.-X.}\ \bibnamefont {Wang}},\ and\ \bibinfo
  {author} {\bibfnamefont {J.}~\bibnamefont {Lu}},\ }\bibfield  {title}
  {\bibinfo {title} {{Nonseparability of multipartite systems in dilaton
  black~hole}},\ }\href {https://doi.org/10.1088/1475-7516/2025/09/030}
  {\bibfield  {journal} {\bibinfo  {journal} {JCAP}\ }\textbf {\bibinfo
  {volume} {09}},\ \bibinfo {pages} {030}},\ \Eprint
  {https://arxiv.org/abs/2503.17923} {arXiv:2503.17923 [gr-qc]} \BibitemShut
  {NoStop}%
\bibitem [{\citenamefont {Li}\ \emph {et~al.}(2025{\natexlab{c}})\citenamefont
  {Li}, \citenamefont {Lu},\ and\ \citenamefont {Wu}}]{Li:2025jlu}%
  \BibitemOpen
  \bibfield  {author} {\bibinfo {author} {\bibfnamefont {W.-M.}\ \bibnamefont
  {Li}}, \bibinfo {author} {\bibfnamefont {J.}~\bibnamefont {Lu}},\ and\
  \bibinfo {author} {\bibfnamefont {S.-M.}\ \bibnamefont {Wu}},\ }\bibfield
  {title} {\bibinfo {title} {{Multiqubit coherence of mixed states near event
  horizon}},\ }\href@noop {} {\  (\bibinfo {year} {2025}{\natexlab{c}})},\
  \Eprint {https://arxiv.org/abs/2505.07476} {arXiv:2505.07476 [gr-qc]}
  \BibitemShut {NoStop}%
\bibitem [{\citenamefont {Liu}\ \emph {et~al.}(2025{\natexlab{i}})\citenamefont
  {Liu}, \citenamefont {Liu},\ and\ \citenamefont {Wu}}]{Liu:2025hcx}%
  \BibitemOpen
  \bibfield  {author} {\bibinfo {author} {\bibfnamefont {X.}~\bibnamefont
  {Liu}}, \bibinfo {author} {\bibfnamefont {W.}~\bibnamefont {Liu}},\ and\
  \bibinfo {author} {\bibfnamefont {S.-M.}\ \bibnamefont {Wu}},\ }\bibfield
  {title} {\bibinfo {title} {{Entanglement degradation of static black holes in
  effective quantum gravity}},\ }\href@noop {} {\  (\bibinfo {year}
  {2025}{\natexlab{i}})},\ \Eprint {https://arxiv.org/abs/2511.12245}
  {arXiv:2511.12245 [gr-qc]} \BibitemShut {NoStop}%
\bibitem [{\citenamefont {Liu}\ \emph {et~al.}(2025{\natexlab{j}})\citenamefont
  {Liu}, \citenamefont {Tian},\ and\ \citenamefont {Jing}}]{Liu:2025bzv}%
  \BibitemOpen
  \bibfield  {author} {\bibinfo {author} {\bibfnamefont {X.}~\bibnamefont
  {Liu}}, \bibinfo {author} {\bibfnamefont {Z.}~\bibnamefont {Tian}},\ and\
  \bibinfo {author} {\bibfnamefont {J.}~\bibnamefont {Jing}},\ }\bibfield
  {title} {\bibinfo {title} {{Dissipation suppression for an Unruh-DeWitt
  battery with a reflecting boundary}},\ }\href
  {https://doi.org/10.1007/s11433-025-2723-7} {\bibfield  {journal} {\bibinfo
  {journal} {Sci. China Phys. Mech. Astron.}\ }\textbf {\bibinfo {volume}
  {68}},\ \bibinfo {pages} {100412} (\bibinfo {year} {2025}{\natexlab{j}})},\
  \Eprint {https://arxiv.org/abs/2509.00875} {arXiv:2509.00875 [hep-th]}
  \BibitemShut {NoStop}%
\bibitem [{\citenamefont {Tang}\ \emph
  {et~al.}(2025{\natexlab{a}})\citenamefont {Tang}, \citenamefont {Liu},\ and\
  \citenamefont {Wang}}]{Tang:2025eew}%
  \BibitemOpen
  \bibfield  {author} {\bibinfo {author} {\bibfnamefont {Y.}~\bibnamefont
  {Tang}}, \bibinfo {author} {\bibfnamefont {W.}~\bibnamefont {Liu}},\ and\
  \bibinfo {author} {\bibfnamefont {J.}~\bibnamefont {Wang}},\ }\bibfield
  {title} {\bibinfo {title} {{Observational signature of Lorentz violation in
  acceleration radiation}},\ }\href
  {https://doi.org/10.1140/epjc/s10052-025-14797-4} {\bibfield  {journal}
  {\bibinfo  {journal} {Eur. Phys. J. C}\ }\textbf {\bibinfo {volume} {85}},\
  \bibinfo {pages} {1108} (\bibinfo {year} {2025}{\natexlab{a}})},\ \Eprint
  {https://arxiv.org/abs/2502.03043} {arXiv:2502.03043 [gr-qc]} \BibitemShut
  {NoStop}%
\bibitem [{\citenamefont {Tang}\ \emph
  {et~al.}(2025{\natexlab{b}})\citenamefont {Tang}, \citenamefont {Liu},
  \citenamefont {Liu},\ and\ \citenamefont {Wang}}]{Tang:2025mtc}%
  \BibitemOpen
  \bibfield  {author} {\bibinfo {author} {\bibfnamefont {Y.}~\bibnamefont
  {Tang}}, \bibinfo {author} {\bibfnamefont {W.}~\bibnamefont {Liu}}, \bibinfo
  {author} {\bibfnamefont {Z.}~\bibnamefont {Liu}},\ and\ \bibinfo {author}
  {\bibfnamefont {J.}~\bibnamefont {Wang}},\ }\bibfield  {title} {\bibinfo
  {title} {{Can the latent signatures of quantum superposition be detected
  through correlation harvesting?}},\ }\href@noop {} {\  (\bibinfo {year}
  {2025}{\natexlab{b}})},\ \Eprint {https://arxiv.org/abs/2508.00292}
  {arXiv:2508.00292 [gr-qc]} \BibitemShut {NoStop}%
\end{thebibliography}

%

\end{document}